%% file: main.tex
\documentclass[12pt]{mitthesis}
\usepackage{lgrind}
\usepackage{amsmath}
\usepackage{graphicx}
  \PassOptionsToPackage{linktocpage}{hyperref}
  \usepackage[numbers]{natbib}
  \usepackage[hyperindex,breaklinks]{hyperref}
\usepackage[a4paper, top=2.5cm, bottom=2.5cm, left=3.5cm, right=2cm]{geometry} 
\hypersetup{linktocpage}
\begin{document}
\include{titulpage}
\include{cover}
\pagestyle{plain}
 \singlespace
\tableofcontents
\newpage
\listoffigures
\newpage
\listoftables

\include{Chapter1}

\include{Chapter4}
\include{Chapter3.1}
\include{Chapter5}
\include{Chapter6}
\include{Chapter3}

\include{Chapter7}
\include{resumeSK}
\appendix
\include{appa}
\bibliographystyle{plain}
\bibliography{references}

\end{document}

%% file: titulpage.tex
\thispagestyle{empty}
    \begin{center}
    \
    {\Large\fontshape{sc}\selectfont \textbf{Comenius University in Bratislava\\Faculty of Mathematics, Physics and Informatics}}
    \vskip 1.5cm

\vskip 1.5cm
\begin{figure}[h!]
\centering
\includegraphics[width=4.5cm,height=4.5cm]{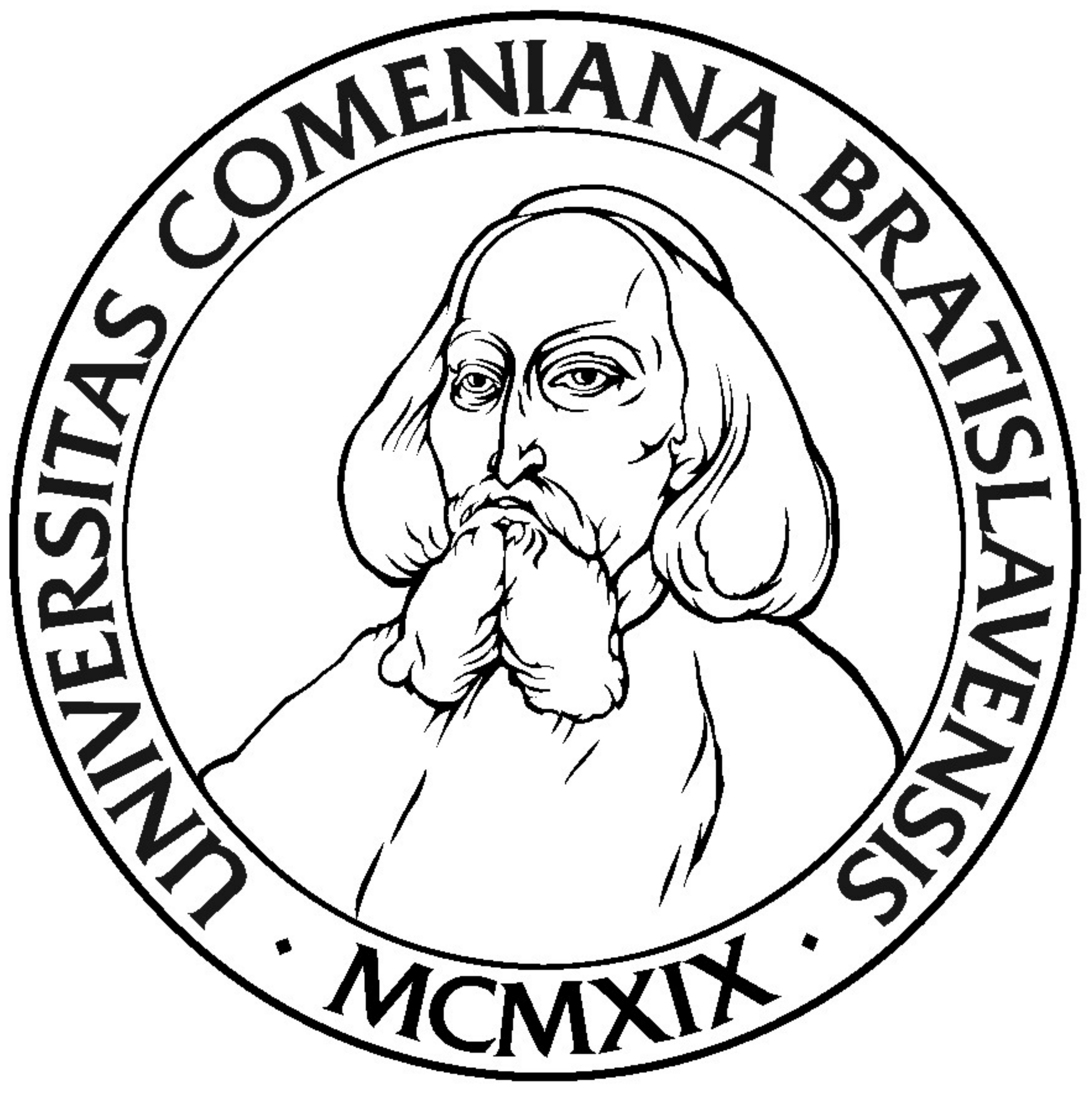}
\end{figure} 
    \vskip 0.3 cm
    {\Huge \textbf{Approaches of near stars to the Sun}}
    \vskip 1.7cm
    \centering {\large \textbf{Master Thesis}}
    \end{center}
    \vskip 7.7cm
    {\Large \textbf{Bratislava 2011}} \hskip 3cm {\Large \textbf{Bc. Jorge Luis Cayao D\'{i}az}}

\newpage
\thispagestyle{empty}
    \begin{center}
    {\Large\fontshape{sc}\selectfont \textbf{Comenius University in Bratislava\\Faculty of Mathematics, Physics and Informatics}}
    \vskip 0.7cm
    \vskip 1cm
  Reference number: 5a178088-2127-4be8-90ad-c41906d8c9ed

    \vskip 0.8cm
\begin{figure}[h!]
\centering
\includegraphics[width=3.8cm,height=3.8cm]{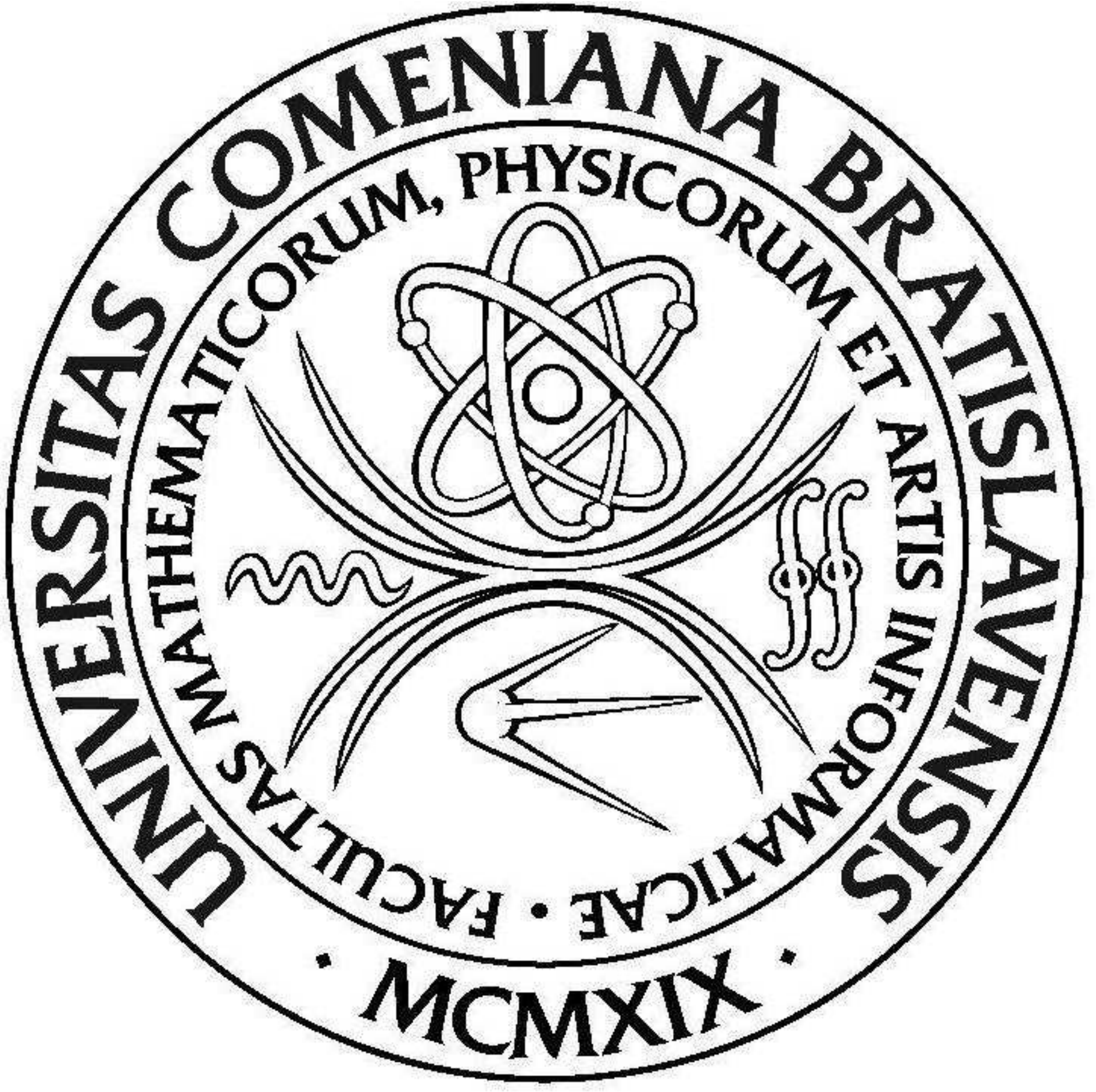}
\end{figure} 
    \Huge\bf{Approaches of near stars to the Sun}\\
 \vskip 1cm

\Large\rm{Master Thesis}
\end{center}
 \vskip 2.5cm
\begin{center}
   \large{Department of Astronomy, Physics of the Earth and Metereology} \\
  \vskip 0.25cm
    Branch of study: Physics\\
  \vskip 0.25cm
    Study programme: 4.1.7 Astronomy and 4.1.8 Astrophysics\\
  \vskip 0.25cm
    Thesis Supervisor: Doc. RNDr. Jozef Kla\v{c}ka, PhD.
\end{center}
     \vskip 0.5cm
    {\Large \textbf{Bratislava 2011}} \hskip 2cm {\Large \textbf{Bc. Jorge Luis Cayao D\'{i}az}}

\newpage
\thispagestyle{empty}
\vskip 3cm
   \begin{figure*}[htbp]
\centering
\includegraphics[width=20cm,height=25cm]{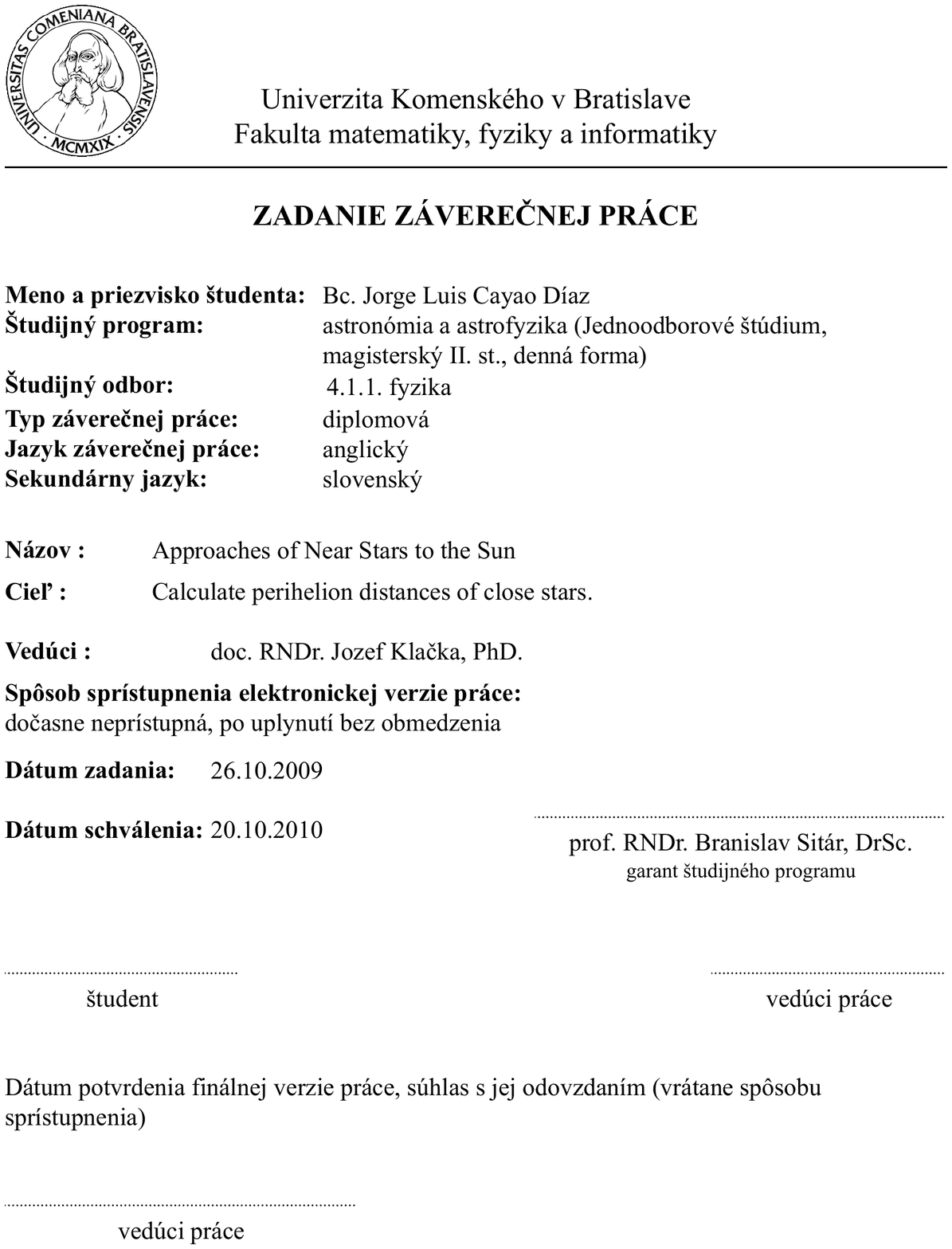}
\end{figure*} 

%% file: cover.tex
%
%
%
%
%
%
%
\title{Approaches of near stars to the Sun}

\author{Jorge Luis Cayao D\'{i}az}
\department{Department of Astronomy, Physics of the Earth and Meteorology}
\degree{Master in Physics, Specialization: Astronomy and Astrophysics}

\degreemonth{June}
\degreeyear{2011}
\thesisdate{May 6, 2011}

\supervisor{Jozef Kla\v{c}ka}{Assoc. Professor of Physics\\ Department of Astronomy, Physics of the Earth and Metereology}

\chairman{ department committee chairman}
 {Professor of Physics\\ Division of Astronomy and Astrophysics}
\maketitle

\newpage
\pagestyle{empty}
\section*{}
\bigskip
\bigskip
\bigskip
\bigskip
\bigskip
\bigskip
\bigskip
\bigskip
\bigskip
\bigskip
\bigskip
\bigskip
\bigskip
\bigskip
\bigskip
\bigskip
\bigskip
\bigskip

I declare that the submitted work is my own under the careful supervision of my adviser  and that I have not used any other than permitted reference
sources or materials nor engaged in any plagiarism. All references and
other sources used by me have been appropriately acknowledged in the work.
\bigskip
\bigskip
\bigskip
\bigskip
\bigskip
\bigskip
\bigskip
\bigskip
\bigskip
\bigskip

\hfill\hfill \quad Jorge Luis Cayao  D\'{i}az



\newpage
\setcounter{savepage}{\thepage}
\pagestyle{empty}
\begin{abstractpage}
\input{abstract}

{\bf Keywords: } perihelion distance, impact parameter, Oort cloud, solar motion
\end{abstractpage}

\setcounter{savepage}{\thepage}
\pagestyle{empty}
\begin{abstraktpage}
\input{abstrakt}
\\
\hspace*{0.25in}{\bf K\v{l}\'{u}\v{c}ov\'{e} slov\'{a}:} perih\'{e}liov\'{a} vzdialenos\v{t}, 
impakt parameter -- z\'{a}mern\'{a} vzdialenos\v{t}, Oortov oblak, pohyb Slnka
\\\\
\v{S}kolitel: Jozef Kla\v{c}ka
\\
Titul: Docent fyziky, Katedra astron\'{o}mie, fyziky Zeme a meteorol\'{o}gie
\end{abstraktpage}

\setcounter{savepage}{\thepage}
\pagestyle{empty}
\begin{resumenpage}
\input{resumen}
\\
\hspace*{0.25in}{\bf Palabras clave: } Distancia del perihelio, par\'{a}metro de impacto, nube de Oort, movimiento solar.
\\\\
Supervisor de tesis: Jozef Kla\v{c}ka, Profesor Asociado de f\'{i}sica, Departamento de Astronom\'{i}a, F\'{i}sica de la Tierra y Meteorolog\'{i}a
\end{resumenpage}
\newpage
\newpage
\pagestyle{empty}
\section*{Acknowledgments}
I gratefully acknowledge financial support from the National Scholarship Programme of the Slovak Republic.
I was not alone on my journey to finishing this Master thesis. First, my thanks go to my parents gave me their love, support and freedom.  
 Second, I would like to express my gratitude to  Doc. Jozef Kla\v{c}ka Ph.D. for his invalueable help and support.
Without his continual assistance this work would not get much farther than to this page.


%% file: abstract.tex
%
%
%
In this thesis, we investigate aspects of computing the perihelion distances and impact parameters for near stars to the Sun, 
and, we analyze the solar motion. The perihelion distances and impact parameters are studied for three different analytical 
approaches. First, we focus on the non-interacting system based on the object's motion with respect to the Sun along a 
straight line. We show that this approach gives results which are in agreement with those published in the literature  
\cite{comet} for Barnard's, Gl 217.1, Gl 729 and GJ 2046 stars. Second, we consider the two-body problem. As we expected,
in this case we obtain less perihelion distances than for the non-interacting system.
Third, we propose a simple model to describe the relative motion Sun-object, where 
in adition to the gravitational effects of the Galaxy, we consider oscillations (including anharmonic) 
of the Sun and the object with respect to the galactic equatorial plane. 
Equation of motion for these anharmonic oscillations is solved analytically. 
We show that some computed perihelion distances in this case 
are greater (totally different) than those given in literature, what is possible since 
anharmonic oscillacions are not considered in the literature \cite{comet, garcia}. 
However, for stars Gl 729, Gl 54.1 and LP 816-60 
this simple model gives reasonable values in comparison with the literature.
Finally, we focus on the solar motion (motion of the Sun with respect to the Local Standard of Rest), 
where the reference frame is given by the nearest stars. We develope a new method for determining the solar
motion taking into account also stellar proper motions and radial velocities. The results are compared with other
solutions given by approximation methods, where some of the
observational parameters are neglected. The direct method is the
best method in determining the solar motion. 
However, the motion of the Sun with respect to the
surrounding interstellar gas and dust yields significantly different
values for the solar motion. The only consistency between the results
of stars and the intestellar gas and dust is the value of  solar motion velocity component 
normal to the galactic equatorial plane $Z_{S}$.
Relevance of the solar motion, in direction perpendicular to the Galactic equator, for
evolution of the orbits of the comets of the Oort cloud, and,
in directions lying in the Galactic equator, for the values of the Oort constants and the shape of the rotation curve and the essence of
the dark matter is pointed out.

%% file: abstrakt.tex
V tejto pr\'{a}ci sme sk\'{u}mali met\'{o}dy na po\v{c}\'{i}tanie perih\'{e}liov\'{y}ch vzdia\-le\-nost\'{i} 
a z\'{a}mern\'{e} vzdia\-le\-nosti pre vybran\'{e}
bl\'{i}zke hviezdy a~ analyzovali sme pohyb Slnka. 
Perih\'{e}liov\'{e} vzdialenosti a z\'{a}mern\'{e} vzdialenosti sme sk\'{u}mali analyticky z troch r\^{o}znych poh\v{l}adov. 
V~ pr\-vom sme sa s\'{u}stredili na neinterak\v{c}n\'{y} syst\'{e}m zalo\v{z}en\'{y} na pohybe objektu 
po priam\-ke ur\v{c}enej po\v{c}iato\v{c}nou polohou a r\'{y}chlos\v{t}ou hviezdy vzh\v{l}adom na Slnko.
V\'{y}sled\-ky, ktor\'{e} sme dosiahli, boli v~ s\'{u}lade s~ literat\'{u}rou \cite{comet} 
pre Barnardovu hviezdu a pre hviezdy Gl 217.1, Gl 729 a GJ 2046. 
V druhom pr\'{i}pade sme sa s\'{u}stredili na probl\'{e}m dvoch telies, ke\v{d} sa uva\v{z}ovala gravit\v{c}n\'{a}
interakcia medzi hviezdou a Slnkom. 
V tomto pr\'{i}pade, ako sme o\v{c}ak\'{a}vali, sme dosiahli men\v{s}ie perih\'{e}lne vzdialenosti ako 
pre neinterak\v{c}n\'{y} syst\'{e}m. 
V tre\v{t}om sp\^{o}sobe sme navrhli jednoduch\'{y} model popisuj\'{u}ci relat\'{i}vny pohyb hviezdy vzh\v{l}adom na Slnko, 
kde sme uva\v{z}ovali gravita\v{c}n\'{e} efekty Galaxie vo forme oscil\'{a}cie Slnka a hviezdy vzh\v{l}adom na galaktick\'{y}
rovn\'{\i}k a aj anharmonick\'{e} oscil\'{a}ci\'{e} v tomto smere. 
Pohybov\'{a} rovnica t\'{y}chto anharmonick\'{y}ch oscil\'{a}ci\'{\i} sa tie\v{z} rie\v{s}i analyticky.
Uk\'{a}zali sme, \v{z}e v tomto pr\'{i}pade s\'{u} niektor\'{e} vypo\v{c}\'{i}tan\'{e} perih\'{e}liov\'{e} vzdialenosti 
v\"{a}\v{c}\v{s}ie (\'{u}plne odli\v{s}n\'{e}) od t\'{y}ch uv\'{a}dzan\'{y}ch v literat\'{u}re.
Toto m\^{o}\v{z}e by\v{t} sp\^{o}soben\'{e} neuva\v{z}ovan\'{i}m (anharmonick\'{y}ch) oscil\'{a}cii v ich pou\v{z}it\'{y}ch 
met\'{o}dach \cite{comet, garcia}. Napriek tomu, pre hviezdy Gl 729, Gl 54.1 and LP 816-60 tento jednoduch\'{y} model 
n\'{a}m d\'{a}va rozumn\'{e} hodnoty v porovnan\'{i} s literat\'{u}rou. 
Nakoniec sme sa s\'{u}stredili na pohyb Slnka (pohyb vzh\v{l}adom na miestny \v{s}tandard pokoja), 
kde sme za n\'{a}\v{s} referen\v{c}n\'{y} syst\'{e}m zvolili najbli\v{z}\v{s}ie hviezdy. 
Vyvinuli sme nov\'{u} met\'{o}du na ur\v{c}ovanie pohybu Slnka, kde sme uva\v{z}ovali aj vlastn\'{e} pohyby hviezd a 
radi\'{a}lne r\'{y}chlosti. V\'{y}sledky sme porovn\'{a}vali s ostatn\'{y}mi rie\v{s}eniami z\'{i}skan\'{y}mi 
z aproxima\v{c}n\'{y}ch met\'{o}d, pri ktor\'{y}ch boli niektor\'{e} z pozorovan\'{y}ch parametrov zanedban\'{e}. 
Z  v\'{y}sledkov n\'{a}m vyplynulo, \v{z}e priama met\'{o}da je najlep\v{s}ia pri  ur\v{c}ovan\'{i} 
pohybu Slnka. Av\v{s}ak, pohyb Slnka, ak berieme do \'{u}vahy okolit\'{y} medzihviezdny plyn a prach, 
d\'{a}va signifikantne odli\v{s}n\'{e} hodnoty pre sol\'{a}rny pohyb. 
Jedin\'{a} spojitos\v{t} medzi v\'{y}sledkom pre hviezdy a medzihviezdny plyn a prach je hodnota 
r\'{y}chlosti sol\'{a}rneho pohybu v smere norm\'{a}ly na galaktick\'{u} rovn\'{i}kov\'{u} rovinu $Z_{S}$. 
D\^{o}le\v{z}itos\v{t} pohybu Slnka, v smere kolmom na Galaktick\'{y} rovn\'{i}k sa zd\^{o}razňuje hlavne 
pri evol\'{u}cii orb\'{\i}t kom\'{e}t Oort-ovho oblaku. Zistenie sol\'{a}rneho pohybu, 
v smere rovnobe\v{z}nom s galaktick\'{y}m rovn\'{i}kom je d\^{o}le\v{z}it\'{e} pre zistenie 
hodn\^{o}t Oortov\'{y}ch kon\v{s}t\'{a}nt, tvaru rota\v{c}nej krivky a podstaty tmavej hmoty.

%% file: resumen.tex
En esta tesis se investigan aspectos del c\'{a}lculo de las distancias del perihelio y los par\'{a}metros de 
impacto para estrellas cercanas, y,
se analiza el movimiento solar. Las distancias del perihelio y los par\'{a}metros de impacto se estudian siguiendo 
tres enfoques diferentes.
En primer lugar, nos centramos en un sistema sin interacci\'{o}n basado en el movimiento del objecto 
con respecto al Sol a lo largo 
de una l\'{i}nea recta. Se demuestra que este m\'{e}todo da resultados de acuerdo con \cite{comet} 
para las estrellas Barnard, Gl 217.1, Gl 729 y GJ 2046. 
En segundo lugar, consideramos el problema de los dos cuerpos. Como era de esperar,
en este caso se obtienen distancias m\'{a}s cortas que para el sistema sin interacci\'{o}n.
En tercer lugar, proponemos un modelo simple para describir el movimiento relativo Sol-objeto, donde
en adici\'{o}n a los efectos gravitacionales de la Galaxia, consideramos oscilaciones (incluyendo anarm\'{o}nicas) 
del Sol y del Objeto con respecto al plano gal\'{a}ctico ecuatorial.
La ecuaci\'{o}n de movimiento de estas oscilaciones anarm\'{o}nicas se resuelve analíticamente. 
Se demuestra, que algunas distancias calculadas en este caso
son mayores (totalmente diferentes) a las que figuran en la literatura, lo que es posible, ya que
oscilaciones anarm\'{o}nicas no se consideran en la literatura \cite{comet, garcia}. 
Sin embargo, para Gl 729, Gl 54.1 y LP 816-60
este modelo da valores razonables en comparacion con \cite{comet, garcia}.
Por \'{u}ltimo, nos centramos en el movimiento solar (movimiento del Sol con respecto al Lugar de reposo estandar), 
donde se toma como marco de referencia las
estrellas m\'{a}s cercanas. Se desarrolla un nuevo m\'{e}todo para la determinaci\'{o}n del 
solar movimiento teniendo en cuenta tambi\'{e}n los movimientos propios estelares
y velocidades radiales. Los resultados se comparan con otras
soluciones dadas por m\'{e}todos de aproximaci\'{o}n, cuando algunos de los
par\'{a}metros de observaci\'{o}n no se toman en cuenta. El m\'{e}todo directo es el
mejor m\'{e}todo para determinar el movimiento solar.
Sin embargo, el movimiento del Sol con respecto al
gas y polvo interestelar circundante da valores significativamente diferentes para el movimiento solar. 
La consistencia s\'{o}lo entre los resultados
de las estrellas y el gas y el polvo intestellar es el valor de la componente de la velocidad de movimiento solar
normal al plano ecuatorial gal\'{a}ctico $ Z_ {S} $.
Se indica la relevancia del movimiento solar en direcci\'{o}n perpendicular al ecuador gal\'{a}ctico para la
evoluci\'{o}n de las \'{o}rbitas de cometas de la nube de Oort, y en direcciones situadas en el ecuador gal\'{a}ctico para
los valores de las constantes de Oort, la forma de la curva de rotaci\'{o}n y la esencia de
la materia oscura.

%% file: Chapter1.tex
\chapter{Introduction}
The outer part of the Solar System, known as the Oort cloud of comets, is under gravitational influence of both the Sun and 
the Galaxy. The effect of the Sun is described by the two-body problem. The effect of the Galaxy is equivalent to 
galactic tide and the physical model is given by  Klacka (2009) \cite{galaxy}. However, besides the regular permanent action 
of the gravity of the Sun and the Galaxy, also random perturbations can play non-negligible role in the orbital evolution 
of the comets in the Oort cloud. These irregular random perturbations are also of gravitational origin. They are generated 
by \emph{close} approaches of stars and interstellar clouds (consisting of gas and dust) to the Sun, or, more correctly, 
to the comets of the Oort cloud \cite{oort}. This Master Thesis deals with the problem of the \emph{close} approach of 
a galactic object to the Sun.
Hence, we want to find the minimal distance between the object and the Sun, i.e., the perihelion distance of the object. 
More generally, position vector of the perihelion. This Thesis presents several approaches to the problem of finding 
the perihelion distance of a galactic object. These approaches treat the problem from an analytical point of view. 
The first case assumes no interaction 
between the Sun and the Galactic object. The second case considers the two-body problem, i.e., gravitational interaction 
between the Sun and the object is taken into account. The third analytical approach considers oscillation of the Sun and 
the object with respect to the galactic equatorial plane. All three cases are treated in an analytical way and the obtained 
results for several stars are summarized in tables. 
By solving the equation of general motion given in Chapter \ref{chapter.2}, where gravity of the Sun and Galaxy 
are considered, our results may be improved.

These results are compared with those published in the literature. Since the oscillatory motion of the Sun 
with respect to the galactic equatorial plane plays an important role in the evolution of the Oort cloud of comets,
Chapter \ref{chapter.5} of this thesis deals with the solar motion. 
The kinematics of stars near to the Sun has long been known
to provide crucial information regarding both the structure and
evolution of the Milky Way \cite{other}.
That is the reason why
we calculate the solar motion in the reference frame
connected with the nearest stars. We identify series of $N$ stars
with heliocentric distances less than $100, 40, 15$ pc and then
determine the velocity of the Sun relative to the mean velocity of
these stars.

\newpage
\section{Outline and Summary of Results}
We study approaches of near stars, objects, to the Sun. 
General motivation for this problem is its application to the evolution of the Oort cloud 
of comets situated at the borders of the Solar System.
The object (a star or interstellar cloud of dust or gas) motion with respect to the Sun 
is calculated on the basis of gravitational forces
acting both on the Sun and the comet. Effect of the Galaxy in the form of galactic tide
is based on the observations. The flat rotation curve is considered together with the
most probable values of the Oort constants. The effect of the galactic bulge is consistent
with the mass distribution given by Maoz (2007) \cite{maoz}. Moreover, motion of the Sun in the Galaxy
is considered, too. We take into account not only the fact that the Sun revolves the center of the Galaxy,
as it is conventionally done. We take into account also the oscillations of the Sun with respect
to the galactic equatorial plane. This model is presented in Kla\v{c}ka (2009) \cite{galaxy}. 
Besides the fixed mass density used as an approximation to reality for the galactic bulge,
we can take into account the decrease of the bulge mass density as a function of the distance
from the galactic equatorial plane. The effect of this simple improvement will be treated.
Great part of the Thesis will be devoted to the effect of close approaches of stars
or interstellar clouds to the Solar System and to the study of the solar motion.

We will first calculate perihelion distance and impact parameter for a star with close
approach to the Solar system. Better value of the perihelion will be found using
the two-body problem (the star and the Sun). After the two-body consideration, we focus on the 
relative motion Sun-object in the field of the Galaxy (here, we consider that 
the Sun and star have a motion with respect to galactic equatorial plane), for which
we propose a simple model for stars nearest to the Sun. 
How can these effects enhance the number of comets entering the inner part of the Solar System?\\
We also devote our reserach to analyze the solar motion.\\
We present the following steps of this thesis in the next six chapters. 
{\bf Chapter \ref{chapter.1}: The impact parameter and the perihelion position vector} gives some definitions found in the literature 
for the impact parameter and for the perihelion position vector. \\
{\bf Chapter \ref{chap.nonint}: The non-interacting system} is based on the motion of 
an object (a star or an interstellar cloud which in reality perturb comets 
in the Oort cloud) which does not interact with the Sun. Here, the object moves along a straight line in an inertial frame of reference (e.g., galactic)\\
{\bf Chapter \ref{chap.twobody}: The two-body system} analyzes the way to compute the path of an object (a star or an interstellar cloud) under 
the solar gravitation, and, finally the impact parameter and the position vector of the perihelion. It is a typical two-body problem.\\
{\bf Chapter \ref{chapter.2}: Galactic tide} investigates the relative motion Sun-object, where 
in addition to the gravitational effects of the Galaxy, we consider anharmonic oscillations for the Sun and object. 
This simple model does not consider gravitational interaction between these two bodies, and it is based on the fact that the relative motion of the object with respect to the Sun depends linearly on time for 
$x$ and $y$ coordinates and takes into account anharmonic oscillations along the $z$-axis (for the Sun and for the Star, we suppose the same anharmonic motion). 
Equation of motion for this anharmonic motion is solved analytically. \\
{\bf Chapter \ref{chapter.3}: Summary of main equations} summarizes the important equations found in previous Chapters. 
{Chapter \ref{chapter.4}: Applications} applies the results to compute the perihelion position vector and the impact parameter for stars, 
which are taken from the literature \cite{comet}.\\
{\bf Chapter \ref{chapter.5}: The solar motion} calculates the solar motion in the reference frame
connected with the nearest stars. The reference frame is given by the nearest stars. We develop a new method for determining the solar
motion taking account also stellar proper motions and radial velocities. The results are compared with other
solutions given by approximation methods, where some of the observational parameters are neglected.
\\\\
Finally, I summarize the thesis in Chapter \ref{chapter.6}, and discuss further research directions. Also, I summarize the thesis in Slovak in Chapter 
\ref{chapter.7}. Additional material can be found at the end of the thesis. Appendix A shows the equations deduced from the Least square method.

%% file: Chapter4.tex
\chapter{The impact parameter and the perihelion position vector} 
\label{chapter.1} 
We are interested in the nearest distance between an approaching star (or interstellar cloud consisting of dust and gas)
and the Sun. This is given by the magnitude of the perihelion position vector. 

The process of scattering is standardly described by the {\bf impact parameter}.
Let us present several definitions of the impact parameter: 
\begin{enumerate}
\item ``The distance of one of the scattering partners from the
asymptote of the trajectory of the other scattering partner is the {\it impact parameter}.'' \cite[p.~189]{strauch}
\item ``The {\it impact parameter} $b$ is defined as the perpendicular distance from the projectile's incoming straight-line
path to a parallel axis through the target's center.'' \cite[p.~558]{taylor}.
\item Consider scattering of a projectile from a target particle. ``The initial velocity of the projectile
is ${\bf v}_{0}$ and it is assumed that in the absence of any interaction it would travel in a straight line
and pass the target at a distance $b$, called the {\it impact parameter}.'' \cite[p.~362]{martin}  
\end{enumerate}  
Thus, we can define the impact parameter $b$ as the perpendicular distance between 
the straight-line of the initial velocity vector of an object (a star or an interstellar cloud)  
and the center of the field $U(r)$ created by the Sun that the object is approaching (see also \cite[p.~35--66]{landau} and \cite[p.~126--143]{symon}). 
{\bf The perihelion} is the point in the solar orbit of an object when it is nearest to the Sun. {\bf The perihelion distance} is the 
distance between the object and the Sun at this point. 
 
When a comet moves around the Sun, they interact and the motion of the comet is affected by the solar gravitation, 
by the galactic tide and other forces. Perturbations from near stars or interstellar clouds can be also important.

\chapter{Non-interacting system}
\label{chap.nonint}
In this Chapter we focus on the system where an object (a star or an interstellar cloud which in reality perturb comets 
in the Oort cloud) does not interact with the Sun. Here, the object moves along a straight line 
in an inertial frame of reference (e.g., galactic). The motion is given by the position vector ${\bf r}$ in the
inertial frame: 
\begin{equation}
\label{eq.3.1}
 {\bf r}={\bf r}_{\odot}+{\bf r}_{0}+{\bf v}_{0}t\quad\Rightarrow \quad {\bf r}-{\bf r}_{\odot}
={\bf r}_{0}+{\bf v}_{0}t,
\end{equation}
where ${\bf r}_{0}=(x_{0},y_{0},z_{0})$ and ${\bf v}_{0}=(v_{x,0},v_{y,0},v_{z,0})$  
are the initial ($t$ $=$ 0) position and velocity vectors of the object with respect to the Sun, 
${\bf r}_{\odot}$ is the position vector of the Sun in the inertial frame of reference.
To find the impact parameter we minimize the magnitude of the second equation in Eqs.~(\ref{eq.3.1})

\begin{equation}
 \label{eq.3.2}
b=\min{|{\bf r}-{\bf r}_{\odot}|}=\min_{t}{|{\bf r}_{0}+{\bf v}_{0}t|}
\end{equation}
or
\begin{equation}
 \label{eq.3.3}
b=\min_{t}{\sqrt{(x_{0}+v_{x,0}t)^{2}+(y_{0}+v_{y,0}t)^{2}+(z_{0}+v_{z,0}t)^{2}}}~.
\end{equation}
Minimizing the previous equation we find the condition

\begin{displaymath}
 (x_{0}+v_{x,0}t_{b})v_{x,0}+(y_{0}+v_{y,0}t_{b})v_{y,0}+(z_{0}+v_{z,0}t_{b})v_{z,0}=0,
\end{displaymath}
from which
\begin{equation}
\label{eq.3.4}
 t_{b}=-\frac{x_{0}v_{x,0}+y_{0}v_{y,0}+z_{0}v_{z,0}}{v_{x,0}^{2}+v_{y,0}^{2}+v_{z,0}^{2}}
\end{equation}

Putting Eq.~(\ref{eq.3.4}) into Eq.~(\ref{eq.3.3}) we get

\begin{eqnarray}
\label{eq.3.5}
 b&=&\frac{1}{v_{0}^{2}}\biggr\{\left[x_{0}\left(v_{y,0}^{2}+v_{z,0}^{2}\right)
-\left(y_{0}v_{y,0}+z_{0}v_{z,0}\right)v_{x,0}\right]^{2}\nonumber\\
&&+\left[y_{0}\left(v_{x,0}^{2}+v_{z,0}^{2}\right)-\left(x_{0}v_{x,0}+z_{0}v_{z,0}\right)v_{y,0}\right]^{2}\nonumber\\
& &+\left[z_{0}\left(v_{x,0}^{2}+v_{y,0}^{2}\right)-\left(x_{0}v_{x,0}+y_{0}v_{y,0}\right)v_{z,0}\right]^{2} \biggr\}^{1/2},
\end{eqnarray}
where $v_{0}^{2}=v_{x,0}^{2}+v_{y,0}^{2}+v_{z,0}^{2}$~.

Then, the vector of perihelion position is 

\begin{equation}
\label{eq.sininteracion}
 {\bf q}={\bf r_{b}}-{\bf r_{\odot}}={{\bf r_{0}}}+{{\bf v}_{0}}t_{b},\quad 
t_{b}= -~\frac{x_{0}v_{x,0}+y_{0}v_{y,0}+z_{0}v_{z,0}}{v_{x,0}^{2}+v_{y,0}^{2}+v_{z,0}^{2}} ~.
\end{equation}

We treated non-interacting system. In this case the impact parameter fulfills the condition $b$ $=$ $| {\bf q} |$. The impact parameter
equals to the perihelion distance.

\chapter{The two-body problem}
\label{chap.twobody}
The classical problem of celestial mechanics focus on the motion of one body with respect to another under the influence of
their mutual gravitation \cite[p.~34]{landau}, \cite[p.~184]{symon}. In its simplest form, this problem is little more than the
generalization of the central force problem\cite[p.~126]{symon}, but in some cases the bodies are of
finite size and are not spherical. This may complicate the problem enormously as
the potential fields of the objects no longer vary as the inverse square of the
distance. This causes orbits to precess and the objects themselves to undergo
gyrational motion \cite[p.~71]{george}. The masses need not be point masses - as long as they are spherically symmetric, they  act 
gravitationally as if they were point masses \cite[p.~71--80]{george}. In this Chapter we compute the path of an object (a star or an interstellar cloud) under the solar gravitation, 
and, finally the impact parameter and the position vector of the perihelion. It is a typical two-body problem.
Angular momentum is conserved. We show two ways how to find the impact parameter. For this, we use some important parameters as the reduced mass which we defined (as is given in the literature) 
in the form of $\mu=mM/(m+M)$ and  
the Universal Gravitational Constant $G$. The value of $G$ is the same anywhere in the Universe, 
and it doest not vary with time \cite{PhysRevLett.48.121, PhysRevD.82.022001}. Here, we adopt 
these assumptions, while noting that it is a legitimate cosmological question to consider what implications there may be if either of them is not so.
$G$ is among those fundamental constants whose numerical value has been determined with least precision. 
Its currently accepted value is $(6.67259\pm 0.00085)\times10^{-11}$ Nm$^{2}$kg$^{-2}$ \cite{PhysRevLett.48.121, PhysRevD.82.022001}. 
It is worth noting that, while the product $GM$ for the Sun is known with very great precision, the mass of 
the Sun is not known to any higher degree of precision than that of the gravitational constant.
\section{Impact parameter}
\begin{figure}[h!]
\includegraphics[width=12cm,height=10cm,bb=0 0 580 480]{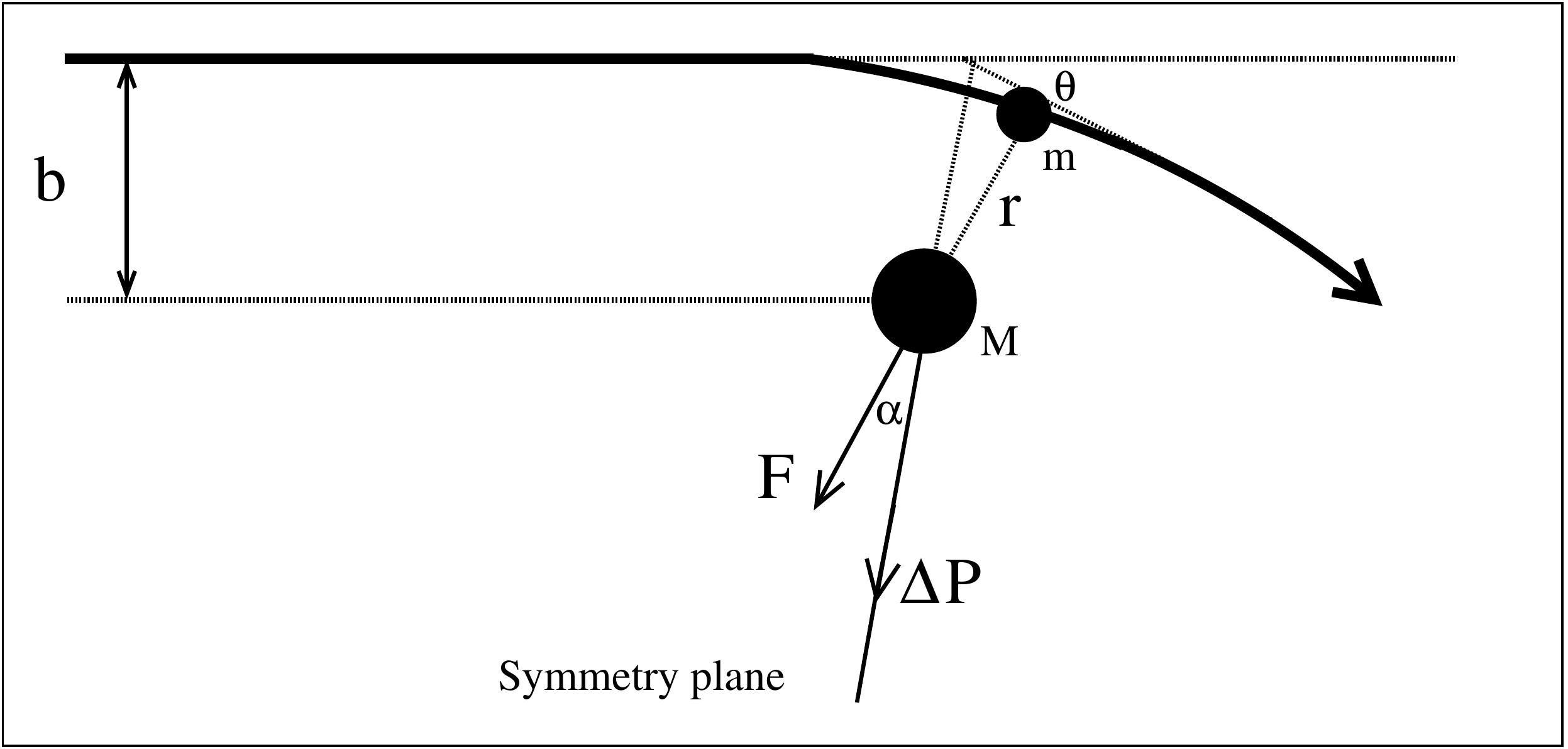}
\caption{{\small Path of the studied object in the field of the Sun.}} 
\label{fig.pp}
\end{figure}

Our problem is represented in Fig.~\ref{fig.pp}.
Conservation of the angular momentum yields $\mu v_{\infty}b=\mu r^{2}\omega$ ($\omega=d\alpha/dt=v_{\infty}b/r^{2}$), 
where $v_{\infty} = | {\bf v_{\infty}} |$ is the magnitude of the velocity vector in the infinity ${\bf v}_{\infty}$ 
of the object with respect to the Sun, $v = | {\bf v} |$ is the magnitude of the velocity vector ${\bf v}$ of the 
object with respect to the Sun
at a given time $t$. We assume that the interaction
between the object and the Sun can be neglected at the initial moment. Here, $m$ is the mass of the object, 
${\bf r}$ is the object position vector at the time $t$, and, $b$ is the impact parameter.
\begin{figure}[htb]
\includegraphics[width=14cm,height=10cm,bb=0 0 580 480]{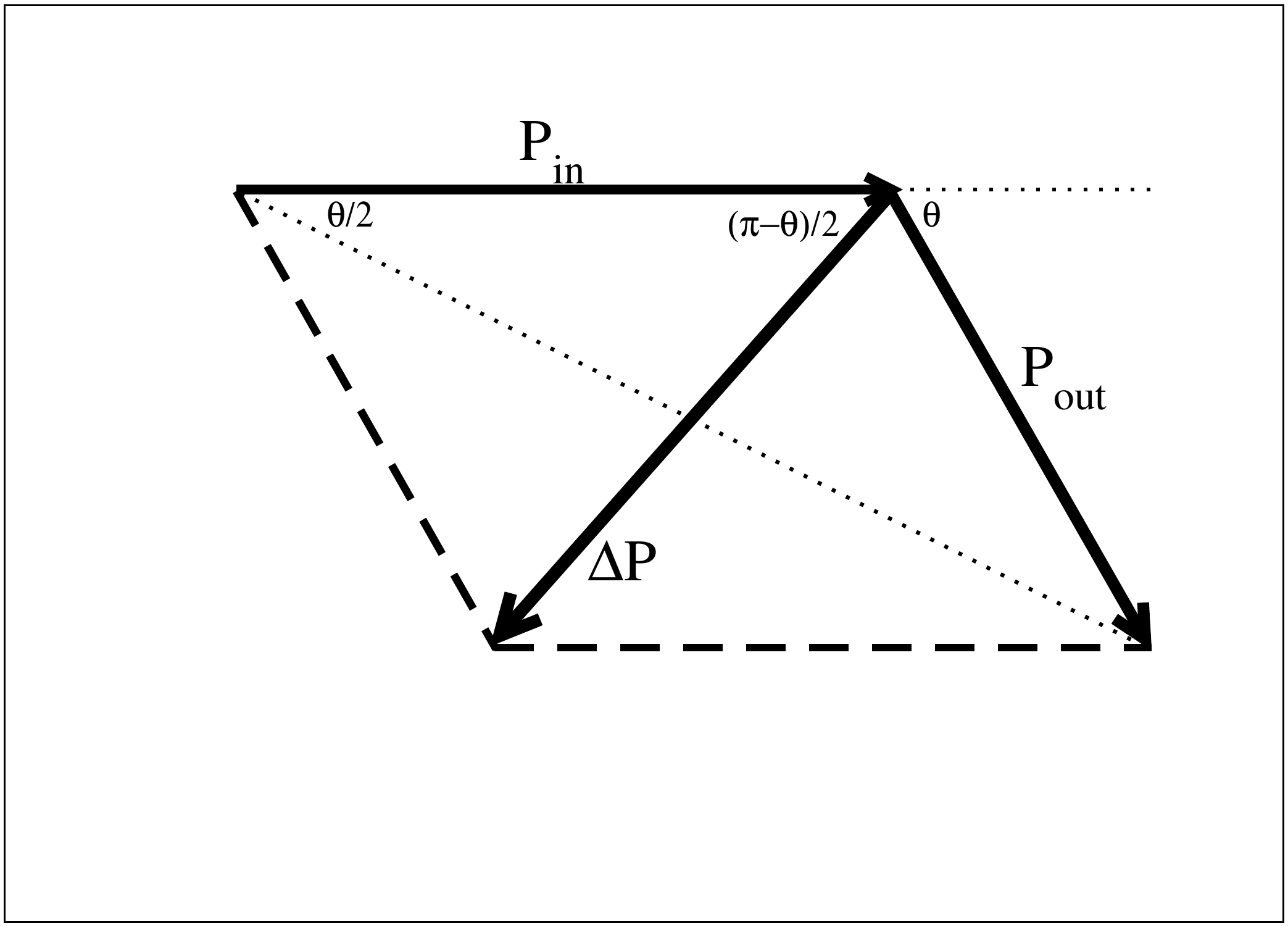}
\caption{{\small Geometric representation of the object's momentum.} }
\label{fig.hh}
\end{figure}

In accordance with a simple geometry, from Fig.~\ref{fig.hh}, where $P=|{\bf P}_{out}|~=~|{\bf P}_{in}|$, we have $\Delta P=2P\sin(\theta/2)$, 
and from the second Newton's law we get
\begin{eqnarray}
 \label{3.7}
 {\bf F}&=&\frac{d{\bf P}}{dt},\nonumber\\
 \Delta P&=&\int F\cos\alpha\, dt\nonumber\\
        &=&\int \frac{GmM}{r^2}\cos\alpha \,dt=GmM\int \frac{\cos\alpha}{r^2}\frac{dt}{d\alpha}d\alpha\nonumber\\
        &=&GmM\int \frac{\cos\alpha}{r^2}\frac{r^2}{v_{\infty}b}d\alpha=\frac{GmM}{v_{\infty}b}\int_{-(\pi-\theta)/2}^{(\pi-\theta)/2}\cos\alpha\, d\alpha\nonumber\\
        \Delta P&=&\frac{GmM}{v_{\infty}b}2\cos(\theta/2)~,
\end{eqnarray}
where $F\cos\alpha$ is the force projection onto $\Delta {\bf P}$. Doing an elementary comparison between the found relations for $\Delta P$ (from geometry and Newton's law) 
we find the impact parameter
\begin{equation}
 \label{eq.3.8}
b=\frac{GmM}{Pv_{\infty}}\cot(\theta/2)~,
\end{equation}   
since, in accordance with Fig. 2-2 and cosine theorem, $| \Delta {\bf P} |^{2} = 2 P^{2} (1 - \cos \theta)$, or
\begin{equation}
 \label{eq.impactone}
b=\frac{G(m+M)}{v_{\infty}^{2}}\cot(\theta/2)~,
\end{equation}   
since $P=\mu v_{\infty}$.
\section{Trajectory}
In this part, we study the motion of an object (e.g. a star) in a central field (solar gravitational field). 
We compute the impact parameter and the perihelion position vector. 
The mutual potential energy of two spherically symmetric distributed masses at a distance $r$ apart, which is the 
work required to bring them to a distance $r$ from an infinite initial separation, is 
\begin{displaymath}
 V(r)=\frac{k}{r}=-\frac{GMm}{r}.
\end{displaymath}
$l=\mu v r \sin\varphi$ is the angular momentum.
The position of $m$ with respect to $M$, in polar coordinates, is given by 
\begin{displaymath}
 (x,y)=|{\bf r}|(\cos\varphi,\sin\varphi)~. 
\end{displaymath}
Notice that $r$ and $\varphi$ are time dependent. The 
kinetic energy is $T=\mu(\dot{r}^{2}+r^{2}\dot{\varphi}^{2})/2$. Therefore the Lagrangian in polar coordinates for this system is 
\begin{equation}
 \label{eq.3.9}
L(r,\dot{r},\varphi,\dot{\varphi})=\frac{\mu}{2}\left(\dot{r}^{2}+r^{2}\dot{\varphi}^{2}\right)-\frac{k}{r},
\end{equation}
from where the equations of motion are given by
\begin{equation}
 \label{eq.3.10}
\frac{d}{d\,t}\frac{\partial\, L}{\partial\, \dot{\varphi}}-\frac{\partial\,L}{\partial\,\varphi}=0, \quad\quad\frac{d}{d\,t}\frac{\partial\, L}{\partial\, \dot{r}}-\frac{\partial\,L}{\partial\,r}=0.
\end{equation}
If the Lagrangian does not contain a given coordinate, in this case $\varphi$, then the coordinate is said to be cyclic and the corresponding conjugate momentum is conserved.
Such quantity is  the angular momentum perpendicular to the plane of motion, i.e.~${\bf l} ={\bf r}\times{\bf p}$, then 
\begin{equation}
\label{eq.3.10.1S}
l=\mu v r \sin\varphi=\mu v_{0} b ~.
\end{equation}
By using Eq.~(\ref{eq.3.10}) we get
\begin{equation}
 \label{eq.3.11}
\frac{d}{d\,t}\frac{\partial\, L}{\partial\, \dot{\varphi}}=0 \quad \Rightarrow\quad \frac{\partial\, L}{\partial\, \dot{\varphi}}=l\quad \Rightarrow\quad \dot{\varphi}=\frac{l}{\mu r^{2}},
\end{equation}
where $l$ is a constant of motion; and for $r$:
\begin{equation}
 \label{eq.3.12}
\mu\ddot{r}-\frac{l}{\mu r^{3}}-\frac{k}{r^{2}}=0.
\end{equation}
The solution of the previous equation gives the orbit of a particle in a central field. To solve Eq.~(\ref{eq.3.12}) we substitute $r=1/u$. Applying the chain 
rule one can verify
\begin{equation}
\label{eq.3.12.1}
 \dot{r}=\dot{\varphi}\frac{d\,r}{d\,u}\frac{d\,u}{d\,\varphi}\quad\Rightarrow\quad \dot{r}=-\frac{l}{\mu}\frac{d\,u}{d\,\varphi}\quad\Rightarrow\quad\ddot{r}=-\frac{l^{2}u^{2}}{\mu^{2}}\frac{d^{2}u}{d\varphi^{2}}.  
\end{equation}
Doing correct substitutions we find
\begin{equation}
\label{eq.3.13}
     \frac{d^{2}u}{d\varphi^{2}} + u \, = \, -\frac{\mu k}{l^2}, 
\end{equation}
which is a nonhomogeneous second order linear equation. The solutions of this equation are of two types: 
along \textbf{unbound} orbits $r\rightarrow \infty$ and hence $u\rightarrow 0$, while on 
\textbf{bound} orbits $r$ and $u$ oscillate between finite limits. The general solution for this equation is
\begin{equation}
\label{eq.solutiongeneral}
 u(\varphi)=\alpha\sin\varphi+\beta\cos\varphi-\frac{\mu k}{l^{2}},
\end{equation}
where the constants $\alpha$ and $\beta$ are given by the initial conditions.
 The initial incoming state (\textit{in} state) is given  (see Fig.~\ref{fig.pp} and polar coordinates of the object) by the condition
 $\varphi\rightarrow \pi$, $\dot{r}~=~-v_{\infty}$ (the minus sign represents the fact that the object comes from the left
and approaches to the central field); 
the final outgoing state (\textit{out} state) is given (see Fig.~\ref{fig.pp} and polar 
coordinates of the object) by the condition $\varphi\rightarrow 2\pi-\theta$, $r\rightarrow\infty$. Thus, \textit{in} state 
yields
\begin{displaymath}
 u(\varphi\rightarrow\pi)\rightarrow 0~,\quad\beta\cos\pi-\frac{\mu k}{l^{2}}=0~\Rightarrow\beta= - \mu k/l^{2}
\end{displaymath}
Moreover, in accordance with (see Eq.~(\ref{eq.3.12.1})) 
\begin{displaymath}
\dot{r}=-\frac{l}{\mu}\frac{du}{d\varphi}~,
\end{displaymath}
we obtain (for $\varphi\rightarrow\pi$), with $\dot{r}=-v_{\infty}$,
\begin{displaymath}
 \alpha=-\frac{\mu v_{\infty}}{l}~.
\end{displaymath}
The \textit{out} state yields
\begin{displaymath}
 u(\varphi\rightarrow2\pi-\theta)\rightarrow0~,\quad-\alpha\sin\theta+\beta\cos\theta-\frac{\mu k}{l^{2}}=0~,
\end{displaymath}
\begin{displaymath}
\frac{\mu v_{0}}{l}\sin\theta-\frac{\mu k}{l^{2}}\cos\theta-\frac{\mu k}{l^{2}}=0~,
\end{displaymath}
therefore
\begin{eqnarray}
 1+\cos\theta&=&\frac{v_{\infty}l}{k}\sin\theta~,\nonumber\\
&=&\frac{v_{\infty}^{2}\mu b}{GmM}\sin\theta~,\nonumber\\
&=&\frac{v_{\infty}^{2}b}{G(m+M)}\sin\theta~.\nonumber\\
\end{eqnarray}
Then, for the impact parameter, we get
\begin{eqnarray}
\label{eq.secondimpact}
 b&=&\frac{G(M+m)}{v_{\infty}^{2}}\frac{1+\cos\theta}{\sin\theta}~,\nonumber\\
&=&\frac{G(m+M)}{v_{\infty}^{2}}\cot\left(\frac{\theta}{2}\right)~.
\end{eqnarray}
which is the same result as Eq.~(\ref{eq.impactone}) found in the previous subsection. 
The velocity in the infinity $v_{\infty}$ is found from conservation of energy in the infinity and in the time where are taken 
initial conditions, then the energy for initial conditions is given by
\begin{equation}
 E_{0}=\frac{\mu}{2}v_{0}^{2}+\frac{k}{r_{0}}~,
\end{equation}
and in the infinity, 
\begin{equation}
 E_{\infty}=\frac{\mu}{2}v_{\infty}^{2}~.
\end{equation}
Since the energy is conserved, $E_{0}=E_{\infty}$, for the speed in the infinity we get
\begin{equation}
 v_{\infty}=~\sqrt{\frac{2}{\mu}\left(\frac{\mu}{2}v_{0}^{2}+\frac{k}{r_{0}} \right)}~.
\end{equation}
From Eq.~{\ref{eq.impactone}}, or from Eq.~{\ref{eq.secondimpact}}, we cannot get $b$, cause we do not know neither $b$ nor $\theta$.
For this reason it requires to use another equation. That equation is then given by the conservation of angular momentum, 
$\mu~ v_{\infty}~b~=\mu~|{\bf r}_{0}\times{\bf v}_{0}|$, from where we get
\begin{equation}
 b=\frac{|{\bf r}_{0}\times{\bf v}_{0}|}{v_{\infty}}~,
\end{equation}
where we assume that the velocity vector in the infinity is perpendicular to the impact parameter, then $|{\bf e}_{r}\times{\bf e}_{t}|_{\infty}=1$. 
Hence, from previous equation we find $b$, and replacing in Eq.~{\ref{eq.impactone}}, or in Eq.~{\ref{eq.secondimpact}} we are able
to find the dispersion angle $\varphi$.\\
We write Eq.~(\ref{eq.solutiongeneral}) in a better and 
more familiar form, or, we just seek for a general solution to Eq.~(\ref{eq.3.13}) in the form of 
$u(\varphi)=D\cos(\varphi-\varphi_{0})-\mu k/l^{2}$, 
where $D~=~\sqrt{\alpha^{2}+\beta^{2}}$, $\cos\varphi_{0}~=~\alpha/D$, and $\sin\varphi_{0}~=~-\beta/D$. 
$\varphi_{0}$ can be computed by using Tab.~\ref{t1.1} given below. 
Therefore, 
\begin{equation}
 \label{eq.xxx}
r(\varphi)=\frac{p}{1+\epsilon~\cos(\varphi-\varphi_{0})}~,
\end{equation}
where $p=|{\bf H}|^{2}/(\mu k) = l^{2}/(k\mu)$ is the parameter (known as semi-latus rectum), $\epsilon=-(l^{2}/k\mu)A=\sqrt{1+2El^{2}/(\mu k^{2})}$ 
is the eccentricity, $E$ is the energy of the system and the angle $\varphi-\varphi_{0}$ is known as the \textbf{true anomaly}. 
The energy of the system is given by 
\begin{displaymath}
E=T+V=\frac{\mu \dot{r}^{2}}{2}+\frac{\mu}{2}r^{2}\dot{\varphi}^{2}+\frac{k}{r}~, 
\end{displaymath}
where the second and the third terms represent the effective potential energy. 
For a circular orbit $E$ is a minimum, $dV'/dr=0$ $\Rightarrow$ $r_{c}=-l^{2}/\mu k$, $v=\sqrt{-k/\mu r_{c}}$
and for the minimal value of the energy we have $E_{min}=-k^{2}\mu/(2l^{2})=k/(2r_{c})$.

For the perihelion ($r_{min}=r_{p}$) and aphelion ($r_{max}=r_{a}$) distances, $\dot{r}=0$ $\Rightarrow$ $E=V'$, $E=l^{2}/(2\mu r^{2}) +k/r$, then
\begin{eqnarray}
 \label{3.15}
 \frac{1}{r_{p}}&=&-\frac{\mu k}{l^{2}}+\sqrt{\frac{\mu^{2} k^{2}}{l^{4}}+\frac{2\mu E}{l^{2}}}~,\nonumber\\
 \frac{1}{r_{a}}&=&-\frac{\mu k}{l^{2}}-\sqrt{\frac{\mu^{2} k^{2}}{l^{4}}+\frac{2\mu E}{l^{2}}}~.
\end{eqnarray}
Now, we define and redefine some expressions for a better understanding: the angular momentum 
${\bf H}= \mu {\bf r}\times {\bf v}$, and the semi-major axis  $a=p/(1-\epsilon^{2})$.
So, for the orbit of the object we have   
\begin{eqnarray}
 \label{eq.3.16}
\frac{p}{r(\varphi)}&=&1+\epsilon~\cos(\varphi-\varphi_{0})~,\nonumber\\
 \frac{1}{r_{p}}&=&-\frac{\mu k}{l^{2}}+\sqrt{\frac{\mu^{2} k^{2}}{l^{4}}+\frac{2\mu E}{l^{2}}}~,\nonumber\\
 \frac{1}{r_{a}}&=&-\frac{\mu k}{l^{2}}-\sqrt{\frac{\mu^{2} k^{2}}{l^{4}}+\frac{2\mu E}{l^{2}}}~.
\end{eqnarray}
where $A=\sqrt{(\mu^{2} k^{2})/l^{4}+(2\mu E)/l^{2}}$, $\varphi_{0}$ is an arbitrary constant.

An orbit for which $\epsilon \geq 1$ is unbound, since  $r\rightarrow \infty$ as ($\varphi -\varphi_{0}$)$\rightarrow\arccos(-1/\epsilon)$; the orbit forms a hyperbola if
$\epsilon > 1$ and a parabola if $\epsilon = 1$. Also, the object's asymptotic speed $v_{0}$ as $r\rightarrow\infty$ is related to $\epsilon$ and $l$.
Orbits for which $\epsilon < 1$ are bound, $r$ is finite for all values of $\varphi$. 
Furthermore, $r$ is now a periodic function of $\varphi$ with 
period $2\pi$, so the object returns to its original radial coordinate after exactly one revolution in $\varphi$. Thus these orbits are closed, and they form ellipses
 with the attracting center at one focus.

To construct the position vector ${\bf r}$ we use the fact that ${\bf r}=r{\bf e}_{r}$, where ${\bf e}_{r}$ is the unit vector onto the radial direction. For 
our case ${\bf e}_{r}$ is defined as
\begin{equation}
 \label{eq.3.17}
{\bf e}_{r}=(\cos\Omega\cos\Theta-\sin\Omega\sin\Theta\cos i,\cos\Theta\sin\Omega+\sin\Theta\cos\Omega\cos i,\sin\Theta\sin i),
\end{equation}
 where $\Omega$ is the longitude of the ascending node, $\Theta=\omega+f$, $\omega$ is the argument of periapsis 
(as an angle measured from the ascending node to the pericenter/periapsis), 
$f=2\Delta\varphi$ is the true anomaly, and $i$ is the inclination with respect to the reference plane, measured at the ascending node (where the orbit passes upward through the reference plane).
It is easy to prove that the minimal distance $q$ is given by $f=0$, $r_{p}=a(1-\epsilon)\equiv q$. 
The angles in Eq.~(\ref{eq.3.17}) can be found by using (Kla\v{c}ka 2004)
\begin{eqnarray}
 \label{eq.3.18}
&&\sin\Omega\sin i =\frac{H_{x}}{|{\bf H}|},\quad-\cos\Omega\sin i=\frac{H_{y}}{|{\bf H}|}, \quad i=\arccos\left(\frac{H_{z}}{|{\bf H}|}\right),\nonumber\\
&&\sin\Theta\sin i=\frac{z}{r}, \quad\cos\Theta\sin i=\frac{yH_{x}-xH_{y}}{r|{\bf H}|}.\\\nonumber
&&\sin(\Theta-\omega)=\frac{{\bf v}\cdot{\bf e}_{r}}{\epsilon\sqrt{G(M+m)/p}}, \quad 
\cos(\Theta-\omega)=\frac{{\bf v}\cdot{\bf e}_{t}}{\epsilon\sqrt{G(M+m)/p}}-\frac{1}{\epsilon},\\\nonumber
\end{eqnarray}
where ${\bf v}$ is the velocity of the object given by the radial ($v_{r}=\sqrt{G(M+m)/p} ~\epsilon\sin f$) and 
transversal ($v_{t}=\sqrt{G(M+m)/p}(1+\epsilon\cos f)$) components,
\begin{equation}
 {\bf v}=v_{r}{\bf e}_{r}+v_{t}{\bf e}_{t},
\end{equation}
and ${\bf e}_{t}$ is the unit vector onto the transversal direction
 \begin{equation}
\label{eq.unittanget}
{\bf e}_{t}=(-\cos\Omega\sin\Theta-\sin\Omega\cos\Theta\cos i,-\sin\Omega\sin\Theta+\cos\Omega\cos\Theta\cos i,\cos\Theta\sin i).
 \end{equation}
By finding the angles in previous equations one can build the object's position vector ${\bf r}$. We focus 
on two special positions, for an initial time (all used parameters with index zero, $0$) and for the time when the object passes 
the perihelion (all used parameters with index $p$). 
For initial conditions ${\bf r}_{0}=(x_{0},y_{0},z_{0})$, 
${\bf v}_{0}=(v_{x,0},v_{y,0},v_{z,0})$ $\Rightarrow$ ${\bf H}_{0}\equiv(H_{x,0},H_{y,0},H_{z,0})$,  ${\bf H}_{0}=\mu {\bf r_{0}}\times{\bf v_{0}}$, 
the angle $i_{0}$ is computed as
\begin{equation}
 \label{eq.3.19.1}
i_{0}=\arccos\left(\frac{H_{z,0}}{|{\bf H}_{0}|}\right),
\end{equation}
the previous relation is a number given by initial conditions. Since we know $i_{0}$, 
in Eqs.~(\ref{eq.3.18}) we have two equations by which we are able
to find $\Omega_{0}$,
\begin{equation}
 \label{eq.3.19.2}
\sin\Omega_{0} =\frac{H_{x,0}}{|{\bf H}_{0}|\sin i_{0}}~,\quad \cos\Omega_{0}= - ~\frac{H_{y,0}}{|{\bf H}_{0}|\sin i_{0}} ~.
\end{equation} 
To choose the right angle quadrant, and therefore the right angle, we have to know the signs of the right sides in previous 
equations. Since we find these problems in several parts of this Thesis, we show a table (which can be used as algorithm) to 
choose the right (quadrant) angle.  Let
\begin{eqnarray}
\label{eq.cuadrante}
\cos \vartheta &=& A ~,
\nonumber \\
\sin\vartheta&=& B ~,
\end{eqnarray}
where $A$ and $B$ are known numbers computed from the right sides of given equations. These numbers can be positive or negative, then 
we need to know which quadrant the angle belongs. It is recommended to remember where $\sin \vartheta$, $\cos \vartheta$ and $\tan\vartheta$ 
are positive (or negative). Table ~\ref{t1.1} allows us to solve problems as the given in Eq.~(\ref{eq.cuadrante}). 
\begin{table}[h]
\begin{center}
  \begin{tabular}{|c||c|c|c|}
	\hline
  & $\sin\vartheta$ $>$ $0$    &   $\sin\vartheta$ $<$ $0$ &$\sin\vartheta$ $=$ $0$ \\
	\hline \hline
$\cos\vartheta$ $>$ $0$   & $\vartheta=\arccos A$ & $\vartheta=2\pi-\arccos A$& $\vartheta=0$\\
	\hline
$\cos\vartheta$ $<$ $0$    & $\vartheta=\arccos A$ & $\vartheta=2\pi-\arccos A$& $\vartheta=\pi$\\
	\hline       
$\cos\vartheta$ $=$ $0$    & $\vartheta=\pi/2$ & $\vartheta=3\pi/2$&  \\
       \hline
\end{tabular}
\end{center}
\caption{{\small Choosing the right angle.}}
\label{t1.1}
\end{table}\\
Doing the same, from Eqs.~(\ref{eq.3.18}), $\Theta_{0}$ is computed as
\begin{equation}
 \label{eq.3.19.3}
\sin\Theta_{0}=\frac{z_{0}}{r_{0} \sin i_{0}}, \quad\cos\Theta_{0}=\frac{y_{0}H_{x,0}-x_{0}H_{y,0}}{r_{0}|{\bf H}_{0}|\sin i_{0}}.
\end{equation} 
In Eq.~(\ref{eq.3.19.3}) we have the same problem as before with the right side signs, and for this we use Table \ref{t1.1}. 
Until now we should know from previous Equations $i_{0}, \Omega_{0}$ and $\Theta_{0}$. The last angle which we have to find is 
$\omega_{0}$, and doing the same as before where in Eqs.~(\ref{eq.3.18}) we also have two equations,
\begin{equation}
 \label{eq.3.19.4}
\sin(\Theta_{0}-\omega_{0})=\frac{{\bf v}_{0}\cdot{\bf e}_{r,0}}{\epsilon\sqrt{G(M+m)/p}}, \quad 
\cos(\Theta_{0}-\omega_{0})=\frac{{\bf v}_{0}\cdot{\bf e}_{t,0}}{\epsilon\sqrt{G(M+m)/p}}-\frac{1}{\epsilon},
\end{equation} 
the right angle is found by using the same way, by Table \ref{t1.1}.
$i_{0}$, $\Omega_{0}$, $\Theta_{0}$ and $\omega_{0}$ are now known values which help us to build the unit radial and 
transversal vectors, 
Eq.~(\ref{eq.3.17}) and Eq.~(\ref{eq.unittanget}), respectively. And hence it helps us to construct the position and velocity vectors.
In Eq.~(\ref{eq.3.19.4}) ${\bf v}_{0}$ and ${\bf e}_{r,0}$ may be found from initial conditions or just by using the known values of the angles 
as ${\bf e}_{t,0}$ is found.
Now, for the perihelion (all used parameters with index $p$) distance $f=0$ and $| {\bf r}_{p}|=|(x_{p},y_{p},y_{p})|=a(1-\epsilon)\equiv q$, then the perihelion vector of position 
is easily computed from ${\bf r}_{p}=r_{p}{\bf e}_{r,p}$. The point here is to calculate ($x_{p},y_{p},z_{p}$), and 
we carry this out by using Eq.~(\ref{eq.3.17}) and the values of angles from Eqs(\ref{eq.3.19.1} - \ref{eq.3.19.4}). Hence 
\begin{eqnarray*}
\label{eq.3.20}
x_{p}&=&r_{p}(\cos\Omega_{0}\cos\Theta_{p}-\sin\Omega_{0}\sin\Theta_{p}\cos i_{0}),\\\nonumber
y_{p}&=&r_{p}(\cos\Theta_{p}\sin\Omega_{0}+\sin\Theta_{p}\cos\Omega_{0}\cos i_{0}),\\\nonumber
z_{p}&=&r_{p} \sin\Theta_{p}\sin i_{0} ~.
\end{eqnarray*}
In the previous equation $r_{p}$, $\Omega_{0}$, $i_{0}$ are known and since $f_{p}=0$ (for the perihelion), 
$\Theta_{p}=\omega_{0}$, then the perihelion position vector ${\bf q}$ is definitely given by
\begin{equation}
\label{eq.3.20.1}
{\bf q}\equiv{\bf r}_{p}=(x_{p},y_{p},z_{p}) = \left\{ \begin{array}{rrl}
x_{p}&=& r_{p}(\cos\Omega_{0}\cos\omega_{0}-\sin\Omega_{0}\sin\omega_{0}\cos i_{0}), \\
 y_{p}&=& r_{p}(\cos\omega_{0}\sin\Omega_{0}+\sin\omega_{0}\cos\Omega_{0}\cos i_{0}),\\
z_{p}&=&r_{p} \sin\omega_{0}\sin i_{0} ~.
\end{array} \right.
\end{equation}
As we can see, the perihelion position vector ${\bf q}$ for the two-body problem is given by Eq.~(\ref{eq.3.20.1}). The angles, 
which allow us to build the position vector, are computed from Eqs.(\ref{eq.3.19.1}), (\ref{eq.3.19.2}), (\ref{eq.3.19.3}) and from Eq.~(\ref{eq.3.19.4}).

%% file: Chapter3.1.tex
\chapter{Galactic tide}
\label{chapter.2}
In this chapter we choose our reference frame in the galactic center (GC). In adition to the gravitational influence of the Sun, 
we consider that the object is perturbed by the gravitational effects of the Galaxy. As before we focus on the relative motion of the object 
with respect to the Sun. Global galactic gravitational field influences the relative motion of two close bodies in the form of galactic tide. 
\section{Model of Galaxy}
\label{sec.3.1}
A spiral galaxy like the Milky Way (Galaxy) has three basic components to its visible matter: the disk (containing the spiral arms), the halo, 
and the nucleus or central bulge \cite{maoz}. Simple models can be construct by neglecting spiral structure of the Galaxy.
There are several galactic models, as of Dauphole \cite{dauphole}, which considers spherical symmetry for galactic bulge and halo and cylindrical 
symmetry for galactic disk \cite{dauphole}. Drawback of this model is that it does not yield values of Oort constants corresponding to $A=(14.2\pm0.5)\,$km s$^{~-~1}$kpc$^{~-~1}$ and 
$B=(-12.4\pm0.5)\,$km s$^{-1}$kpc$^{-1}$. Another disadvantage is that the model cannot be used as a realistic model of Galaxy for galactocentric distances 
larger than $40$ kpc, since the model produces a decreasing rotation curve for these distances.

We consider the model given in \cite{galaxy}, better consistent with values of the Oort constants and with flat rotation curve of the Galaxy. 
This model considers galactic bulge (index $b$) of Dauphole \cite{dauphole} and its gravitational potential is 
\begin{eqnarray}
\label{eq.3.3.1}
 \varPhi_{b}(r)&=&~-~\frac{G M_{b}}{\sqrt{r^{2}~+~b_{b}^{2}}},\nonumber\\
M_{b}&=&1.3955\times10^{10}M_{\odot},\nonumber\\
b_{b}&=&0.35\, kpc,
\end{eqnarray}
where $G$ is the gravitational constant. 

Galactic disk in this model \cite{galaxy, maoz} is represented by mass density function given by
\begin{equation}
\label{eq.3.3.2}
 \varrho_{d}(R,z)=\varrho_{0}\left[\exp\left(\frac{-R}{R_{d}}\right)\exp\left(\frac{-|z|}{h_{d}}\right)\right],
\end{equation}
where $R_{d}=(3.5\pm0.5)\,$kpc is the scale length of the disk\footnote{ Notice that at $R=8\,$kpc the Sun is in outer regions of the galactic disk}, $h_{d}=330\,$pc is the characteristic scale hight for the 
lower-mass (older) objects in the disk and 
$h_{d}=160\,$pc for the gas-dust disk. 

Finally, the simple model of galactic halo is given by a flat rotation curve. The rotation curve is given by circular speed $v_{h}(r)$ for spherical halo as
\begin{eqnarray}
\label{eq.3.3.3}
v_{h}^{2}(r)&=&v_{H}^{2}\biggr\{1-\alpha\frac{a_{H}}{r}\arctan\left(\frac{r}{a_{H}}\right)~-~(1~-~\alpha)\exp\left[~-~\left(\frac{r}{b_{H}}\right)\right]\biggr\}\nonumber\\
v_{H}&=&220\, km\, s^{-1},\nonumber\\
\alpha&=&0.174,\nonumber\\
a_{H}&=&0.04383\, kpc,\nonumber\\
b_{H}&=&37.3760\, kpc.
\end{eqnarray}
 For more details about the model, see \cite{galaxy}.  
\subsection{Motion in Galaxy near to the galactic equator}
The model described in \cite{galaxy} considers an approximation when global galactic gravitational field is described by cylindrically 
symmetric potential $\varPhi(R,z)$, where $R$ is the distance from the axis of rotation and $z$ the coordiante of a body above/below the galactic 
plane ($z=0$ corresponds to the galactic equatorial plane; right-handed system $x-y-z$ has its origin at the center of the Galaxy, $z$ is positively oriented 
toward the north pole of the Galaxy; $R=\sqrt{x^{2}+y^{y}}$). The galactic gravitational potential is generated by mass distribution within the Galaxy.
Then, bulge, halo and disk contribute to the total gravitational field. The bulge contribution is computed from Eq.~(\ref{eq.3.3.1}), for the 
disk contribution is given by the solution to the Poisson's equation with right side given in Eq.~(\ref{eq.3.3.2}), and, finally for the halo is given by using 
Eq.~(\ref{eq.3.3.3}). Acceleration of the body is found from the total gravitational field.

We focus on the relative motion of a object with respect to the Sun. The Sun is moving in a distance $R_{\odot}=8\,$kpc from the galactic center. Nowadays, 
the Sun is situated $30\,$pc above the galactic equatorial plane ($Z_{\odot}=30\,$pc). The Sun has rotational motion with speed $(A~-~B)R_{\odot}$ and vertical 
(in the normal direction to the galactic plane) with speed $7.3\,$km/s. As we said, our system of coordinates is the galactic center. The position vector of 
the Sun is given by $(X_{\odot},Y_{\odot},Z_{\odot})$, of the object by $(X_{\star},Y_{\star},Z_{\star})$, and the relative positon vector Sun~-~object by ${\bf r}=(\xi,\eta,\zeta)$.

The total action of all galactic components  explained in \cite{galaxy}, can be summarized in the following equations
\begin{eqnarray}
 \label{eq.x}
\frac{d^{2}X_{\star}}{dt^{2}}&=&-~\frac{v_{0}^{2}}{R_{\odot}^{2}}\biggr\{X_{\odot}~+~\xi+2\left(R_{\odot}\frac{v_{0}^{\prime}}{v_{0}}~-~1\right)\left[\left(\frac{X_{\odot}}{R_{\odot}}\right)^{2}\xi ~+~\frac{X_{\odot}Y_{\odot}}{R_{\odot}^{2}}\eta\right]\nonumber\\
&&-~X_{\odot}[\varGamma_{1}(Z_{\odot}^{2}~+~2~Z_{\odot}~\zeta)~-~\frac{1}{2}~\varGamma_{2}(Z_{\odot}^{4}~+~4 ~Z_{\odot}^{3}~\zeta)]\biggr\}~,\nonumber\\
\frac{d^{2}Y_{\star}}{dt^{2}}&=&-~\frac{v_{0}^{2}}{R_{\odot}^{2}}\biggr\{Y_{\odot}~+~\eta~+~2\left(R_{\odot}\frac{v_{0}^{\prime}}{v_{0}}~-~1\right)\left[\frac{X_{\odot}Y_{\odot}}{R_{\odot}^{2}}\xi ~+~ \left(\frac{Y_{\odot}}{R_{\odot}}\right)^{2}\eta \right]\nonumber\\
&&-~Y_{\odot}[\varGamma_{1}(Z_{\odot}^{2}~+~2~Z_{\odot}~\zeta)~-~\frac{1}{2}~\varGamma_{2}(Z_{\odot}^{4}~+~4 ~Z_{\odot}^{3}~\zeta)]\biggr\}~,\nonumber\\
\frac{d^{2}Z_{\star}}{dt^{2}}&=&-~\biggr\{4\pi G\left[\varrho_{d}\left(1~-~\frac{u}{2}|Z_{\star}|\right)~+~\varrho_{h}\right]~+~2\left(A^{2}~-~B^{2}\right)\biggr\}Z_{\star}\nonumber\\
&&-~4\pi G\biggr\{\frac{X_{\odot}}{R_{\odot}}(X_{\star}~-~X_{\odot}~+~\frac{Y_{\odot}}{R_{\odot}}(Y_{\star}~-~Y_{\odot})\biggr\}\times\nonumber\\
&&\times~\biggr\{\varrho_{d}^{\prime}\left(1~-~\frac{u}{2}|Z_{\star}|\right)~+~\varrho_{h}^{\prime}\biggr\}Z_{\star}~,
\end{eqnarray}
where higher order in $\xi$, $\eta$, $\zeta$ are neglected and 
$\varrho^{\prime}=\varrho_{d}^{\prime}~+~\varrho_{h}^{\prime}$, 
$\varrho=\varrho_{d}~+~\varrho_{h}$. For the galactic plane $[v(R)]^{2}=v_{0}^{2}\{1+2(v_{0}^{\prime}/v_{0})(X_{\odot}\xi+Y_{\odot}\eta)/R_{\odot}\}$, 
where the prime denotes differentiation with respect to $R$, $(v_{0}\equiv v_{R_{0}}$, $v_{0}^{\prime}\equiv [dv(R)/dR]_{R_{0}})$ and, again, higher orders in 
$\xi$ and $\eta$ are neglected. We dealing only with $|z|\ll 1\,kpc$.

Equations of motion for the relative motion of the object with respect to the Sun are in detail described, explained and deduced in \cite{galaxy}. From  where we have
\begin{eqnarray}
\label{eq.3.3.4}
 \frac{d^{2}\xi}{dt^{2}}&=&-~\frac{G(M_{\odot}+m)}{r^{3}}\xi~+~(A~-~B)[A~+~B ~+~ 2A\cos(2\omega_{0}t)]\xi\nonumber\\
&&-~2A(A~-~B)\sin(2\omega_{0}t)\eta~+~2(A~-~B)^{2}(\varGamma_{1}~-~\varGamma_{2} Z_{\odot}^{2})R_{\odot}Z_{\odot}\cos(\omega_{0}t)\zeta~,\nonumber\\
 \frac{d^{2}\eta}{dt^{2}}&=&-~\frac{G(M_{\odot}+m)}{r^{3}}\eta~+~(A~-~B)[A~+~B ~-~ 2A\cos(2\omega_{0}t)]\eta\nonumber\\
&&-~2A(A~-~B)\sin(2\omega_{0}t)\xi~-~2(A~-~B)^{2}(\varGamma_{1}~-~\varGamma_{2} Z_{\odot}^{2})R_{\odot}Z_{\odot}\sin(\omega_{0}t)\zeta~,\nonumber\\
\frac{d^{2}\zeta}{dt^{2}}&=&-~\frac{G(M_{\odot}+m)}{r^{3}}\zeta~-~4\pi G\Big\{\varrho\zeta~-~\varrho_{d}\frac{u}{2}\left[|Z_{\odot}~+~\zeta|(Z_{\odot}~+~\zeta)~-~|Z_{\odot}|Z_{\odot}\right]\Big\}\nonumber\\
&&-~2\left(A^{2}~-~B^{2}\right)\zeta\nonumber\\
&&-~4\pi G\left(\varrho^{\prime}~-~\varrho_{d}^{\prime}\frac{u}{2}|Z_{\odot}~+~\zeta|\right)(Z_{\odot}~+~\zeta)\left[\cos(\omega_{0}t)\xi~-~\sin(\omega_{0}t)\eta\right],\nonumber\\ 
r&=&\sqrt{\xi^{2}~+~\eta^{2}~+~\zeta^{2}}~,\nonumber\\
\omega_{0}&=&A~-~B~,
\end{eqnarray}
where the first term in each equation (for each coordinate) represents the gravitational interaction between them,
and for $z$ coordinates (in application to the Sun and star --or gas cloud),
\begin{eqnarray}
\label{eq.ahnarmonicoscillation}
\frac{d^{2}Z_{\odot}}{dt^{2}}&=&-~\biggr\{4\pi G\left[\varrho_{d}\left(1~-~\frac{u}{2}|Z_{\odot}|\right)~+~\varrho_{h}\right]~+~2\left(A^{2}~-~B^{2}\right)\biggr\}Z_{\odot}~,\nonumber\\
\frac{d^{2}Z_{\star}}{dt^{2}}&=&-~\biggr\{4\pi G\left[\varrho_{d}\left(1~-~\frac{u}{2}|Z_{\star}|\right)~+~\varrho_{h}\right]~+~2\left(A^{2}~-~B^{2}\right)\biggr\}Z_{\star}\nonumber\\
&&-~4\pi G\biggr\{\frac{X_{\odot}}{R_{\odot}}(X_{\star}~-~X_{\odot})~+~\frac{Y_{\odot}}{R_{\odot}}(Y_{\star}~-~Y_{\odot})\biggr\}\times\nonumber\\
&&\times~\biggr\{\varrho_{d}^{\prime}\left(1~-~\frac{u}{2}|Z_{\star}|\right)~+~\varrho_{h}^{\prime}\biggr\}Z_{\star}~,
\end{eqnarray}
where $\varrho^{\prime}=\varrho_{d}^{\prime}~+~\varrho_{h}^{\prime}$, 
$\varrho=\varrho_{d}~+~\varrho_{h}$, $X_{\odot}=R_{\odot}\cos(-\omega_{0}t)$, $Y_{\odot}=R_{\odot}\sin(-\omega_{0}t)$ \footnote{Here, the sign minus 
at angular velocity ($-\omega_{0}$) denotes negative orientation of the galactic rotation (clockwise orientation/direction of the solar motion with respect to 
the center of the Galaxy)}, $G$ is the gravitational constant, $M_{\odot}$ is the mass of the Sun and the numerical values of the used quantities are
\begin{eqnarray}
\label{eq.3.3.5}
 A&=&14.2\, km\, s^{-1} kpc^{-1}~,\nonumber\\
 B&=&-~12.4\,km\, s^{-1} kpc^{-1}~,\nonumber\\
\varGamma_{1}&=&0.124\, kpc^{-2}~,\nonumber\\
\varGamma_{2}&=&1.586\, kpc^{-4}~,\nonumber\\
\varrho_{d}&=&0.126\, M_{\odot}pc^{-3}~,\nonumber\\
\varrho_{h}&=&0.004\, M_{\odot}pc^{-3}~,\nonumber\\
\varrho_{d}^{\prime}&=&-~0.0360\, M_{\odot}pc^{-3}kpc^{-1}~,\nonumber\\
\varrho_{h}^{\prime}&=&-~ 0.0006\, M_{\odot}pc^{-3}kpc^{-1}~,\nonumber\\
u&=&3.3\,kpc^{~-~1}~.
\end{eqnarray}
In our computations it is important to consider two cases in the density: $u=0\,$kpc$^{-1}$ and $u=3.3\,$kpc$^{-1}$, 
and we expect better results for the second case. In next sections we solve the equations of motion for any value of $u$. 
As first step we analytically solve the equation for the Sun, which represents vertical motion.
\subsection[Solar oscillations along the z-axis]{Analytical solution to the solar equation of motion along the z-axis}
\label{subsec.3.1.1}
In this subsection we give an analytical solution to the solar equation of motion for the $z$ direction, which is given by
\begin{equation}
\label{eq.3.3.6}
 \frac{d^{2}Z_{\odot}}{dt^{2}}=-~\Big\{4\pi G\left[\varrho_{d}\left(1~-~\frac{u}{2}|Z_{\odot}|\right)~+~\varrho_{h}\right]~+~2\left(A^{2}~-~B^{2}\right)\Big\}Z_{\odot}~,
\end{equation}
 which represents the solar anharmonic oscillations along the $z$-axis. 
Initial conditions for this equation are given by currently (at the time $t=0$) observed values: 
\begin{equation}
 \label{eq.initialconditions}
Z_{\odot}(0)=30\,pc,\quad\quad\dot{Z}_{\odot}(0)=7.3\,km\,s^{-1}~.
\end{equation}
Eq.~(\ref{eq.3.3.6}) can be written as
\begin{equation}
 \label{eq.3.3.7}
 \frac{d^{2}Z_{\odot}}{dt^{2}}=-~\gamma_{0}^{2}Z_{\odot}~+~\kappa|Z_{\odot}|Z_{\odot}~,
\end{equation}
where 
\begin{eqnarray}
 \label{eq.parameterssecondecuadion}
\gamma^{2}&=&4\pi G(\varrho_{d}~+~\varrho_{h})~+~2\left(A^{2}~-~B^{2}\right)~,\nonumber\\
\kappa&=&2\pi Gu\varrho_{d}~.
\end{eqnarray}
For $u=0$, $\kappa=0$, 
Eq.~(\ref{eq.3.3.6}) describes the known simple harmonic oscillations. Taking into account
the non-linear term ($u\neq 0$), we solve Eq.~(\ref{eq.3.3.7}) considering the method of succesive approximations \cite[p.~102]{landau}. 
Here, we suppose a general solution in the form of 
\begin{equation}
 \label{eq.3.3.7.1}
Z_{\odot}(t)=Z_{\odot}^{(1)}(t)~+~Z_{\odot}^{(2)}(t)~,
\end{equation}
where initial conditions for $Z_{\odot}^{(1)}$ are given by Eq.~(\ref{eq.initialconditions}) and $Z_{\odot}^{(2)}(0)=0\,pc$, 
and $\dot{Z}_{\odot}^{(2)}(0)=0\,km\,s^{-1}$ satisfy the general initial conditions, Eq.~(\ref{eq.initialconditions}). 
Also, $|Z_{\odot}^{(2)}|\ll |Z_{\odot}^{(1)}|$, and  functions $Z_{\odot}^{(1)}$ satisfy the non-perturbative equation
\begin{equation}
 \label{eq.3.3.8}
 \frac{d^{2}Z_{\odot}^{(1)}}{dt^{2}}~+~\gamma_{0}^{2}Z_{\odot}^{(1)}=0~.
\end{equation}
Eq.~(\ref{eq.3.3.8}) represents ordinary harmonic oscillations 
\begin{equation}
 \label{eq.3.3.9}
Z_{\odot}^{(1)} (t) = a_{\odot}\cos(\gamma t~+~\phi_{\odot})~,
\end{equation}
where $a_{\odot}$ and $\phi_{\odot}$ are found from initial conditions given by Eq.~(\ref{eq.initialconditions}). Eq.~(\ref{eq.3.3.9}) can be written as 
$Z_{\odot}^{(1)}=c_{1}\cos(\gamma t)~+~c_{2}\sin(\gamma t)$, then by using Eq.~(\ref{eq.initialconditions}) we get 
$c_{1}=Z_{\odot}(t=0)$ and 
$c_{2}=\dot{Z}_{\odot}(t=0)/\gamma $.  Also, it implies that $a_{\odot}=\sqrt{c_{1}^{2}~+~c_{2}^{2}}$, $\cos\phi_{\odot}=c_{1}/a_{\odot}$ 
and $\sin\phi_{\odot}=~-~c_{2}/a_{\odot}$. The angle $\phi_{\odot}$ 
is then found by using  the Tab.~\ref{t1.1}.
We look for $\gamma$ as $\gamma=\gamma_{0}~+~\gamma^{(1)}~+~\dots\,$, and we write Eq.~(\ref{eq.3.3.7}) in a convenient equivalent 
form for negative values of $Z_{\odot}$\footnote{$|Z_{\odot}|=~-~Z_{\odot}$ for negative $Z_{\odot}$}
\begin{equation}
\label{eq.3.3.10}
 \frac{\gamma_{0}^{2}}{\gamma^{2}}\ddot{Z}_{\odot}~+~\gamma_{0}^{2}Z_{\odot}=-~\kappa Z_{0}^{2}~-~\left(1~-~\frac{\gamma_{0}^{2}}{\gamma^{2}}\right)\ddot{Z}_{\odot}~.
\end{equation}
The next step is to plug Eq.~(\ref{eq.3.3.7.1}) into Eq.~(\ref{eq.3.3.10})
 \begin{eqnarray}
\label{eq.3.3.11}
LHS&=&RHS~,\nonumber\\
LHS&\equiv& \frac{\gamma_{0}^{2}}{\gamma^{2}}\ddot{Z}_{\odot}^{(1)}~+~\gamma_{0}^{2}Z_{\odot}^{(1)}~+~\frac{\gamma_{0}^{2}}{\gamma^{2}}\ddot{Z}_{\odot}^{(2)}~+~\gamma_{0}^{2}Z_{\odot}^{(2)}~,\nonumber\\
RHS&\equiv&-~\kappa \left(Z_{\odot}^{(1)}~+~Z_{\odot}^{(2)}\right)^{2}~-~\left(1~-~\frac{\gamma_{0}^{2}}{\gamma^{2}}\right)\left(\ddot{Z}_{\odot}^{(1)}~+~\ddot{Z}_{\odot}^{(2)}\right)~.
\end{eqnarray}
Taking into account first order terms and needed derivatives of $Z_{\odot}^{(1)}$, we have
\begin{equation}
\label{eq.3.3.12}
 \ddot{Z}_{\odot}^{(2)}~+~\gamma_{0}^{2}Z_{\odot}^{(2)}=-~\kappa a_{\odot}^{2}\cos^{2}(\gamma t~+~\phi_{\odot})~-~\left(1~-~\frac{\gamma_{0}^{2}}{\gamma^{2}}\right)\left(~-~a_{\odot}\gamma^{2}\cos(\gamma t~+~\phi_{\odot})\right)~,
\end{equation}
where $\gamma > 0$, $\cos^{2}(\gamma t~+~\phi_{\odot})=1/2~+~[\cos(2\gamma t~+~2\phi_{\odot})]/2$, and 
$\gamma^{2}~-~\gamma_{0}^{2}\approx [\gamma^{(1)}]^{2} ~+~2\gamma_{0}\gamma^{(1)}$. Neglecting second order terms, 
$\gamma^{2}~-~\gamma_{0}^{2}\approx2\gamma_{0}\gamma^{(1)}$, we get
\begin{equation}
\label{eq.3.3.13}
 \ddot{Z}_{\odot}^{(2)}~+~\gamma_{0}^{2}Z_{\odot}^{(2)}=-~\frac{\kappa a_{\odot}^{2}}{2}~-~\kappa a_{\odot}^{2}\cos(2\gamma t~+~2\phi_{\odot})~+~ 2\gamma_{0}^{2}\gamma^{(1)}\cos(\gamma t~+~\phi_{\odot})~.
\end{equation}
We assume that there is no resonance, then the amplitude of $\cos(\gamma t+\phi_{\odot})$ (on the right side) have to be zero, 
then for $\gamma^{(1)}=0$, Eq.~(\ref{eq.3.3.13}) gives
\begin{equation}
\label{eq.3.3.14}
 \ddot{Z}_{\odot}^{(2)}~+~\gamma_{0}^{2}Z_{\odot}^{(2)}=-~\frac{\kappa a_{\odot}^{2}}{2}~-~\kappa a_{\odot}^{2}\cos(2\gamma t~+~2\phi_{\odot})~,\quad \gamma=\gamma_{0}~,
\end{equation}
with initial conditions given by $Z_{\odot}^{(2)}(0)=0\,pc$, and $\dot{Z}_{\odot}^{(2)}(0)=0\,km\,s^{-1}$.
Eq.~(\ref{eq.3.3.14}) is a nonhomogeneous second order linear equation. Recall that the general solution is given by
$Z_{\odot}^{(2)}=Z_{0,h}^{(2)}~+~Z_{0,p}^{(2)}$, where $Z_{0,h}^{(2)}$ is the solution of the associated homogeneous equation 
(without right side) and $Z_{0,p}^{(2)}$ is a particular solution of Eq.~(\ref{eq.3.3.14}). 

Now, the solution of the homogeneous equation is well known and we look for it in the form of
\begin{equation}
 \label{eq.3.3.15}
Z_{0,h}^{(2)}(t)=A\cos(\gamma t~+~\phi^{\prime})=c_{1}^{\prime}\cos(\gamma t)~+~c_{2}^{\prime}\sin(\gamma t)~,
\end{equation}
where $A$ and $\phi^{\prime}$ are found from initial conditions and can be 
computed by using $c_{1,2}^{\prime}$ as $A=\sqrt{c_{1}^{\prime 2}~+~c_{1}^{\prime 2}}$, $\cos\phi^{\prime}=c_{1}^{\prime}/A$ and $\sin\phi^{\prime}=~-~c_{2}^{\prime}/A$. 
The angle $\phi^{\prime}$ may be computed by using  the Tab.~\ref{t1.1}, or we just use the second expression of Eq.~(\ref{eq.3.3.15}).

We guess the particular solution to be
\begin{equation}
 \label{eq.3.3.16}
Z_{0,p}^{(2)}(t)=b\cos(2\gamma t~+~2\phi_{\odot})~+~c~.
\end{equation}
Then, putting the previous equation and its respective derivatives into Eq.~(\ref{eq.3.3.14}) we find $b$ and $c$. 
For a better description of negative and positive values of $Z_{\odot}$, we define the parameter $q$ as
\[
  q_{\odot} = \left\{
  \begin{array}{l l}
    ~+~1, & \quad Z_{\odot} > 0\\
    ~-~1, & \quad Z_{\odot} < 0\\
  \end{array} \right.
\]

Hence, for both negative and positive values of $Z_{\odot}$, the particular solution of Eq.~(\ref{eq.3.3.14}) is given by
\begin{equation}
 \label{eq.3.3.17}
Z_{0,p}^{(2)}(t)=-~q_{\odot}\frac{\kappa a_{\odot}^{2}}{2\gamma_{0}^{2}}\left[\frac{\cos(2\gamma t~+~2\phi_{\odot})}{3}~~-~~1\right]
\end{equation}
As we wrote above, $Z_{\odot}^{(2)}=Z_{0,h}^{(2)}~+~Z_{0,p}^{(2)}$, then
\begin{equation}
 \label{eq.3.3.18}
Z_{\odot}^{(2)}(t)=c_{1}^{\prime}\cos(\gamma t)~+~c_{2}^{\prime}\sin(\gamma t)~-~q_{\odot}\frac{\kappa a_{\odot}^{2}}{2\gamma_{0}^{2}}\left[\frac{\cos(2\gamma t~+~2\phi_{\odot})}{3}~-~1\right]~.
\end{equation}
As last step, we have to evaluate the solution in initial conditions to find $c_{1}^{\prime}$ and $c_{2}^{\prime}$
\begin{equation}
\label{eq.3.3.19}
 c_{1}^{\prime}=q_{\odot}\frac{\kappa a_{\odot}^{2}}{2\gamma^{2}}\left[\frac{\cos(2\phi_{\odot})}{3}~-~1\right]~,\quad
 c_{2}^{\prime}=-~q_{\odot}\frac{\kappa a_{\odot}^{2}}{\gamma^{2}}\frac{\cos(2\phi_{\odot})}{3}~.
\end{equation}
Then,
\begin{eqnarray}
 \label{eq.3.3.20}
Z_{\odot}^{(2)}(t)&=&q_{\odot}\frac{\kappa a_{\odot}^{2}}{2\gamma^{2}}\left[\frac{\cos(2\phi_{\odot})}{3}~-~1\right]\cos(\gamma t)
~-~q_{\odot}\frac{\kappa a_{\odot}^{2}}{\gamma^{2}}\frac{\cos(2\phi_{\odot})}{3}\sin(\gamma t)\nonumber\\
&&-~q_{\odot}\frac{\kappa a_{\odot}^{2}}{2\gamma_{0}^{2}}\left[\frac{\cos(2\gamma t~+~2\phi_{\odot})}{3}~-~1\right]~.
\end{eqnarray}
Putting back Eq.~(\ref{eq.3.3.20}) and Eq.~(\ref{eq.3.3.9}) into Eq.~(\ref{eq.3.3.7.1}) we get
\begin{eqnarray}
 \label{eq.3.3.21}
Z_{\odot}&=&Z_{\odot}^{(1)}~+~Z_{\odot}^{(2)}~,\nonumber\\
&=&Z_{\odot}(0)\,\cos(\gamma t)~+~\frac{\dot{Z}_{\odot}(0)}{\gamma}\sin(\gamma t)\nonumber\\
&&+~q_{\odot}\frac{\kappa a_{\odot}^{2}}{2\gamma^{2}}\left[\frac{\cos(2\phi_{\odot})}{3}~-~1\right]\cos(\gamma t)
~-~q_{\odot}\frac{\kappa a_{\odot}^{2}}{\gamma^{2}}\frac{\cos(2\phi_{\odot})}{3}\sin(\gamma t)\nonumber\\
&&-~q_{\odot}\frac{\kappa a_{\odot}^{2}}{2\gamma_{0}^{2}}\left[\frac{\cos(2\gamma t~+~2\phi_{\odot})}{3}~-~1\right]~.
\end{eqnarray}
Previous equation represents anharmonic oscillations of the Sun along the z~-~axis across the galactic equator.  
\section{Simple model}
\label{sec.3.2}
For a better understanding we propose a simple model in the relative motion of the Sun-object, e.g. a star or an interstellar gas cloud. In this simple model 
we suppose a closest motion of the object with respect to the Sun and we do not consider gravitational interaction 
between these two bodies. Since we focus on the closest stars  with respect to the Sun, it is a good approach to consider that the 
relative motion of the object with respect to the Sun depends linearly on time for 
$x$ and $y$ coordinates and that there are anharmonic oscillations along the $z$-axis. Then,
\begin{eqnarray}
 \label{eq.simple.1}
\Delta X(t)&=&\Delta \dot{X}_{0}(t_{0})~(t-t_{0}) ~+~ \Delta X_{0}(t_{0})~,\nonumber\\
\Delta Y(t)&=&\Delta \dot{Y}_{0}(t_{0})~(t-t_{0}) ~+~ \Delta Y_{0}(t_{0})~,\nonumber\\
\Delta Z(t)&=&Z_{\star}(t)~-~Z_{\odot}(t),
\end{eqnarray}
where initial conditions at $t_{0}$, $\Delta X_{0}(t_{0})$, $\Delta \dot{X}_{0}(t_{0})$, $\Delta Y_{0}(t_{0})$, 
$\Delta \dot{Y}_{0}(t_{0})$, $\Delta Z_{0}(t_{0})$ and $\Delta \dot{Z}_{0}(t_{0})$ are given by observational data that we measure with respect to the Sun. 
For a given time $t_{0}$, we have,
\begin{eqnarray}
 \label{eq.rel1}
\Delta X_{0}(t_{0})&=&X_{\star}(t_{0}~-~r_{0}/c)~-~X_{\odot}(t_{0})~,\nonumber\\ 
\Delta Y_{0}(t_{0})&=&Y_{\star}(t_{0}~-~r_{0}/c)~-~Y_{\odot}(t_{0})~,\nonumber\\
\Delta Z_{0}(t_{0})&=&Z_{\star}(t_{0}~-~r_{0}/c)~-~Z_{\odot}(t_{0})~,
\end{eqnarray}
and
\begin{eqnarray}
 \label{eq.rel2}
\Delta \dot{X}_{0}(t_{0})&=&\dot{X}_{\star}(t_{0}~-~r_{0}/c)~-~\dot{X}_{\odot}(t_{0})~,\nonumber\\
\Delta \dot{Y}_{0}(t_{0})&=&\dot{Y}_{\star}(t_{0}~-~r_{0}/c)~-~\dot{Y}_{\odot}(t_{0})~,\nonumber\\
\Delta \dot{Z}_{0}(t_{0})&=&\dot{Z}_{\star}(t_{0}~-~r_{0}/c)~-~\dot{Z}_{\odot}(t_{0})~,
\end{eqnarray}
where physics is respected. The speed of light, $c$, may play an important role for the signals coming from the 
studied object. For this reason we subtract the travelled time of the signal $|{\bf r}_{0}|/c$. $|{\bf r}_{0}|=r_{0}$ 
is the heliocentric distance of the object at the time $t_{0}$.

For the motion along the $z$-axis  we consider Eqs.~(\ref{eq.ahnarmonicoscillation}).   
Since we consider objects which are close to the Sun, it is a good approach to describe the object's motion by the first part of the first equation in 
Eqs.~(\ref{eq.ahnarmonicoscillation}) (the same equation as for the Sun, first equation in Eqs.~(\ref{eq.ahnarmonicoscillation}), 
the difference lies in initial conditions, only). In reality, the approximation represented by Eqs.~(\ref{eq.simple.1}) 
is not required in general case.

\subsection{Star's oscillation along the z-axis}
\label{subsec.3.2.1}
In this subsection we describe the oscillations of the object along the $z$-axis, where we do not take into account 
the gravitational interaction Sun-object. As we said, we consider that the motion 
of the object is described by the same equation as for the Sun (first part of the second equation given in Eqs.~(\ref{eq.ahnarmonicoscillation})). 
Then, 
\begin{equation}
\label{eq.cometoscillation}
 \frac{d^{2}Z_{\star}}{dt^{2}}=~-~\Big\{4\pi G\left[\varrho_{d}\left(1~-~\frac{u}{2}|Z_{\star}|\right)~+~\varrho_{h}\right]~+~2\left(A^{2}~-~B^{2}\right)\Big\}Z_{\star},
\end{equation}
which represents the object's anharmonic oscillations along the $z$~-~axis. 
We get initial conditions from the third equation of ~(\ref{eq.rel1}) and of ~(\ref{eq.rel2}),
\begin{equation}
 \label{eq.initialconditionscomet}
Z_{\star}(t_{0}~-~r_{0}/c)=\Delta Z_{0}(t_{0}) ~+~ Z_{\odot}(t_{0}) ,\quad\quad\dot{Z}_{\star}(t_{0}~-~r_{0}/c)=\Delta \dot{Z}_{0}(t_{0}) ~+~ \dot{Z}_{\odot}(t_{0}).
\end{equation}
where $\Delta Z_{0}(t_{0})$ and $\Delta \dot{Z}_{0}(t_{0})$ are known values from observations. Then, doing the same as for the solar motion along the $z$-axis, we get
\begin{eqnarray}
 \label{eq.3.3.22}
Z_{\star}(t)&=&Z_{\star}^{(1)}(t)~+~Z_{\star}^{(2)}(t)~,\nonumber\\
&=&Z_{\star}(0)\,\cos(\gamma t)~+~\frac{\dot{Z}_{\star}(0)}{\gamma}\sin(\gamma t)\nonumber\\
&&~+~q_{\star}\frac{\kappa a_{\star}^{2}}{2\gamma^{2}}\left[\frac{\cos(2\phi_{\star})}{3}~-~1\right]\cos(\gamma t)
~-~q_{\star}\frac{\kappa a_{\star}^{2}}{\gamma^{2}}\frac{\cos(2\phi_{\star})}{3}\sin(\gamma t)\nonumber\\
&&~-~q_{\star}\frac{\kappa a_{\star}^{2}}{2\gamma_{0}^{2}}\left[\frac{\cos(2\gamma t~+~2\phi_{\star})}{3}~-~1\right]~.
\end{eqnarray}
where,
\[
  q_{\star} = \left\{
  \begin{array}{l l}
    ~+~1, & \quad Z_{\star} > 0\\
    ~-~1, & \quad Z_{\star} < 0\\
  \end{array} \right.
\]
Homogeneous equation for the star, $Z_{\star}^{(1)}=a_{\star}\cos(\gamma t +\phi_{\star})$, can be written as 
$Z_{\star}^{(1)}=c_{1}\cos(\gamma t)~+~c_{2}\sin(\gamma t)$, then by using initial conditions at $t=0$ we get 
$c_{1}=Z_{\star}(t=0)$ and $c_{2}=\dot{Z}_{\star}(t=0)/\gamma $.  Also, it implies that $a_{\star}=\sqrt{c_{1}^{2}~+~c_{2}^{2}}$, 
$\cos\phi_{\star}=c_{1}/a_{\star}$ and $\sin\phi_{\star}=~-~c_{2}/a_{\star}$. 
The angle $\phi_{\star}$ is then found by using  the Tab.~\ref{t1.1}.

In the next subsection we show the way how to obtain 
initial conditions, $Z_{\star}(0)$ and $\dot{Z}_{\star}(0)$, for the case when $u=0$. We also show how to find these initial conditions 
for others values of $u$.
\subsection{Initial conditions for the motion along the $z$-axis ($u=0$)}
Here, we show a way to find initial conditions for the motion along the $z$-axis when $u=0$. 
From Eq.~(\ref{eq.3.3.22}), the $z$-component of the star's position  at $t_{0}=0$ is
\begin{eqnarray}
 \label{eq.3.3.22.1}
Z_{\star}\left(-\frac{r_{0}}{c}\right)&=&Z_{\star}(0)\,\cos\left(\gamma \frac{r_{0}}{c}\right)~-~\frac{\dot{Z}_{\star}(0)}{\gamma}\sin\left(\gamma \frac{r_{0}}{c}\right)
\end{eqnarray}
and the $z$-component of the star's velocity  at $t_{0}=0$ is
\begin{eqnarray}
 \label{eq.3.3.22.2}
\dot{Z}_{\star}\left(-\frac{r_{0}}{c}\right)&=&Z_{\star}(0)\,\gamma\,\sin\left(\gamma \frac{r_{0}}{c}\right)~+~\dot{Z}_{\star}(0)\cos\left(\gamma \frac{r_{0}}{c}\right)\nonumber\\
\end{eqnarray}
where, 
From Eq.~(\ref{eq.initialconditionscomet}) we compute the left side of Eq.~(\ref{eq.3.3.22.1}) and of Eq.~(\ref{eq.3.3.22.2}). 
Then, $Z_{\star}(-r_{0}/c)$ and $\dot{Z}_{\star}(-r_{0}/c)$ are given by
\begin{eqnarray}
 \label{eq.other}
Z_{\star}\left(-\frac{r_{0}}{c}\right)&=&\Delta Z_{0}(0) ~+~ Z_{\odot}(0)~,\nonumber\\
\dot{Z}_{\star}\left(-\frac{r_{0}}{c}\right)&=&\Delta \dot{Z}_{0}(0) ~+~ \dot{Z}_{\odot}(0)~.
\end{eqnarray}
In Eqs.~(\ref{eq.other}), $\Delta Z_{0}(0)$ and $\Delta \dot{Z}_{0}(0)$ are given from measured data, and, $Z_{\odot}(0)$ and $\dot{Z}_{\odot}$ 
are known values from Eqs.~(\ref{eq.initialconditions}), then $Z_{\star}(-r_{0}/c)$ and $\dot{Z}_{\star}(-r_{0}/c)$ are known values. 
Since in Eq.~(\ref{eq.3.3.22.1}) and in Eq.~(\ref{eq.3.3.22.2}) the unique unknown values are 
$Z_{\star}(0)$ and $\dot{Z}_{\star}(0)$, we have a nonhomogeneous system of equations. Putting to the right side all known values for the system, we get
\begin{eqnarray}
 \label{eq.systemadeeq}
Z_{\star}(0)\,\cos\left(\gamma \frac{r_{0}}{c}\right)~-~\frac{\dot{Z}_{\star}(0)}{\gamma}\sin\left(\gamma \frac{r_{0}}{c}\right)&=&g_{1}~,\nonumber\\
Z_{\star}(0)\,\gamma\,\sin\left(\gamma \frac{r_{0}}{c}\right)~+~\dot{Z}_{\star}(0)\cos\left(\gamma \frac{r_{0}}{c}\right)&=&g_{2}~,
\end{eqnarray}
where $g_{1}$ and $g_{2}$ are known values given by
\begin{eqnarray}
 \label{eq.aab}
g_{1}&=&Z_{\star}\left(-\frac{r_{0}}{c}\right)~,\\
g_{2}&=&\dot{Z}_{\star}\left(-\frac{r_{0}}{c}\right)~.
\end{eqnarray}
We rewrite Eqs.~(\ref{eq.systemadeeq}) as a matrix equation, $Ax=B$,
\begin{equation}
\label{eq.matrix}
\left[ \begin{array}{cc}
\cos\left(\gamma r_{0}/c\right) & -~\sin\left(\gamma r_{0}/c\right)/\gamma  \\
\gamma\sin\left(\gamma r_{0}/c\right) & \cos\left(\gamma r_{0}/c\right)\end{array} \right]\left[ \begin{array}{c}
Z_{\star}(0) \\
\dot{Z}_{\star}(0) \end{array} \right]=\left[ \begin{array}{c}
g_{1}  \\
g_{2} \end{array} \right]
\end{equation}
Here, we look for $(Z_{\star}(0),\dot{Z}_{\star}(0))^{t}$. Since $\det(A)=\cos^{2}(\gamma r_{0}/c)+\sin^{2}(\gamma r_{0}/c)
= 1 \neq 0$, 
the nonhomogeneous system of linear equations (\ref{eq.systemadeeq}) has a unique non-trivial solution. Then, by using the Cramer's rule, 
we get
\begin{equation}
\label{eq.determinante}
 Z_{\star}(0)=\left| \begin{array}{cc}
g_{1} & -~\sin\left(\gamma r_{0}/c\right)/\gamma  \\
g_{2} & \cos\left(\gamma r_{0}/c\right) \end{array} \right|~,
\end{equation}

\begin{equation}
\label{eq.determinante2}
 \dot{Z}_{\star}(0)=\left| \begin{array}{cc}
\cos\left(\gamma r_{0}/c\right) & g_{1}  \\
\gamma\sin\left(\gamma r_{0}/c\right) & g_{2} \end{array} \right|~.
\end{equation}
Thus, the solution for the star's oscillations along the $z$-axis when $u=0$ is then given by 
\begin{equation}
 \label{eq.uzerostar}
Z_{\star}(t)=Z_{\star}(0)\,\cos(\gamma t)~+~\frac{\dot{Z}_{\star}(0)}{\gamma}\sin(\gamma t)~,
\end{equation}
with initial conditions 
$Z_{\star}(0)$ and $\dot{Z}_{\star}(0)$, which are given by Eq.~(\ref{eq.determinante}) and by Eq.~(\ref{eq.determinante2}), respectively,  together with 
observational data represented by Eqs.~(\ref{eq.other}).
\subsection{Initial conditions for the object's motion along $x$, $y$ and $z$ directions ($u\neq 0$)}
Because we already showed how to find initial conditions at $t_{0}=0$ for the motion of the object along the $z$-axis for the case when oscillations 
represent harmonic motion ($u=0$), in this section we give an description 
to find initial conditions at $t_{0}$ for the object's motion along the $x$, $y$ and $z$ axis. From measurements we are able to 
know 
\begin{eqnarray}
 \label{eq.summary00}
\Delta X_{0}(t_{0})&=&X_{\star}(t_{0}~-~r_{0}/c)~-~X_{\odot}(t_{0})~,\nonumber\\ 
\Delta Y_{0}(t_{0})&=&Y_{\star}(t_{0}~-~r_{0}/c)~-~Y_{\odot}(t_{0})~,\nonumber\\ 
\Delta Z_{0}(t_{0})&=&Z_{\star}(t_{0}~-~r_{0}/c)~-~Z_{\odot}(t_{0})~,
\end{eqnarray}
and
\begin{eqnarray}
 \label{eq.summary01}
\Delta \dot{X}_{0}(t_{0})&=&\dot{X}_{\star}(t_{0}~-~r_{0}/c)~-~\dot{X}_{\odot}(t_{0})~,\nonumber\\
\Delta \dot{Y}_{0}(t_{0})&=&\dot{Y}_{\star}(t_{0}~-~r_{0}/c)~-~\dot{Y}_{\odot}(t_{0})~,\nonumber\\ 
\Delta \dot{Z}_{0}(t_{0})&=&\dot{Z}_{\star}(t_{0}~-~r_{0}/c)~-~\dot{Z}_{\odot}(t_{0})~.
\end{eqnarray}
In right sides of previous equations are represented the differences that we measure from observational data (from equatorial coordinates). 
Now, in addition to the left sides of previous equations that are known from observational data, we suppose we know the coordinates of the Sun for the time $t_{0}$. 
Then, we solve previous equation where the only unknown values are $X_{\star}(t_{0})$, $Y_{\star}(t_{0})$, $Z_{\star}(t_{0})$, $\dot{X}_{\star}(t_{0})$, 
$\dot{Y}_{\star}(t_{0})$ and $\dot{Z}_{\star}(t_{0})$.
We also suppose that the velocity and position vector of the object are infinitely differentiable in a neighborhood of $t_{0}$, 
then, considering the Taylor expansion for these object's functions we get
\begin{eqnarray}
 \label{eq.summary001}
\Delta X_{0}(t_{0})&=&X_{\star}(t_{0})~-~X_{\odot}(t_{0})~-~\dot{X}_{\star}(t_{0})\frac{r_{0}}{c}~+~\frac{1}{2}\ddot{X}_{\star}(t_{0})\left(\frac{r_{0}}{c}\right)^{2}\ldots~,\nonumber\\ 
\Delta Y_{0}(t_{0})&=&Y_{\star}(t_{0})~-~Y_{\odot}(t_{0})~-~\dot{Y}_{\star}(t_{0})\frac{r_{0}}{c}~+~\frac{1}{2}\ddot{Y}_{\star}(t_{0})\left(\frac{r_{0}}{c}\right)^{2}\ldots~,\nonumber\\
\Delta Z_{0}(t_{0})&=&Z_{\star}(t_{0})~-~Z_{\odot}(t_{0})~-~\dot{Z}_{\star}(t_{0})\frac{r_{0}}{c}~+~\frac{1}{2}\ddot{Y}_{\star}(t_{0})\left(\frac{r_{0}}{c}\right)^{2}\ldots~,
\end{eqnarray}
and
\begin{eqnarray}
 \label{eq.summary011}
\Delta \dot{X}_{0}(t_{0})&=&\dot{X}_{\star}(t_{0})~-~\dot{X}_{\odot}(t_{0})~-~\ddot{X}_{\star}(t_{0})\frac{r_{0}}{c}~+~\ldots~,\nonumber\\
\Delta \dot{Y}_{0}(t_{0})&=&\dot{Y}_{\star}(t_{0})~-~\dot{Y}_{\odot}(t_{0})~-~\ddot{Y}_{\star}(t_{0})\frac{r_{0}}{c}~+~\ldots~,\nonumber\\
\Delta \dot{Z}_{0}(t_{0})&=&\dot{Z}_{\star}(t_{0})~-~\dot{Z}_{\odot}(t_{0})~-~\ddot{Z}_{\star}(t_{0})\frac{r_{0}}{c}~+~\ldots~,
\end{eqnarray}
where,
\begin{eqnarray}
 \label{eq.summary012}
{\bf r}_{0}(t_{0})&=&[\Delta X_{0}(t_{0}),~ \Delta Y_{0}(t_{0}),~\Delta Z_{0}(t_{0})]~,\nonumber\\
{\bf v}_{0}(t_{0})&=&[\Delta \dot{X}_{0}(t_{0}),~\Delta \dot{Y}_{0}(t_{0}),~\Delta \dot{Z}_{0}(t_{0})]~,\nonumber\\
{\bf r}_{\star\odot}(t_{0})&=&[X_{\star}(t_{0})~-~X_{\odot}(t_{0}),~Y_{\star}(t_{0})~-~Y_{\odot}(t_{0}),~Z_{\star}(t_{0})~-~Z_{\odot}(t_{0})]~,\nonumber\\
{\bf v}_{\star\odot}(t_{0})&=&[\dot{X}_{\star}(t_{0})~-~\dot{X}_{\odot}(t_{0}),~\dot{Y}_{\star}(t_{0})~-~\dot{Y}_{\odot}(t_{0}),~\dot{Z}_{\star}(t_{0})~-~\dot{Z}_{\odot}(t_{0})]~.
\end{eqnarray}
From observational data we measure ${\bf r}_{0}(t_{0})$ and ${\bf v}_{0}(t_{0})$.
$\ddot{X}_{\star}(t_{0})$ and $\ddot{Y}_{\star}(t_{0})$ are given by 
\begin{eqnarray}
 \label{eq.xyt0}
\ddot{X}_{\star}(t_{0})&=&-~\frac{v_{0}^{2}}{R_{\odot}^{2}}\biggr\{X_{\odot}(t_{0})~+~\xi(t_{0})+2\left(R_{\odot}\frac{v_{0}^{\prime}}{v_{0}}~-~1\right)\times\nonumber\\
&&\times\Bigg[\left(\frac{X_{\odot}(t_{0})}{R_{\odot}}\right)^{2}~\xi(t_{0}) ~+~\frac{X_{\odot}(t_{0})~Y_{\odot}(t_{0})}{R_{\odot}^{2}}~\eta(t_{0})\Bigg]\nonumber\\
&&-~X_{\odot}(t_{0})[\varGamma_{1}(Z_{\odot}^{2}(t_{0})~+~2~Z_{\odot}(t_{0})~\zeta(t_{0}))\nonumber\\
&&-~\frac{1}{2}~\varGamma_{2}(Z_{\odot}^{4}(t_{0})~+~4 ~Z_{\odot}^{3}(t_{0})~\zeta(t_{0}))]\biggr\}~,\nonumber\\
&&\nonumber\\
\ddot{Y}_{\star}(t_{0})&=&-~\frac{v_{0}^{2}}{R_{\odot}^{2}}\biggr\{Y_{\odot}(t_{0})~+~\eta(t_{0})~+~2\left(R_{\odot}\frac{v_{0}^{\prime}}{v_{0}}~-~1\right)\times\nonumber\\
 &&\times\Bigg[\frac{X_{\odot}(t_{0})Y_{\odot}(t_{0})}{R_{\odot}^{2}}~\xi(t_{0})~+~\left(\frac{Y_{\odot}(t_{0})}{R_{\odot}}\right)^{2}~\eta(t_{0}) \Bigg]\nonumber\\
&&-~Y_{\odot}(t_{0})[\varGamma_{1}(Z_{\odot}^{2}(t_{0})~+~2~Z_{\odot}(t_{0})~\zeta(t_{0}))\nonumber\\
&&-~\frac{1}{2}~\varGamma_{2}(Z_{\odot}^{4}(t_{0})~+~4 ~Z_{\odot}^{3}(t_{0})~\zeta(t_{0}))]\biggr\}~,\nonumber\\
\ddot{Z}_{\star}(t_{0})&=&-\gamma^{2}Z_{\star}(t_{0})+\kappa Z_{\star}(t_{0})|Z_{\star}(t_{0})|~,
\end{eqnarray}
where $\xi(t_{0})=X_{\star}(t_{0})-X_{\odot}(t_{0})$, $\eta(t_{0})=Y_{\star}(t_{0})-Y_{\odot}(t_{0})$ $\zeta(t_{0})=Z_{\star}(t_{0})-Z_{\odot}(t_{0})$.
The last equation is equivalent to Eqs.~(\ref{eq.3.3.4}) without the two-body problem terms.
To have a better understanding of the motion and to 
know the role of the light velocity  it is important to find initial conditions for the motion along the $x$ and $y$ 
directions. 
Considering that measured values $\Delta \dot{X}_{0}(t_{0})$, $\Delta \dot{Y}_{0}(t_{0})$ and $\Delta \dot{Z}_{0}(t_{0})$ 
are given by first and second derivatives (the first two terms of the Taylor expansion), we may neglect higher 
terms in Eqs.~(\ref{eq.summary011}). Hence, 
\begin{eqnarray}
 \label{eq.summary013}
\Delta \dot{X}_{0}(t_{0})&=&\dot{X}_{\star}(t_{0})~-~\dot{X}_{\odot}(t_{0})~-~\ddot{X}_{\star}(t_{0})\left(\frac{r_{0}}{c}\right)~,\nonumber\\
\Delta \dot{Y}_{0}(t_{0})&=&\dot{Y}_{\star}(t_{0})~-~\dot{Y}_{\odot}(t_{0})~-~\ddot{Y}_{\star}(t_{0})\left(\frac{r_{0}}{c}\right)~,\nonumber\\
\Delta \dot{Z}_{0}(t_{0})&=&\dot{Z}_{\star}(t_{0})~-~\dot{Z}_{\odot}(t_{0})~-~\ddot{Z}_{\star}(t_{0})\left(\frac{r_{0}}{c}\right),
\end{eqnarray}
therefore, $\dot{X}_{\star}(t_{0})$, $\dot{Z}_{\star}(t_{0})$ and $\dot{Z}_{\star}(t_{0})$ are given by
\begin{eqnarray}
 \label{eq.summary014}
\dot{X}_{\star}(t_{0})&=&\Delta \dot{X}_{0}(t_{0})~+~\dot{X}_{\odot}(t_{0})~+~\ddot{X}_{\star}(t_{0})\left(\frac{r_{0}}{c}\right)~,\nonumber\\
\dot{Y}_{\star}(t_{0})&=&\Delta \dot{Y}_{0}(t_{0})~+~\dot{Y}_{\odot}(t_{0})~+~\ddot{Y}_{\star}(t_{0})\left(\frac{r_{0}}{c}\right)~,\nonumber\\
\dot{Z}_{\star}(t_{0})&=&\Delta \dot{Z}_{0}(t_{0})~+~\dot{Z}_{\odot}(t_{0})~+~\ddot{Z}_{\star}(t_{0})\left(\frac{r_{0}}{c}\right)~.
\end{eqnarray}
Putting back Eq.~(\ref{eq.summary014}) into Eq.~(\ref{eq.summary001}) we get
\begin{eqnarray}
 \label{eq.summary015}
\Delta X_{0}(t_{0})&=&X_{\star}(t_{0})~-~X_{\odot}(t_{0})\nonumber\\
&&-~\Big[\Delta \dot{X}_{0}(t_{0})+\dot{X}_{\odot}(t_{0})~+~\ddot{X}_{\star}(t_{0})\left(\frac{r_{0}}{c}\right)\Big]\frac{r_{0}}{c}\nonumber\\ 
&&+~\frac{1}{2}\ddot{X}_{\star}(t_{0})\left(\frac{r_{0}}{c}\right)^{2}~,\nonumber\\
&&\nonumber\\
\Delta Y_{0}(t_{0})&=&Y_{\star}(t_{0})~-~Y_{\odot}(t_{0})\nonumber\\
&&-~\Big[\Delta \dot{Y}_{0}(t_{0})~+~\dot{Y}_{\odot}(t_{0})+\ddot{Y}_{\star}(t_{0})\left(\frac{r_{0}}{c}\right)\Big]\frac{r_{0}}{c}\nonumber\\
&&+~\frac{1}{2}\ddot{Y}_{\star}(t_{0})\left(\frac{r_{0}}{c}\right)^{2}~,\nonumber\\
\Delta Z_{0}(t_{0})&=&Z_{\star}(t_{0})~-~Y_{\odot}(t_{0})\nonumber\\
&&-~\Big[\Delta \dot{Z}_{0}(t_{0})~+~\dot{Y}_{\odot}(t_{0})+\ddot{Z}_{\star}(t_{0})\left(\frac{r_{0}}{c}\right)\Big]\frac{r_{0}}{c}\nonumber\\
&&+~\frac{1}{2}\ddot{Z}_{\star}(t_{0})\left(\frac{r_{0}}{c}\right)^{2}~,
\end{eqnarray}
then,
\begin{eqnarray}
 \label{eq.initxy}
X_{\star}(t_{0})&=&\Delta X_{0}(t_{0})~+~X_{\odot}(t_{0})\nonumber\\
&&+~\Big[\Delta \dot{X}_{0}(t_{0})~+~\dot{X}_{\odot}(t_{0})~+~\frac{1}{2}\ddot{X}_{\star}(t_{0})\left(\frac{r_{0}}{c}\right)\Big]\frac{r_{0}}{c}~,\nonumber\\ 
&&\nonumber\\
Y_{\star}(t_{0})&=&\Delta Y_{0}(t_{0})~+~Y_{\odot}(t_{0})\nonumber\\
&&+~\Big[\Delta \dot{Y}_{0}(t_{0})~+~\dot{Y}_{\odot}(t_{0})~+~\frac{1}{2}\ddot{Y}_{\star}(t_{0})\left(\frac{r_{0}}{c}\right)\Big]\frac{r_{0}}{c}~,\nonumber\\
Z_{\star}(t_{0})&=&\Delta Z_{0}(t_{0})~+~Y_{\odot}(t_{0})\nonumber\\
&&+~\Big[\Delta \dot{Z}_{0}(t_{0})~+~\dot{Z}_{\odot}(t_{0})~+~\frac{1}{2}\ddot{Z}_{\star}(t_{0})\left(\frac{r_{0}}{c}\right)\Big]\frac{r_{0}}{c}~.
\end{eqnarray}
Previous equations represent the initial conditions at time $t_{0}$ for the object's motion along the $x$, $y$ and $z$ directions. 
To describe the relative motion along $x$, $y$ and $z$ directions, we have to know the initial conditions for each motion.
First, we find the initial conditions for the motion along $z$, then we put this into first and second equations of Eqs.~(\ref{eq.xyt0}), 
and therefore we put these two equations for $\ddot{X}_{\star}(t_{0})$ and $\ddot{Y}_{\star}(t_{0})$ into first and second equations of Eqs.~(\ref{eq.initxy}).
From this we get a system of two equations for $X_{\star}(t_{0})$ and $Y_{\star}(t_{0})$. By solving this system we find the initial conditions for the motion
along $x$ and $y$. 
Since in this Chapter we focus on the solar and star's motion along the $z$-axis, we give a complete expression for initial conditions of the motion along $z$. 
We solve the second order equation given by
\begin{equation}
 \label{eq.secondorderEq}
a~Z_{\star}^{2}(t_{0})~+~b~Z_{\star}(t_{0})~+~c=0~,
\end{equation}
then, we get 
\begin{equation}
 Z_{\star}(t_{0})=\frac{-~b~\pm~\sqrt{b^{2}-4~a~c}}{2~a}~,
\end{equation}
where,
\begin{eqnarray}
 \label{eq.seconorderEqseoncd}
a&=&\left[\kappa\left(\frac{r_{0}}{c}\right)^{2}~+~\frac{\kappa}{2}\left(\frac{r_{0}}{c}\right)^{2}\right]=\frac{3\kappa}{2}\left(\frac{r_{0}}{c}\right)^{2}~,\nonumber\\
b&=&\left[1~+~\gamma^{2}\left(\frac{r_{0}}{c}\right)^{2}~-~\frac{\gamma^{2}}{2}\left(\frac{r_{0}}{c}\right)^{2}\right]=\left[1~+~\frac{\gamma^{2}}{2}\left(\frac{r_{0}}{c}\right)^{2}\right]~,\nonumber\\
c&=&-~\left[Z_{\odot}(t_{0})~+~\dot{Z}_{\odot}(t_{0})\left(\frac{r_{0}}{c}\right)~+~\Delta Z_{0}(t_{0})~+~\Delta \dot{Z}_{0}(t_{0})\left(\frac{r_{0}}{c}\right)\right]~,
\end{eqnarray}
are computed values from known parameters described in this chapter. To compute the initial condition of the velocity at $t_{0}$
we can use the third equation of Eqs.~(\ref{eq.summary014}), from where we get
\begin{equation}
 \dot{Z}_{\star}(t_{0})=\Delta \dot{Z}_{0}(t_{0})~+~\dot{Z}_{\odot}(t_{0})~+~\left[-\gamma^{2}Z_{\star}(t_{0})+\kappa Z_{\star}(t_{0})|Z_{\star}(t_{0})|\right]\left(\frac{r_{0}}{c}\right)~,
\end{equation}
 where we replaced the value of $\ddot{Z}_{\star}(t_{0})$ by the third equation of Eqs.~(\ref{eq.xyt0}).
Here, $\gamma$ and $\kappa$ are computed from Eqs.~(\ref{eq.parameterssecondecuadion}), 
$r_{0}$, $\Delta Z_{0}(t_{0})$, $\Delta\dot{Z}_{0}(t_{0})$ are given from observational data by Eqs.~(\ref{eq.summary00},\ref{eq.summary01}), 
$c$ is the speed of light, $Z_{\odot}(t_{0})$ and $\dot{Z}_{\odot}(t_{0})$ are given by Eqs.~(\ref{eq.initialconditions}). 
\\
As we explained in this chapter, we focus on the relative motion Sun-object, where objects are considered to be close to the Sun. 
In our computations we show results for objects which are close to the Sun. Since at this moment the solar motion along the $z$-axis 
is oriented toward up, and in adition the Sun is situated $30\,pc$ above the galactic equatorial plane, we focus our study on positive 
values of $Z_{\star}(t_{0})$. Also, in our work we carry out computations assuming that $t_{0}=0$.
\subsection{The perihelion position vector}
In this subsection we compute the perihelion position vector ${\bf q}$ for this simple case.
We proceed as for the non-interacting system. We minimize the following equation
\begin{equation}
\label{eq.minimizing}
 r(t)~=~\sqrt{[\Delta X(t)]^{2}~+~[\Delta Y(t)]^{2}~+~[\Delta Z(t)]^{2}}~,
\end{equation}
where $\Delta X$, $\Delta Y$ and $\Delta Z$ are given in Eqs.~(\ref{eq.simple.1}). For $\Delta Z = Z_{\star}(t)~-~Z_{\odot}(t)$ we get
\begin{eqnarray}
\label{eq.deltaz}
\Delta Z(t)&=&Z_{\star}(t)~-~Z_{\odot}(t)~,\nonumber\\
&=&k_{1}\sin(\gamma t)~+~k_{2}\cos(\gamma t)~-~k_{3}\cos(2\gamma t)\nonumber\\
&&+~k_{4}\sin(2 \gamma t)~+~k_{5}~,
\end{eqnarray}
where, 
\begin{eqnarray}
\label{eq.stark1k2k3k4k5}
k_{1}&=&\frac{\dot{Z}_{\star}(0)~-~\dot{Z}_{\odot}(0)}{\gamma}~-~\left[q_{\star}a_{\star}^{2}\cos(2\phi_{\star})~-~q_{\odot}a_{\odot}^{2}\cos(2\phi_{\odot})\right]\frac{\kappa}{3\gamma^{2}}~,\nonumber\\
k_{2}&=&\left[Z_{\star}(0)~-~Z_{\odot}(0)\right]+\bigg\{q_{\star}a_{\star}^{2}\Big[\frac{\cos(2\phi_{\star})}{3}~-~1\Big]~-~q_{\odot}a_{\odot}^{2}\Big[\frac{\cos(2\phi_{\odot})}{3}~-~1\Big]\bigg\}\frac{\kappa}{2\gamma^{2}}~,\nonumber\\
k_{3}&=&\left[q_{\star}a_{\star}^{2}\cos(2\phi_{\star})~-~q_{\odot}a_{\odot}^{2}\cos(2\phi_{\odot})\right]\frac{\kappa}{6\gamma_{0}^{2}}~,\nonumber\\
k_{4}&=&\left[q_{\star}a_{\star}^{2}\sin(2\phi_{\star})~-~q_{\odot}a_{\odot}^{2}\sin(2\phi_{\odot})\right]\frac{\kappa}{6\gamma_{0}^{2}}~,\nonumber\\
k_{5}&=&\left(q_{\star}a_{\star}^{2}~-~q_{\odot}a_{\odot}^{2}\right)~,
\end{eqnarray}
are known values.
The point is easy, to find $(r)_{min}=r_{min}$, at first we look for $(t)_{min}=t_{min}$, and that is all. Minimizing 
Eq.~(\ref{eq.minimizing}), we find
\begin{eqnarray}
 \label{eq.timezero}
0&=&\left[\Delta \dot{X}_{0}(0)~t~+~\Delta X_{0}(0)\right]\Delta \dot{X}_{0}(0)~+~\left[\Delta \dot{Y}_{0}(0)~t~+~\Delta Y_{0}(0)\right]\Delta \dot{Y}_{0}(0)\nonumber\\
&&+\left[k_{1}\sin(\gamma t)+k_{2}\cos(\gamma t)-k_{3}\cos(2\gamma t)+k_{4}\sin(2\gamma t)+k_{5}\right]\times\nonumber\\
&&\times\left[\gamma k_{1}\cos(\gamma t)-\gamma k_{2}\sin(\gamma t)+2\gamma k_{3}\sin(2\gamma t)+2\gamma k_{4}\cos(2\gamma t)\right]~.
\end{eqnarray}
Previous equation may be written as
\begin{eqnarray}
 \label{eq.timegalac0}
0&=&\left[\Delta \dot{X}_{0}(0)~t~+~\Delta X_{0}(0)\right]\Delta \dot{X}_{0}(0)~+~\left[\Delta \dot{Y}_{0}(0)~t~+~\Delta Y_{0}(0)\right]\Delta \dot{Y}_{0}(0)\nonumber\\
&&+~\sum_{i=1}^{4}\,f_{i}~\sin(i \gamma t)~+~\sum_{j=1}^{4}\,h_{j}~\cos(j \gamma t)~.
\end{eqnarray}
Previous equation has to be solved numerically, but by using Taylor expansions for sine and cosine we can discover the role of 
higher terms. As a good approach, for short times, we can consider: $\sin(i\gamma t)\approx i\gamma t$ and 
$\cos(j\gamma t)\approx 1$, then
\begin{eqnarray}
 \label{eq.timegalac}
0&=&\Big\{ \left[\Delta \dot{X}_{0}(0)\right]^{2}~+~\left[\Delta \dot{Y}_{0}(0)\right]^{2}~+~\sum_{i=1}^{4}\,if_{i}\gamma\Big\}~t\nonumber\\
&&+~\sum_{j=1}^{4}\,h_{j}~+~\Delta X_{0}(0)~\Delta \dot{X}_{0}(0)~+~\Delta Y_{0}(0)~\Delta \dot{Y}_{0}(0)~.
\end{eqnarray}
Now, the time $t$, corresponds to $t_{min}$, then
\begin{equation}
 \label{eq.timefirstapprox}
t_{min}=~-~\frac{\Delta X_{0}(0)\Delta \dot{X}_{0}(0)~+~\Delta Y_{0}(0)\Delta \dot{Y}_{0}(0)~+~\sum_{j=1}^{4}\,h_{j}}{[\Delta \dot{X}_{0}(0)]^{2}~+~[\Delta \dot{Y}_{0}(0)]^{2}~+~\sum_{i=1}^{4}\,if_{i}\gamma}~,
\end{equation}
where, 
\begin{eqnarray}
h_{1}&=&\left(\frac{k_{2}k_{4}~+~k_{1}k_{3}}{2}~+~k_{1}k_{5}\right)\gamma~,\nonumber\\
h_{2}&=&(k_{1}k_{2}~+~2~k_{4}k_{5})\gamma~,\nonumber\\
h_{3}&=&(2~k_{2}k_{4}~-~k_{1}k_{3})\frac{\gamma}{2}~,\nonumber\\
h_{4}&=&-~\left(\frac{2~k_{1}k_{3}~+~k_{2}k_{4}}{2}~-~2~k_{3}k_{4}\right)\gamma~,
\end{eqnarray}
and
\begin{eqnarray}
f_{1}&=&\left(\frac{k_{2}k_{3}~-~k_{1}k_{4}}{2}~-~k_{2}k_{5}\right)\gamma~,\nonumber\\
f_{2}&=&\left(\frac{k_{1}^{2}~-~k_{2}^{2}}{2}~+~2~k_{3}k_{5}\right)\gamma~,\nonumber\\
f_{3}&=&3~(k_{1}k_{4}~+~k_{2}k_{3})\frac{\gamma}{2}~,\nonumber\\
f_{4}&=&\left(k_{4}^{2}~-~k_{3}^{2}\right)\gamma~.
\end{eqnarray}
Hence,
\begin{equation}
 \label{eq.rmin}
 r(t_{min})~=~\sqrt{[\Delta X(t_{min})]^{2}~+~[\Delta Y(t_{min})]^{2}~+~[\Delta Z(t_{min})]^{2}}~,
\end{equation}
represents the perihelion distance for this simple case assuming anharmonic oscillations along the $z$-axis. The perihelion position vector is given 
by Eq.~(\ref{eq.simple.1}), where for short times $t=t_{min}$ is a good approach, then
\begin{eqnarray}
 \label{eq.perihsimplemodel}
{\bf q}&=&(\Delta X(t_{min}),\Delta Y(t_{min}),\Delta Z(t_{min}))~,\nonumber\\\nonumber\\
t_{min}&=&-~\frac{\Delta X_{0}(0)\Delta \dot{X}_{0}(0)~+~\Delta Y_{0}(0)\Delta \dot{Y}_{0}(0)~+~\sum_{j=1}^{4}\,h_{j}}{[\Delta \dot{X}_{0}(0)]^{2}~+~[\Delta \dot{Y}_{0}(0)]^{2}~+~\sum_{i=1}^{4}\,if_{i}\gamma}~.
\end{eqnarray}
Solving Eq.~(\ref{eq.timegalac0}) numerically we obtain results not only for short times. 
As we said, we use observational data for our computations. These data are measured in equatorial coordinates, and since we focus on motions 
in the Galaxy, we will use the galactic coordinate system. 

\section{The Equatorial and galactic Coordinates}
\label{sec.gal}
The most frequently used such system is the equatorial coordinate system which is still related 
to planet Earth and thus convenient for observers as is used in Eqs.~(\ref{eq.rel1}, \ref{eq.rel2}) and other equations in this Thesis. As we know the first 
coordinate in the equatorial system, corresponding to the latitude, is called Declination, $\delta$, 
and is the angle between the position of an object and the celestial equator, and 
 the longitudinal coordinate, called Right Ascension, $\alpha$, that is the angle in degrees(or hours) measured from 
the Vernal Equinox along the celestial equator toward the east to the foot of the hour circle which passes through the object. 
Dealing with motions in the Galaxy, we will use the
galactic coordinate system with the galactic longitude $l$ and
the galactic latitude $b$. We consider a spherical coordinate system, with its center being at the location of the Sun. The galactic plane if the plane
of the galactic disk, i.e., it is parallel to the band of the Galaxy. The two galactic coordinates $l$ and $b$ are angular coordinates on the sphere. Here, 
$b$ denotes the galactic latitute, the angular distance of a source from the galactic plane, with $b\in[-90^{\circ}, ~+~90^{\circ}]$. The great circle 
$b=0^{\circ}$ is then located in the plane of the galactic disk and denotes the North galactic Pole, while $b=-90^{\circ}$ marks the direction to the South 
galactic Pole. The second angular coordinate is the galactic longitude $l$, with $l\in[0^{\circ}, 360^{\circ}]$. It measures the angular separation between 
the position of a source, projected perpendicularly onto the galactic disk, and the galactic center, which itself has angular coordinates $b=0^{\circ}$ and $l=0^{\circ}$.
Given $l$ and $b$ for a source, its location on the sky is fully specified. In order to specify its three-dimensional location, the distance of that source 
from us is also needed. The conversion of the positions of sources given in equatorial coordinates $(\alpha,\delta)$ to that in galactic coordinates is 
obtained from the rotation between these two coordinate systems, and is described by spherical trigonometry. In the next lines we give the necesary 
formulae to carry out this transformation.
The coordinates of a star at $t_{0}=0$ with respect to the Sun in the galactic coordinate system are
\begin{eqnarray}
 \label{eq.transcoordinates}
 \Delta X_{0}(0) &=& r~\cos l~\cos b~,\nonumber\\
 \Delta Y_{0}(0) &=& r~\sin l~\cos b~,\nonumber\\
 \Delta Z_{0}(0) &=& r~ \sin b~,
\end{eqnarray}
where $l$ and $b$ are computed from equatorial coordinates $\alpha$ and $\delta$.
The velocity of the star at $t_{0}=0$ with respect to the Sun in galactic coordinates is
\begin{eqnarray}\label{eq.333333}
\Delta \dot{X}_{0}(0) &=& \dot{r} ~\cos l ~\cos b
\nonumber \\
& & ~-~ 4.74 ~r \left \{  ( \mu_{l} \cos b ) ~\sin l
~+~ \mu_{b} ~\cos l ~\sin b \right \} ~,
\nonumber \\
\Delta \dot{Y}_{0}(0) &=& \dot{r} ~\sin l ~\cos b
\nonumber \\
& & ~+~ 4.74 ~r \left \{ ( \mu_{l} ~\cos b )
~\cos l ~-~ \mu_{b} ~\sin l ~\sin b \right \} ~,
\nonumber \\
\Delta \dot{Z}_{0}(0) &=& \dot{r} ~\sin b ~+~ 4.74 ~r ~\mu_{b} ~\cos b ~,
\end{eqnarray}
where $\dot{r}$ is the radial velocity and $\mu_{l}$, $\mu_{b}$ are the proper motions in the  galactic longitude and latitude, respectively.
Since the units used in our calculations are
$[v]$ $=$ $[\dot{r}]$ $=$ $km~s^{-1}$, $ [r] = pc$, 
$[\mu_{l}]$ $=$ $[\mu_{b}]$ $=$ $''~yr^{-1}$,
the numerical factor in Eq.~(\ref{eq.3}) is $4.74$. To get Eq.~(\ref{eq.333333}) and Eq.~(\ref{eq.transcoordinates}) we have to know the proper motions in 
$l$ and $b$ directions, and, the galactic coordinates $(l, b)$. For this reason in the next subsections we show how to carry this out.
\subsection{The conversion from $(\alpha,\delta)$ to $(l,b)$}
From observational data we know the equatorial coordinates $(\alpha,\delta)$, then the galactic coordinates $(l,b)$ can be obtained from
\begin{eqnarray}
\label{eq.4.18}
\cos b ~\cos ( l -l_{0} ) &=& \cos \delta ~\sin ( \alpha - \alpha_{0} ) ~,
\nonumber \\
\cos b ~\sin ( l - l_{0} ) &=& \sin \delta ~\cos \delta_{0}
~-~ \cos \delta ~\sin \delta_{0} ~\cos ( \alpha - \alpha_{0} ) ~,
\nonumber \\
\sin b &=& \sin \delta ~\sin \delta_{0}
~+~ \cos \delta ~\cos \delta_{0} ~\cos ( \alpha - \alpha_{0}) ~,
\end{eqnarray}
where $l_{0}$ $=$ 33.932$^{\circ}$,
$\alpha_{0}$ $=$ 192.85948$^{\circ}$, and $\delta_{0}$ $=$ 27.17825$^{\circ}$
for the Epoch J2000.0. 
If $\alpha$ and $\delta$ are given, then Eqs.~(\ref{eq.4.18}) offer
\begin{eqnarray}
\label{eq.4.19}
 b &=& \arcsin(\sin \delta ~\sin \delta_{0}~+~ \cos \delta ~\cos \delta_{0} ~\cos ( \alpha - \alpha_{0})) ~,\nonumber\\
\cos ( l -l_{0} ) &=& \frac{1}{\cos b}\cos \delta ~\sin ( \alpha - \alpha_{0} ) ~,\nonumber \\
\sin ( l - l_{0} ) &=& \frac{1}{\cos b}\sin \delta ~\cos \delta_{0}~-~ \cos \delta ~\sin \delta_{0} ~\cos ( \alpha - \alpha_{0} ) ~.
\end{eqnarray}
In Eqs.~(\ref{eq.4.19}): from the first equation we compute $b$, second and third equations allow us to 
calculate $l$ by using Tab.~\ref{t1.1}.
\subsection{Proper motions}
Stars are moving relative to us or, more precisely, relative to the Sun. To study the kinematics
of the Galaxy we need to be able to measure the velocities of stars. 
To know the components of the velocity in Eqs.~(\ref{eq.3}), besides  $v_{r}$ which is a known value for each star, 
the values of proper motions $(\mu_{l}\cos b)$ and $\mu_{b}$ are also required. It is carried out by doing the transformation from equatorial to galactic coordinates.
$(\mu_{l}\cos b)$ and $\mu_{b}$ are computed from
\begin{eqnarray}
 \label{eq.transuno}
\mu_{l}\cos b &=& \mu_{\alpha}\cos \delta \cos \psi +\mu_{\delta} \sin \psi~,\nonumber\\
\mu_{b} &=& -\mu_{\alpha}\cos\delta \sin \psi +\mu_{\delta} \cos\psi.
\end{eqnarray}
The angle $\psi$ is found by using Tab.~\ref{t1.1} from the $\sin \psi$ and $\cos \psi$ relations given in the following equations,
\begin{eqnarray}
 \label{eq.transunouno}
\cos b ~\sin \psi &=&\cos\delta_{0}~\sin(\alpha-\alpha_{0})~,\nonumber\\
\cos b ~\cos \psi &=&\sin\delta_{0}~\cos(\alpha-\alpha_{0})~-~\cos\delta_{0}~\sin\delta~\cos(\alpha-\alpha_{0})~,
\end{eqnarray}
or just by replacing in Eqs.~(\ref{eq.transuno}) the values of $\sin \psi$ and $\cos \psi$ given by
\begin{eqnarray}
 \label{eq.transunounodos}
\sin \psi &=&\frac{1}{\cos b }[\cos\delta_{0}~\sin(\alpha-\alpha_{0})]~,\nonumber\\
\cos \psi &=&\frac{1}{\cos b}[\sin\delta_{0}~\cos(\alpha-\alpha_{0})~-~\cos\delta_{0}~\sin\delta~\cos(\alpha-\alpha_{0})]~,
\end{eqnarray}
then, the proper motions in galactic coordinates to be used are:
\begin{eqnarray}
 \label{eq.transunoone}
\mu_{l}\cos b &=& \mu_{\alpha}\cos \delta \frac{1}{\cos b}[\sin\delta_{0}~\cos(\alpha-\alpha_{0})~-~\cos\delta_{0}~\sin\delta~\cos(\alpha-\alpha_{0})]\nonumber\\
&&+~\mu_{\delta} \frac{1}{\cos b }[\cos\delta_{0}~\sin(\alpha-\alpha_{0})]~,\nonumber\\
\mu_{b} &=& -~\mu_{\alpha}\cos\delta \frac{1}{\cos b }[\cos\delta_{0}~\sin(\alpha-\alpha_{0})]\nonumber\\
&&+~\mu_{\delta} \frac{1}{\cos b}[\sin\delta_{0}~\cos(\alpha-\alpha_{0})~-~\cos\delta_{0}~\sin\delta~\cos(\alpha-\alpha_{0})]~.
\end{eqnarray}

%% file: Chapter5.tex
\chapter{Summary of main equations}
\label{chapter.3}
In this chapter we  summarize the theoretical results to find the perihelion position vector ${\bf q}$ and the impact parameter $b$. We show this in 
three cases: the non-interacting system, the two-body problem, and,  the simple model, where in addition to the gravitational influence of the Sun, 
we consider that the object is perturbed by the gravitational effects of the Galaxy (the relative motion of the object with respect to the Sun depends 
linearly on time for $x$ and $y$ coordinates and that there are anharmonic oscillations along the $z$-axis).
Here, equations for the perihelion position vector and for the impact parameter  are computed from known observational data.
\section{Observational data}
In this section we describe what exactly is measured, and how it is transformed to carry out our computations.
For a given star we know from observational data the equatorial coordinates $(\alpha, \delta)$, then by using next equation 
we find the galactic coordinates $(l, b)$,
 \begin{eqnarray}
\label{eq.summary11}
 b &=& \arcsin(\sin \delta ~\sin \delta_{0}~+~ \cos \delta ~\cos \delta_{0} ~\cos ( \alpha - \alpha_{0})) ~,\nonumber\\
\cos ( l -l_{0} ) &=& \frac{1}{\cos b}\cos \delta ~\sin ( \alpha - \alpha_{0} ) ~,\nonumber \\
\sin ( l - l_{0} ) &=& \frac{1}{\cos b}\sin \delta ~\cos \delta_{0}~-~ \cos \delta ~\sin \delta_{0} ~\cos ( \alpha - \alpha_{0} ) ~.
\end{eqnarray}
The correct angle, $l$, is found by using Tab.~\ref{t1.2}.
Once that is done, we  are able to find
\begin{eqnarray}
\label{eq.summary12}
 \Delta X_{0}(t_{0}) &=& r_{0}~\cos l~\cos b~,\nonumber\\
 \Delta Y_{0}(t_{0}) &=& r_{0}~\sin l~\cos b~,\nonumber\\
 \Delta Z_{0}(t_{0}) &=& r_{0}~ \sin b~,
\end{eqnarray}
which represents the position coordinates at $t_{0}=0$  and,
\begin{eqnarray}
\label{eq.summary13}
\Delta \dot{X}_{0}(t_{0}) &=& \dot{r}_{0} ~\cos l ~\cos b
\nonumber \\
& & ~-~ 4.74 ~r_{0} \left \{  ( \mu_{l} \cos b ) ~\sin l
~+~ \mu_{b} ~\cos l ~\sin b \right \} ~,
\nonumber \\
\Delta \dot{Y}_{0}(t_{0}) &=& \dot{r}_{0} ~\sin l ~\cos b
\nonumber \\
& & ~+~ 4.74 ~r_{0} \left \{ ( \mu_{l} ~\cos b )
~\cos l ~-~ \mu_{b} ~\sin l ~\sin b \right \} ~,
\nonumber \\
\Delta \dot{Z}_{0}(t_{0}) &=& \dot{r}_{0} ~\sin b ~+~ 4.74 ~r_{0} ~\mu_{b} ~\cos b ~,
\end{eqnarray}
which allow us to compute the velocity at $t_{0}$, $r_{0}$ is the heliocentric distance of the object,
$\dot{r}_{0}$ is the radial velocity of the object with respect to the Sun, and $\mu_{l}$, $\mu_{b}$ are the proper motions in the  
galactic longitude and latitude at $t_{0}=0$, respectively. The units used in our calculations are: $[v]$ $=$ $[\dot{r}]$ $=$ $km~s^{-1}$, $ [r] = pc$, 
$[\mu_{l}]$ $=$ $[\mu_{b}]$ $=$ $''~yr^{-1}$. Also, here we notice that
\begin{eqnarray}
 \label{eq.summary0000}
{\bf r}_{0}(t_{0})&=&[\Delta X_{0}(t_{0}),~ \Delta Y_{0}(t_{0}),~\Delta Z_{0}(t_{0})]~,\nonumber\\
{\bf v}_{0}(t_{0})&=&[\Delta \dot{X}_{0}(t_{0}),~\Delta \dot{Y}_{0}(t_{0}),~\Delta \dot{Z}_{0}(t_{0})]~,\nonumber\\
\end{eqnarray}
are the measured initial velocity and position vectors of the object with respect to the Sun, which we compute from equatorial coordinates. 
In the following sections we label (as we also did in the previous section)
${\bf r}_{0}=(x_{0},y_{0},z_{0})$ and ${\bf v}_{0}=(v_{x,0},v_{y,0},v_{z,0})$.
\section{Non-interacting system}
From observational data we know ${\bf r}_{0}=(x_{0},y_{0},z_{0})$ and ${\bf v}_{0}=(v_{x,0},v_{y,0},v_{z,0})$, which  
are the initial ($t$ $=$ $0$) position and velocity vectors of the object with respect to the Sun. The impact parameter is computed as
\begin{eqnarray}
\label{eq.summary1}
 b&=&\frac{1}{v_{0}^{2}}\biggr\{\left[x_{0}\left(v_{y,0}^{2}+v_{z,0}^{2}\right)
-\left(y_{0}v_{y,0}+z_{0}v_{z,0}\right)v_{x,0}\right]^{2}\nonumber\\
&&+\left[y_{0}\left(v_{x,0}^{2}+v_{z,0}^{2}\right)-\left(x_{0}v_{x,0}+z_{0}v_{z,0}\right)v_{y,0}\right]^{2}\nonumber\\
& &+\left[z_{0}\left(v_{x,0}^{2}+v_{y,0}^{2}\right)-\left(x_{0}v_{x,0}+y_{0}v_{y,0}\right)v_{z,0}\right]^{2} \biggr\}^{1/2},
\end{eqnarray}
where $v_{0}^{2}=v_{x,0}^{2}+v_{y,0}^{2}+v_{z,0}^{2}~$.
and the perihelion position vector is 
\begin{equation}
\label{eq.summary2}
 {\bf q}={\bf r}_{b}-{\bf r_{\odot}}={\bf r}_{0}+{\bf v}_{0}t_{b},\quad 
t_{b}= -~\frac{x_{0}v_{x,0}+y_{0}v_{y,0}+z_{0}v_{z,0}}{v_{x,0}^{2}+v_{y,0}^{2}+v_{z,0}^{2}} ~.
\end{equation}
\section{The two-body problem}
Initial conditions ${\bf r}_{0}=(x_{0},y_{0},z_{0})$, ${\bf v}_{0}=(v_{x,0},v_{y,0},v_{z,0})$ $\Rightarrow$ ${\bf H}_{0}\equiv(H_{x,0},H_{y,0},H_{z,0})$, 
 ${\bf H}_{0}=\mu {\bf r_{0}}\times{\bf v_{0}}$ are given from observational data. The impact parameter, we get from 
\begin{eqnarray}
\label{eq.summary3}
 b&=&\frac{G(M+m)}{v_{\infty}^{2}}\frac{1+\cos\theta}{\sin\theta}~,\nonumber\\
&=&\frac{G(m+M)}{v_{\infty}^{2}}\cot\left(\frac{\theta}{2}\right)~,
\end{eqnarray}
where, $\theta=2\pi-\varphi$, $m$ is the mass of the object, $M$ is the solar mass, and $v_{\infty}$ is the magnitude of the velocity vector in the infinity
that is found from conservation of energy and given by
\begin{equation}
 v_{\infty}=~\sqrt{\frac{2}{\mu}\left(\frac{\mu}{2}v_{0}^{2}+\frac{k}{r_{0}} \right)}~.
\end{equation}
The perihelion position vector,
\begin{equation}
\label{eq.summary4}
{\bf q}\equiv{\bf r}_{p}=(x_{p},y_{p},z_{p}) = \left\{ \begin{array}{rrl}
x_{p}&=& r_{p}(\cos\Omega_{0}\cos\omega_{0}-\sin\Omega_{0}\sin\omega_{0}\cos i_{0}), \\
 y_{p}&=& r_{p}(\cos\omega_{0}\sin\Omega_{0}+\sin\omega_{0}\cos\Omega_{0}\cos i_{0}),\\
z_{p}&=&r_{p} \sin\omega_{0}\sin i_{0} ~.
\end{array} \right.
\end{equation}
where $|{\bf r}_{p}|\equiv|(x_{p},y_{p},y_{p})|=a(1-\epsilon)$, $i_{0}$ is computed from
\begin{equation}
 \label{eq.summary5}
i_{0}=\arccos\left(\frac{H_{z,0}}{|{\bf H}_{0}|}\right),
\end{equation}
$\Omega_{0}$ from
\begin{equation}
 \label{eq.summary6}
\sin\Omega_{0} =\frac{H_{x,0}}{|{\bf H}_{0}|\sin i_{0}}~,\quad \cos\Omega_{0}= - ~\frac{H_{y,0}}{|{\bf H}_{0}|\sin i_{0}} ~.
\end{equation}
The correct value of $\Omega_{0}$ is computed by using Tab.~\ref{t1.2}. 
$\Theta_{0}$ is computed as
\begin{equation}
 \label{eq.summary7}
\sin\Theta_{0}=\frac{z_{0}}{r_{0} \sin i_{0}}, \quad\cos\Theta_{0}=\frac{y_{0}H_{x,0}-x_{0}H_{y,0}}{r_{0}|{\bf H}_{0}|\sin i_{0}},
\end{equation} 
where, we also use Tab.~\ref{t1.2} to find the right value of $\Theta_{0}$.
And the last one, $\omega_{0}$, we compute from,
\begin{equation}
 \label{eq.summary8}
\sin(\Theta_{0}-\omega_{0})=\frac{{\bf v}_{0}\cdot{\bf e}_{r,0}}{\epsilon\sqrt{G(M+m)/p}}, \quad 
\cos(\Theta_{0}-\omega_{0})=\frac{{\bf v}_{0}\cdot{\bf e}_{t,0}}{\epsilon\sqrt{G(M+m)/p}}-\frac{1}{\epsilon},
\end{equation}
where as for the previous cases, here, we obtain the right angle by using Tab.~\ref{t1.2}. ${\bf e}_{r}$ is defined as
\begin{equation}
 \label{eq.summary9}
{\bf e}_{r}=(\cos\Omega\cos\Theta-\sin\Omega\sin\Theta\cos i,\cos\Theta\sin\Omega+\sin\Theta\cos\Omega\cos i,\sin\Theta\sin i),
\end{equation}
and, ${\bf e}_{t}$ is the unit vector onto the transversal direction
 \begin{equation}
\label{eq.summary10}
{\bf e}_{t}=(-\cos\Omega\sin\Theta-\sin\Omega\cos\Theta\cos i,-\sin\Omega\sin\Theta+\cos\Omega\cos\Theta\cos i,\cos\Theta\sin i).
 \end{equation}
At initial time $t=0$, ${\bf e}_{r,0}$ and ${\bf e}_{t,0}$ are given by angles $i_{0}$, $\Omega_{0}$ and by $\Theta_{0}$ from equations 
(\ref{eq.summary5}), (\ref{eq.summary6}), (\ref{eq.summary7}), respectively. 
For the time when the object passes the perihelion position, $\Theta_{0}$ changes its value to $\Theta_{p}$, 
it means that we also need to know $\Theta_{p}$, which can be computed assuming that $\Theta_{0}=\omega_{0}$, and $\omega_{0}$ is computed from Eq.~(\ref{eq.summary8}).
Here, $v_{0}$, ${\bf e}_{r,0}$ and ${\bf e}_{t,0}$ are computed 
from initial conditions, since ${\bf e}_{r,0}={\bf r}_{0}/|{\bf r}_{0}|$, or just by using equations ~(\ref{eq.summary5}), ~(\ref{eq.summary6}), ~(\ref{eq.summary7}).
\subsection{Choosing the right angle}
Let
\begin{eqnarray}
\label{eq.cuadrante}
\cos \vartheta &=& A ~,
\nonumber \\
\sin\vartheta&=& B ~,
\end{eqnarray}
where $A$ and $B$ are known numbers computed from the right sides of given equations. These numbers can be positive or negative, then 
we need to know which quadrant the angle belongs. It is recommended to remember where $\sin \vartheta$, $\cos \vartheta$ and $\tan\vartheta$ 
are positive (or negative). Table ~\ref{t1.1} allows us to solve problems as the given in Eq.~(\ref{eq.cuadrante}). 
\begin{table}[h]
\begin{center}
  \begin{tabular}{|c||c|c|c|}
	\hline
  & $\sin\vartheta$ $>$ $0$    &   $\sin\vartheta$ $<$ $0$ &$\sin\vartheta$ $=$ $0$ \\
	\hline \hline
$\cos\vartheta$ $>$ $0$   & $\vartheta=\arccos A$ & $\vartheta=2\pi-\arccos A$& $\vartheta=0$\\
	\hline
$\cos\vartheta$ $<$ $0$    & $\vartheta=\arccos A$ & $\vartheta=2\pi-\arccos A$& $\vartheta=\pi$\\
	\hline       
$\cos\vartheta$ $=$ $0$    & $\vartheta=\pi/2$ & $\vartheta=3\pi/2$&  \\
       \hline
\end{tabular}
\end{center}
\caption{{\small Choosing the right angle}}
\label{t1.2}
\end{table}
\section{Galactic tide}
The third equations of Eqs.~(\ref{eq.summary12}, \ref{eq.summary13}) are important in 
equations given by
\begin{eqnarray}
\label{eq.summary14}
Z_{\star}\left(-\frac{r_{0}}{c}\right)&=&\Delta Z_{0}(0) ~+~ Z_{\odot}(0)~,\nonumber\\
\dot{Z}_{\star}\left(-\frac{r_{0}}{c}\right)&=&\Delta \dot{Z}_{0}(0) ~+~ \dot{Z}_{\odot}(0)~.
\end{eqnarray}
which permits us to find $Z_{\star}(0)$ and $\dot{Z}_{\star}(0)$ for the case when $u=0$ given by
\begin{equation}
\label{eq.summary15}
 Z_{\star}(0)=\left| \begin{array}{cc}
g_{1} & -~\sin\left(\gamma r_{0}/c\right)/\gamma  \\
g_{2} & \cos\left(\gamma r_{0}/c\right) \end{array} \right|~,
\end{equation}
\begin{equation}
\label{eq.summary16}
 \dot{Z}_{\star}(0)=\left| \begin{array}{cc}
\cos\left(\gamma r_{0}/c\right) & g_{1}  \\
\sin\left(\gamma r_{0}/c\right) & g_{2} \end{array} \right|~,
\end{equation}
where $g_{1}$ and $g_{2}$ are known values given by
\begin{equation}
\label{eq.summary17}
g_{1}=Z_{\star}\left(-\frac{r_{0}}{c}\right)~,\quad \quad g_{2}=\dot{Z}_{\star}\left(-\frac{r_{0}}{c}\right)~.
\end{equation}
In Eqs.~(\ref{eq.summary14}), $\Delta Z_{0}(0)$ and $\Delta \dot{Z}_{0}(0)$ are given from measured (observational) data, from 
Eqs.~(\ref{eq.summary12}, \ref{eq.summary13}), and, $Z_{\odot}(0)$ and $\dot{Z}_{\odot}$ are given by
\begin{equation}
\label{eq.summary19}
Z_{\odot}(0)=30\,pc,\quad\quad\dot{Z}_{\odot}(0)=7.3\,km\,s^{-1}.
\end{equation}
\\
The initial conditions for the motion along the $z$-axis when $u\neq 0$, $Z_{\star}(t_{0})$, are given by
\begin{eqnarray}
 Z_{\star}(t_{0})&=&\frac{-~b~\pm~\sqrt{b^{2}-4~a~c}}{2~a}~,\nonumber\\
 \dot{Z}_{\star}(t_{0})&=&\Delta \dot{Z}_{0}(t_{0})~+~\dot{Z}_{\odot}(t_{0})~+~\left[-\gamma^{2}Z_{\star}(t_{0})+\kappa Z_{\star}(t_{0})|Z_{\star}(t_{0})|\right]\left(\frac{r_{0}}{c}\right)~,
\end{eqnarray}
where,
\begin{eqnarray}
 \label{eq.seconorderEqseoncd1}
a&=&\left[\kappa\left(\frac{r_{0}}{c}\right)^{2}~+~\frac{\kappa}{2}\left(\frac{r_{0}}{c}\right)^{2}\right]=\frac{3\kappa}{2}\left(\frac{r_{0}}{c}\right)^{2}~,\nonumber\\
b&=&\left[1~+~\gamma^{2}\left(\frac{r_{0}}{c}\right)^{2}~-~\frac{\gamma^{2}}{2}\left(\frac{r_{0}}{c}\right)^{2}\right]=\left[1~+~\frac{\gamma^{2}}{2}\left(\frac{r_{0}}{c}\right)^{2}\right]~,\nonumber\\
c&=&-~\left[Z_{\odot}(t_{0})~+~\dot{Z}_{\odot}(t_{0})\left(\frac{r_{0}}{c}\right)~+~\Delta Z_{0}(t_{0})~+~\Delta \dot{Z}_{0}(t_{0})\left(\frac{r_{0}}{c}\right)\right]~,
\end{eqnarray}
are computed values from known parameters described in the previous chapter. Here, $\gamma$ and $\kappa$ are computed from 
\begin{eqnarray}
 \label{eq.parameterssecondecuadion1}
\gamma^{2}&=&4\pi G(\varrho_{d}~+~\varrho_{h})~+~2\left(A^{2}~-~B^{2}\right)~,\nonumber\\
\kappa&=&2\pi Gu\varrho_{d}~.
\end{eqnarray} 
$r_{0}$, $\Delta Z_{0}(t_{0})$, $\dot{Z}_{0}(t_{0})$ are given from observational data, $c$ is the speed of light, 
and, $Z_{\odot}(t_{0})$ and $\dot{Z}_{\odot}(t_{0})$ from Eqs.~(\ref{eq.summary19}). As initial conditions for the star's motion along 
the $z$-axis we consider positive values, only.  
Now, we are able to compute $\Delta Z (t)$ for any value of $u$.
To find the perihelion position vector we minimize
\begin{equation}
\label{eq.minimizing2}
 r(t)~=~\sqrt{[\Delta X(t)]^{2}~+~[\Delta Y(t)]^{2}~+~[\Delta Z(t)]^{2}}~,
\end{equation}
where $\Delta X$, $\Delta Y$ and $\Delta Z$ are given in Eqs.~(\ref{eq.simple.1}). For $\Delta Z = Z_{\star}(t)~-~Z_{\odot}(t)$ we get
\begin{eqnarray}
\label{eq.deltaz}
\Delta Z(t)&=&Z_{\star}(t)~-~Z_{\odot}(t)~,\nonumber\\
&=&k_{1}\sin(\gamma t)~+~k_{2}\cos(\gamma t)~-~k_{3}\cos(2\gamma t)\nonumber\\
&&+~k_{4}\sin(2 \gamma t)~+~k_{5}~,
\end{eqnarray}
where, 
\begin{eqnarray}
\label{eq.stark1k2k3k4k5}
k_{1}&=&\frac{\dot{Z}_{\star}(0)~-~\dot{Z}_{\odot}(0)}{\gamma}~-~\left[q_{\star}a_{\star}^{2}\cos(2\phi_{\star})~-~q_{\odot}a_{\odot}^{2}\cos(2\phi_{\odot})\right]\frac{\kappa}{3\gamma^{2}}~,\nonumber\\
k_{2}&=&\left[Z_{\star}(0)~-~Z_{\odot}(0)\right]+\bigg\{q_{\star}a_{\star}^{2}\Big[\frac{\cos(2\phi_{\star})}{3}~-~1\Big]~-~q_{\odot}a_{\odot}^{2}\Big[\frac{\cos(2\phi_{\odot})}{3}~-~1\Big]\bigg\}\frac{\kappa}{2\gamma^{2}}~,\nonumber\\
k_{3}&=&\left[q_{\star}a_{\star}^{2}\cos(2\phi_{\star})~-~q_{\odot}a_{\odot}^{2}\cos(2\phi_{\odot})\right]\frac{\kappa}{6\gamma_{0}^{2}}~,\nonumber\\
k_{4}&=&\left[q_{\star}a_{\star}^{2}\sin(2\phi_{\star})~-~q_{\odot}a_{\odot}^{2}\sin(2\phi_{\odot})\right]\frac{\kappa}{6\gamma_{0}^{2}}~,\nonumber\\
k_{5}&=&\left(q_{\star}a_{\star}^{2}~-~q_{\odot}a_{\odot}^{2}\right)~,
\end{eqnarray}
are known values. Then we get
\begin{eqnarray}
 \label{eq.timegalac0}
0&=&\left[\Delta \dot{X}_{0}(0)~t~+~\Delta X_{0}(0)\right]\Delta \dot{X}_{0}(0)~+~\left[\Delta \dot{Y}_{0}(0)~t~+~\Delta Y_{0}(0)\right]\Delta \dot{Y}_{0}(0)\nonumber\\
&&+~\sum_{i=1}^{4}\,f_{i}~\sin(i \gamma t)~+~\sum_{j=1}^{4}\,h_{j}~\cos(j \gamma t)~.
\end{eqnarray}
Previous equation has to be solved numerically, but for short times we just consider the first terms of Taylor expansions for sine and cosine, then
\begin{eqnarray}
0&=&\Big\{ \left[\Delta \dot{X}_{0}(0)\right]^{2}~+~\left[\Delta \dot{Y}_{0}(0)\right]^{2}~+~\sum_{i=1}^{4}\,if_{i}\gamma\Big\}~t\nonumber\\
&&+~\sum_{j=1}^{4}\,h_{j}~+~\Delta X_{0}(0)~\Delta \dot{X}_{0}(0)~+~\Delta Y_{0}(0)~\Delta \dot{Y}_{0}(0)~.
\end{eqnarray}
Now, the time $t$, corresponds to $t_{min}$, then
\begin{equation}
t_{min}=~-~\frac{\Delta X_{0}(0)\Delta \dot{X}_{0}(0)~+~\Delta Y_{0}(0)\Delta \dot{Y}_{0}(0)~+~\sum_{j=1}^{4}\,h_{j}}{[\Delta \dot{X}_{0}(0)]^{2}~+~[\Delta \dot{Y}_{0}(0)]^{2}~+~\sum_{i=1}^{4}\,if_{i}\gamma}~,
\end{equation}
where, 
\begin{eqnarray}
h_{1}&=&\left(\frac{k_{2}k_{4}~+~k_{1}k_{3}}{2}~+~k_{1}k_{5}\right)\gamma~,\nonumber\\
h_{2}&=&(k_{1}k_{2}~+~2~k_{4}k_{5})\gamma~,\nonumber\\
h_{3}&=&(2~k_{2}k_{4}~-~k_{1}k_{3})\frac{\gamma}{2}~,\nonumber\\
h_{4}&=&-~\left(\frac{2~k_{1}k_{3}~+~k_{2}k_{4}}{2}~-~2~k_{3}k_{4}\right)\gamma~,
\end{eqnarray}
and
\begin{eqnarray}
f_{1}&=&\left(\frac{k_{2}k_{3}~-~k_{1}k_{4}}{2}~-~k_{2}k_{5}\right)\gamma~,\nonumber\\
f_{2}&=&\left(\frac{k_{1}^{2}~-~k_{2}^{2}}{2}~+~2~k_{3}k_{5}\right)\gamma~,\nonumber\\
f_{3}&=&3~(k_{1}k_{4}~+~k_{2}k_{3})\frac{\gamma}{2}~,\nonumber\\
f_{4}&=&\left(k_{4}^{2}~-~k_{3}^{2}\right)\gamma~.
\end{eqnarray}
Hence,
\begin{equation}
 \label{eq.rmin}
 r(t_{min})~=~\sqrt{[\Delta X(t_{min})]^{2}~+~[\Delta Y(t_{min})]^{2}~+~[\Delta Z(t_{min})]^{2}}~,
\end{equation}
represents the perihelion distance for this simple case assuming anharmonic oscillations along the $z$-axis. The perihelion position vector is then given 
by Eq.~(\ref{eq.simple.1}), where for short times $t=t_{min}$ is a good approach, then
\begin{eqnarray}
 \label{eq.perihsimplemodel}
{\bf q}&=&(\Delta X(t_{min}),\Delta Y(t_{min}),\Delta Z(t_{min}))~,\nonumber\\\nonumber\\
t_{min}&=&-~\frac{\Delta X_{0}(0)\Delta \dot{X}_{0}(0)~+~\Delta Y_{0}(0)\Delta \dot{Y}_{0}(0)~+~\sum_{j=1}^{4}\,h_{j}}{[\Delta \dot{X}_{0}(0)]^{2}~+~[\Delta \dot{Y}_{0}(0)]^{2}~+~\sum_{i=1}^{4}\,if_{i}\gamma}~.
\end{eqnarray}
Solving Eq.~(\ref{eq.timegalac0}) numerically we obtain results not only for short times. 

%% file: Chapter6.tex
\chapter{Applications}
\label{chapter.4}
In this chapter we apply our results to specific stars. As we said, we focus on stars with close passages to the Sun. 
Studied stars were selected in \cite{comet} and some of them in \cite{garcia}. Data from \cite{comet} 
are in Tab.~\ref{t0.1}. Computed perihelion distances by Garcia \cite{garcia} may differ in more than 60 \% 
from those given by \cite{comet}. Although \cite{garcia} present results for two methods of calculations
(integrated orbits and rectilinear motion), \cite{comet} does not present the method of calculation. 
The results presented in Table 2 by \cite{garcia} show that both methods (integrated orbits and rectilinear motion)
are consistent for almost all considered stars: the relative error is less than 1 \% for many stars.
The data published by \cite{comet} are used to compare with our results. 
\begin{table}[ht]
\begin{center}
  \begin{tabular}{|c||c|}
	\hline
  star & $|{\bf q}_{Dyb}|$ [$pc$] \\
	\hline \hline
 Gl 710 & 0.209 \\\hline
 Gl 127.1A & 0.803 \\\hline
 Gl 445 & 1.071 \\\hline
 Barnard's Star &1.146 \\\hline
 Gl 217.1 & 1.316 \\\hline
 Gl 729 &1.988 \\\hline
 GJ 2046 & 2.008  \\\hline
 Gl 54.1 & 2.425 \\\hline
 LP 816-60 & 2.485 \\\hline
\end{tabular}
\end{center}
\caption{{\small Computed perihelion distances by Dybczy\'{n}ski ($|{\bf q}|$)\cite{comet}}}
\label{t0.1}
\end{table}
In Tab.~\ref{t1.2} we show all observed parameters used in our calculations. These were taken from \emph{ARICNS} \cite{ari}, 
from \emph{NStED Data Base} \cite{Stauffer} and from \emph{SIMBAD Data Base} \cite{simbad}. Radial velocity measurements were 
obtained from the astronomical literature, in particular of Wilson (1953) \cite{wilson}, and from other sources as \cite{garcia, ari}.
\begin{table}[h!]
\begin{center}
  \begin{tabular}{|c|c|c|c|c|c|c|c|}
	\hline
  Star & $\alpha$ [$^\circ$] & $\delta$ [$^\circ$] &$\mu_{\alpha}$& $\mu_{\delta}$& $r_{0}$ [pc]& $\dot{r}_{0}$ [km/s]& $m$ [$M_{\odot}$] \\
	\hline \hline
 Gl 710 & 274.963 & -1.990 & -0.130 &-0.05 & 19 & -13.800 & 0.8 \\\hline
 Gl 127.1A & 47.629 & -68.600 & 0.037 &-0.103 & 10.118 & 33.8 & 0.63 \\\hline
 Gl 445 & 176.922 & 78.691 & 0.744 &0.478 & 5.396 & -119 & 0.24 \\\hline
 Barnard's Star & 269.452 & 4.693 & -0.799 & 10.277 & 1.834 & -110.6 & 0.16 \\\hline
 Gl 217.1 & 86.739& -14.822 & -0.015 & -0.001 & 21.5 & 22.35 & 2.0 \\\hline
 Gl 729 & 282.456 & -23.836 & 0.637 & -0.192& 2.97 & -10.7 & 0.17\\\hline
 GJ 2046 & 88.518 & -60.023 & -0.052 & -0.060 & 12.83 & 30 & 0.75\\\hline
 Gl 54.1 & 18.128  &-16.999 & 1.209 & 0.641 & 3.7 & 28.2 & 0.085\\\hline
 LP 816-60 & 313.138 & -16.975 & -0.338 & 0.069 & 5.500 & 15.8 & 0.19\\\hline
\end{tabular}
\end{center}
\caption{{\small Parameters of some nearest stars studied in this thesis.}}
\label{t1.2}
\end{table}
By using Table \ref{t1.2} we compute $l,b, \mu_{b}$ and $\mu_{l}\cos(b)$. These values are given in Table \ref{t1.22}. These measurements allow us to 
compute the respective position and velocity vectors in galactic coordinates.
\begin{table}[h!]
\begin{center}
  \begin{tabular}{|c|c|c|c|c|}
	\hline
  Star & $l$ [$^\circ$] & $b$ [$^\circ$] &$\mu_{l}$& $\mu_{l}\cos(b)$ \\
	\hline \hline
 Gl 710 & 28.575&6.125&0.112&-0.053 \\\hline
 Gl 127.1A & 287.181&-43.781&0.160&0.053\\\hline
 Gl 445 & 127.85&37.998&-0.197&-0.222\\\hline
 Barnard's Star & 31.996 & 14.061 & 1.654 & 9.095 \\\hline
 Gl 217.1 & 220.381 & -20.832 & -0.013 & 0.005 \\\hline
 Gl 729 & 12.316 & -10.303 & -0.528 &-0.170\\\hline
 GJ 2046 & 269.797 & -30.382 & -0.005 &0.069\\\hline
 Gl 54.1 &277.271 & -78.715 & -1.855 & -4.483\\\hline
 LP 816-60 & 31.197 & -34.218 & 0.270 & 0.205\\\hline
\end{tabular}
\end{center}
\caption{{\small Computed galactic coordinates}}
\label{t1.22}
\end{table}
\newpage
\section{Non-interacting calculations }
In this section we present results for the non-interacting system. We use equations described in Ch.~\ref{chap.nonint}.
\begin{table}[ht]
\begin{center}
  \begin{tabular}{|c|c|c|c|}
	\hline
  star & ${\bf q}$ [$10^{17}\,m$] & $|{\bf q}|=b$ [$10^{17}\,m$] \\
	\hline \hline
 Gl 710 & (2.366, -0.111, 2.821) & 3.683  \\\hline
 Gl 127.1A & (-0.312, 0.262, -0.601) & 0.726  \\\hline
 Gl 445 & (0.033, 0.087, -0.051) & 0.106 \\\hline
 Barnard's Star & (0.011, 0.319, 0.094) & 0.326 \\\hline
 Gl 217.1 & (0.211, 0.001,0.350)  &0.409   \\\hline
 Gl 729 & (0.262,-0.080,-0.466) & 0.540 \\\hline
 GJ 2046 & (-0.535, -0.0796, -0.007) & 0.541 \\\hline
 Gl 54.1 & (0.354, -0.287, -0.983) & 1.083 \\\hline
 LP 816-60 & (0.073, -0.297, -0.403) & 0.506 \\\hline
\end{tabular}
\end{center}
\caption{{\small Computed impact parameters and perihelion position vectors for the non-interacting system}}
\label{t2.2}
\end{table}
In Tables ~\ref{t2.2} we show the computed the perihelion position vectors in SI units. Tab.~\ref{t2.22} shows these calculations in parsecs. For this approach, 
we can see that the Gl 445 will have the closest approach. This results is totally different in compression with results of Dybczy\'{n}ski (2006) \cite{comet}. 
Only computations for Barnard's, Gl 217.1, Gl 729 and for GJ 2046 are relatively similar to the presented by Dybczy\'{n}ski (2006) in \cite{comet}.
\begin{table}[ht]
\begin{center}
  \begin{tabular}{|c|c|c|c|}
	\hline
  star  & $|{\bf q}|=b$ [$pc$]& $r_{0}$ \\
	\hline \hline
 Gl 710       & 11.936 & 19.000\\\hline
 Gl 127.1A    & 2.353  &10.120\\\hline
 Gl 445       & 0.344  & 5.400\\\hline
 Barnard's Star &1.058 & 1.830\\\hline
 Gl 217.1     & 1.325  &21.500\\\hline
 Gl 729       & 1.751  & 2.970\\\hline
 GJ 2046      &1.753   &12.830\\\hline
 Gl 54.1      & 3.512  & 3.700\\\hline
 LP 816-60    & 1.641  &5.500\\\hline
\end{tabular}
\end{center}
\caption{{\small Computed impact parameters and perihelion distances for the non-interacting system in parsecs, respectively. $r_{0}$ represents the  initial distance of the object with respect to the Sun.}}
\label{t2.22}
\end{table}
\newpage
\section{Two-body calculations}
Here, Tab.~\ref{t3.2} and Tab.~\ref{t3.22} show computed perihelion position vectors, impact parameters and dispersion angles for selected nearest stars.
These computations were carried out by using steps described in Ch.~\ref{chap.twobody}.
\begin{table}[h!]
\begin{center}
  \begin{tabular}{|c|c|c|c|c|}
	\hline
  star & ${\bf q}$ [$10^{17}\,m$] &$|{\bf q}|$ [$10^{17}\,m$]& $b$ [$10^{17}\,m$] &$\theta$ [$^\circ$] \\
	\hline \hline
 Gl 710  &(0.951, 0.288, 0.532) & 1.127 & 3.226 & 58.088 \\\hline
 Gl 127.1A &(0.012, -0.079, -0.125) & 0.148 & 1.887 & 81.140\\\hline
 Gl 445 & (0.020, -0.028, -0.025) & 0.042 & 0.114& 49.639 \\\hline
 Barnard's Star & (0.249, 0.143, 0.070) & 0.296 & 0.395 & 16.380 \\\hline
 Gl 217.1 & (0.006, 0.006,0.005) & 0.011 & 0.346& 86.521\\\hline
 Gl 729 &(0.016,0.003,-0.004) & 0.017 & 0.126 & 75.365\\\hline
 GJ 2046 &(0.059, 0.014, 0.004) & 0.061 & 1.147 & 83.981 \\\hline
 Gl 54.1  &(0.421, -0.300, -0.941) & 1.073 & 1.309& 11.298\\\hline
 LP 816-60 & (0.001, 0.052, 0.051) & 0.073 & 0.475 & 73.390\\\hline
\end{tabular}
\end{center}
\caption{{\small Computed impact parameters, perihelion position vectors and dispersion angles for the two-body system}}
\label{t3.2}
\end{table}
As we expected, for the two-body system we get smaller perihelion distances than for the system without interaction. 
Here, the star with the closest approach is Gl 217.1. Comparing with the results given in \cite{comet}, only computations for Barnard's star are in agreement with \cite{comet}, but this does not mean that 
our results are wrong. It is possible that in \cite{comet} and in other papers are taken into account others phenomena in addition to the two-body problem 
that we have not included in our computations. Another reason may be the values of used parameters to compute the initial conditions. In several sources (data bases) we found 
that for a parameter of the same star are given different values.
\begin{table}[ht]
\begin{center}
  \begin{tabular}{|c|c|c|c|c|}
	\hline
  star &$|{\bf q}|$ [$pc$]& $b$ [$pc$] &$\theta$ [$^\circ$] \\
	\hline \hline
 Gl 710    & 3.652  & 10.045     & 58.088\\\hline
 Gl 127.1A & 0.480  & 6.115      & 80.341 \\\hline
 Gl 445    & 0.136  &0.371      & 49.639 \\\hline
 Barnard's Star& 0.958 & 1.281   & 16.380 \\\hline
 Gl 217.1  & 0.034 & 1.122    & 86.521\\\hline
 Gl 729    &  0.054 & 0.409   & 75.365 \\\hline
 GJ 2046   &  0.197 & 3.719    & 83.981  \\\hline
 Gl 54.1   &   3.477& 4.241    & 11.298\\\hline
 LP 816-60 &  0.235 & 1.540    & 73.390\\\hline
\end{tabular}
\end{center}
\caption{{\small Computed impact parameters, perihelion distances (in parsecs)  and dispersion angles for the two-body system}}
\label{t3.22}
\end{table}

\newpage
\section{Simple model}
In this section we show and discuss results for the perihelion distance computed by using our proposed simple model. Results are given for 
$u=0$, and for $u\neq 0$. Tab.~\ref{t4.1} shows perihelion distances for zero value of $u$. 
\begin{table}[ht]
\begin{center}
  \begin{tabular}{|c|c|}
	\hline
  star &$|{\bf q}|$ [$pc$]\\
	\hline \hline
 Gl 710   & 16.540\\\hline
 Gl 127.1A & 6.460 \\\hline
 Gl 445    & 2.735\\\hline
 Barnard's Star& 1.567 \\\hline
 Gl 217.1  & 18.112\\\hline
 Gl 729    & 1.241  \\\hline
 GJ 2046   & 8.866   \\\hline
 Gl 54.1   &2.329  \\\hline
 LP 816-60 & 2.720 \\\hline
\end{tabular}
\end{center}
\caption{{\small Computed perihelion distances (in parsecs) for our proposed simple model ($u=0$) }}
\label{t4.1}
\end{table}
As we already know, $u$ represents anharmonic oscillations 
along the $z$-axis, from where, we can say that this anharmonicity in the motion slightly acts on the computation of the perihelion distances, since the computed
perihelion distances for nonzero $u$ are slightly greater than for the case when $u=0$.
\begin{table}[ht]
\begin{center}
  \begin{tabular}{|c|c|}
	\hline
  star &$|{\bf q}|$ [$pc$]\\
	\hline \hline
 Gl 710   & 17.387\\\hline
 Gl 127.1A & 8.394 \\\hline
 Gl 445    & 3.237\\\hline
 Barnard's Star& 1.830 \\\hline
 Gl 217.1  & 20.375\\\hline
 Gl 729    & 1.925 \\\hline
 GJ 2046   &8.961   \\\hline
 Gl 54.1   &3.273  \\\hline
 LP 816-60 & 5.490\\\hline
\end{tabular}
\end{center}
\caption{{\small Computed perihelion distances (in parsecs) for our proposed simple model ($u\neq0$) }}
\label{t4.2}
\end{table}
We get greater perihelion distances, because we have not taken into account the two-body problem, and we separately investigate the role of anharmonic oscillations along 
the $z$-axis. For a better approach it is really recommended to study the role of the galactic tide and the two-body interaction. This can be carry out by 
solving Eqs.~(\ref{eq.3.3.4}). Numerical computation  may improve our presented results. 
Also, it is important to notice that our computations were done teoretically, and we used observational data to apply them, only.
Note that anharmonic oscillations are not considered in the literature, and, this may be the reason of the difference in our 
values for the perihelion distances in comparison with results given in \cite{comet}.

%% file: Chapter3.tex
\chapter{The solar motion} 
\label{chapter.5} 
Note: \emph{In this Chapter we present part of a paper in preparation. This contribution was carried out with the participation of Klacka J., Nagy R., 
{\bf Cayao J.}, Komar L. and M. Jurci.}\\

The kinematics of stars near to the Sun has long been known
to provide crucial information regarding both the structure and
evolution of the Milky Way \cite{other}.
That is the reason why
we calculate the solar motion in the reference frame
connected with the nearest stars. We identify series of $N$ stars
with heliocentric distances less than $100, 40, 15$ pc and then
determine the velocity of the Sun relative to the mean velocity of
these stars. {\it The mean motion of all stars in the volume
element cosidered is clearly the physically most meaningful when
considered in the framework of the Galaxy. A point possessing this
motion defines what is known as the Local Standard of Rest (LSR)} \cite{mihalas}. 
That is the LSR is a point in space that has a
galactic velocity equal to the average velocity of stars in the
solar neighborhood, including the Sun.

\section{The solar motion}\label{sec.22}
 We consider a set of $N$ stars. The coordinates of $i$-star in the
 equatorial coordinate system are
 $x_{i}$ $=$ $r_{i}$ $\cos \alpha_{i}$ $\cos \delta_{i}$,
 $y_{i}$ $=$ $r_{i}$ $\sin \alpha_{i}$ $\cos \delta_{i}$,
 $z_{i}$ $=$ $r_{i}$ $\sin \delta_{i}$,
 where $r_i$ is the heliocentric distance,
 $\alpha_i$ is the right ascention and $\delta_{i}$ is the declination
 of the $i$-star. The velocity of the $i$-th star with respect to the Sun is
\begin{eqnarray}\label{eq.1}
{\bf v}_{i} = {\bf V}_{i} ~-~ {\bf V}_{S} ~,
\end{eqnarray}
where ${\bf V}_{i}$ and ${ \bf V}_{S}$ are the velocities of the
$i-$th star and the Sun in the frame of the LSR. The velocities
${\bf V}_{i}$, ${\bf V}_{S}$ and ${\bf v}_{i}$ can be written as:
\begin{eqnarray}\label{eq.2}
{\bf V}_{S} &=& ( X_{S}, Y_{S}, Z_{S} ) ~,
\nonumber \\
{\bf V}_{i} &=& ( X_{i}, Y_{i}, Z_{i} ) ~,
\end{eqnarray}
and,
\begin{eqnarray}\label{eq.3}
v_{x, i} &=& \dot{r}_{i} ~\cos \alpha_{i} ~\cos \delta_{i}
\nonumber \\
& & -~ 4.74 ~r_{i} \left \{  ( \mu_{\alpha, i} \cos \delta_{i} ) ~\sin \alpha_{i}
+~ \mu_{\delta,i} ~\cos \alpha_{i} ~\sin \delta_{i} \right \} ~,
\nonumber \\
v_{y,i} &=& \dot{r}_{i} ~\sin \alpha_{i} ~\cos \delta_{i}
\nonumber \\
& & +~ 4.74 ~r_{i} \left \{ ( \mu_{\alpha,i} ~\cos \delta_{i} )
~\cos \alpha_{i} ~-~ \mu_{\delta,i} ~\sin \alpha_{i} ~\sin \delta_{i} \right \} ~,
\nonumber \\
v_{z,i} &=& \dot{r}_{i} ~\sin \delta_{i} ~+~ 4.74 ~r_{i} ~\mu_{\delta, i} ~\cos \delta_{i} ~,
\end{eqnarray}
where $\dot{r}_{i}$ is the radial velocity and $\mu_{\alpha,i}$,
$\mu_{\delta,i}$ are the proper motions in the right ascention and
declination. Since the units used in our calculations are
$[v]$ $=$ $[\dot{r}]$ $=$ $km~s^{-1}$, $ [r] = pc$, 
$[\mu_{\alpha}]$ $=$ $[\mu_{\delta}]$ $=$ $''~yr^{-1}$,
the numerical factor in Eq.~(\ref{eq.3}) is $4.74$.

Proper motions, radial velocities (calculated from redshift),
and heliocentric distances (calculated from parallax)
are observational data and can be used to describe the solar motion.

There are a few ways to calculate the solar motion with respect to the LSR.
If we have all parameters of stars in our set, we will
find the solar motion using direct calculation.
Otherwise, we can approximate the solution using the least square method.

\subsection{Determining ${\bf V_{S}}$ by direct calculation}
\label{sec.priama}
The LSR is given by
\begin{equation}\label{eq.4}
\sum_{i=1}^{N}  ~{\bf V}_{i}  = 0 ~,
\end{equation}
where ${\bf V}_{i}$ is given by  Eq.~(\ref{eq.2}).
The average value of Eq.~(\ref{eq.1}) is
\begin{equation}\label{eq.5}
\frac{1}{N} ~\sum_{i=1}^{N} ~ {\bf v}_{i} =
\frac{1}{N} ~\sum_{i=1}^{N} ~{\bf V}_{i} ~-~ {\bf V}_{S} ~.
\end{equation}
Rewriting Eq.~(\ref{eq.5}) by using Eq.~(\ref{eq.4}) we get
\begin {equation}\label{eq.6}
\frac{1}{N}~\sum_{i=1}^{N} ~{\bf v}_{i} = - ~ {\bf V}_{S} ~, 
\end{equation}
or,
\begin {eqnarray}\label{eq.7}
X_{S} &=& -~ \frac{1}{N} ~\sum_{i=1}^N ~\dot{r}_{i} ~\cos \alpha_{i} ~ \cos \delta_{i}  
\nonumber \\ 
& & +~ 4.74 ~\frac{1}{N}~\sum_{i=1}^N ~r_{i} ~ ( \mu_{\alpha,i} ~\cos \delta_{i} ) ~\sin \alpha_{i}  
\nonumber \\
& & +~ 4.74 ~\frac{1}{N} ~\sum_{i=1}^N ~r_{i} ~ \mu_{\delta,i} ~ \cos \alpha_{i} ~\sin \delta_{i}~,
\nonumber \\
Y_{S} &=& - ~ \frac{1}{N} ~\sum_{i=1}^N ~ \dot{r}_{i} ~\sin \alpha_{i} ~\cos \delta_{i} 
\nonumber \\
& & -~ 4.74 ~ \frac{1}{N} ~\sum_{i=1}^N{r_i} ~ ( \mu_{\alpha,i} ~\cos \delta_{i} ) ~\cos \alpha_{i} 
\nonumber \\
& & +~ 4.74 ~ \frac{1}{N} ~\sum_{i=1}^N ~r_{i} ~ \mu_{\delta,i} ~\sin \alpha_{i} ~\sin \delta_{i}~,
\nonumber \\
Z_{S} &=& -~ \frac{1}{N} ~\sum_{i=1}^N ~\dot{r}_{i} ~\sin \delta_{i} ~-~4.74~\frac{1}{N} ~ 
\sum_{i=1}^N ~r_{i} ~\mu_{\delta,i} ~\cos \delta_{i}~.
\end{eqnarray}
Previous equations uniquely determine the solar motion in respect of the LSR.

\subsection{Determining ${\bf V_S}$ by Least square method}
\label{sec.najmestv}
It is not immediately obvious from equations ~(\ref{eq.1}) and ~(\ref{eq.3}) how ${\bf V_S}$ should be determined from a given body of data. 
We address this problem for the case of radial velocities and proper motions data. By using the Least square method we find ${\bf V_{S}}$
as a generalization of the results presented by, e.g., \cite{mihalas, binney}. 
Let 3 $N$ orthogonal unit vectors
($i$ $=$ 1, 2, \ldots, $N$) are
\begin{eqnarray}\label{eq.8}
{\bf e_{r,i}} &=& ( \cos \alpha_{i} ~\cos \delta_{i}, \sin \alpha_{i} ~\cos \delta_{i}, \sin \delta_{i} ) ~,
\nonumber \\
{\bf e_{\alpha,i}}&=& ( -~\sin \alpha_{i}, \cos \alpha_{i} ,0 ) ~,
\nonumber \\
{\bf e_{\delta,i}} &=& ( - ~\cos \alpha_{i} ~\sin \delta_{i}, -~\sin \alpha_{i} ~\sin \delta_{i}, ~\cos \delta_{i} ) ~,
\end{eqnarray}
where ${\bf e_{r,i}}$ are radial vectors and ${\bf e_{\alpha,i}}$, ${\bf e_{\delta,i}}$ are vectors in the direction of the right 
ascention and declination of the $i-$th star.
We define the general unit vector:
\begin{eqnarray}\label{eq.9}
{\bf e_{i} }&=& {\bf e_{\alpha,i}} ~\cos\phi  ~\sin \theta +~ {\bf e_{\delta,i}} ~ \sin \phi ~ \sin \theta
~+~ {\bf e_{r,i}} ~\cos\theta ~,
\nonumber \\
& & \phi \in \langle 0, 2 \pi ) ~,~~~ \theta \in \langle 0, \pi \rangle ~.
\end{eqnarray}
The projection of the vector ${\bf v_{i}}$ onto the direction of the vector ${\bf e_{i}}$ is given by 
${\bf v}_{i} \cdot {\bf e_{i}}$ $=$ $({\bf V_{i}} ~-~ {\bf V_{S}})$ 
$\cdot$ ${\bf e_{i}}$. 
Therefore, using Eqs.~(\ref{eq.3}) we get:
\begin{eqnarray}\label{eq.10}
({\bf V_{i}} ~-~ {\bf V_{S}}) \cdot  {\bf e_{i}} = \dot{r}_{i} ~\cos \theta ~+~
4.74 ~r_{i} ~  [ ( \mu_{\alpha,i} ~\cos \delta_{i} ) ~\cos \phi  ~+~ \mu_{\delta,i} ~\sin \phi ] ~\sin \theta ~.
\end{eqnarray}
We know 
\begin{eqnarray}\label{eq.11}
( {\bf V_{i}} ~-~ {\bf V_{S}} ) \cdot {\bf e_{i}} &=& (X_i ~-~ X_S) ~(-~\sin \alpha_{i} ~\cos\phi ~
\sin \theta 
\nonumber \\
& & -~ \cos \alpha_{i} ~\sin \delta_{i} ~\sin \phi ~\sin \theta ~+~ \cos \alpha_{i} ~\cos \delta_{i} ~\cos \theta )
\nonumber \\
& & +~ (Y_i ~-~Y_S) ~(\cos \alpha_{i} ~\cos \phi ~\sin \theta 
\nonumber \\
& & -~ \sin \alpha_{i} ~\sin \delta_{i} ~\sin \phi ~\sin \theta 
+~\sin \alpha_{i} ~\cos \delta_{i} ~\ cos \theta ~)
\nonumber \\
& & +~( Z_i ~-~ Z_S )~ ( \cos \delta_{i} ~\sin \phi ~\sin \theta 
~+~ \sin \delta_{i} ~\cos\theta )~.
\end{eqnarray}
From equations ~(\ref{eq.10}) and ~(\ref{eq.11}) follows
\begin{eqnarray}\label{eq.12}
p_{i} & \equiv&  \dot{r_i} ~\cos \theta ~+~4.74 ~r_i  [ (\mu_{\alpha,i}~\cos \delta_{i} ) ~ \cos \phi ~+~ \mu_{\delta,i}~\sin \phi ] ~\sin \theta
\nonumber\\
& & -~(X_i~-~X_S)~(-~\sin \alpha_{i} ~\cos \phi ~\sin \theta ~-~\cos \alpha_{i} ~\sin \delta_{i} 
~\sin \phi ~\sin \theta
\nonumber \\
& & +~\cos \alpha_{i} ~\cos \delta_{i} ~\cos\theta ) 
\nonumber \\
& & -~ ( Y_i ~-~Y_S ) ~( \cos \alpha_{i} ~\cos \phi ~\sin \theta
~-~ \sin \alpha_{i} ~\sin \delta_{i} ~\sin \phi ~\sin \theta 
\nonumber \\
& & +~ \sin \alpha_{i} ~\cos \delta_{i} ~\cos \theta )
\nonumber \\
& & -~ ( Z_{i} ~-~ Z_{S} ) ~ ( \cos \delta_{i} ~\sin \phi ~\sin \theta ~+~ \sin \delta_{i} ~\cos \theta ) ~,
\end{eqnarray}
where $p_i$ is the residual  and $p_i=0$, exactly. 
But we admit $p_i$ $\neq$ 0 and we use the Least square method for finding $X_S$, $Y_S$, $Z_S$. Defining the sum of squared residuals
\begin{eqnarray}\label{eq.13}
S (X_{S}, Y_{S}, Z_{S} ) &=& \sum_{i=1}^N [p_{i}]^{2} ~,
\end{eqnarray}
we minimize it:
\begin{eqnarray}\label{eq.14}
\frac{\partial S}{\partial X_S} = \frac{\partial S}{\partial Y_S} = \frac{\partial S}{\partial Z_S} = 0 ~.
\end{eqnarray}
We assume
\begin{eqnarray}\label{eq.15}
\langle {\bf V_{i}}~\cos^k \alpha_{i} ~\sin^l \alpha_{i}~ \cos^m \delta_{i} ~\sin^n \delta_{i} \rangle = 0 
\end{eqnarray}
for arbitrary $k,l,m,n$ \cite{mihalas}. We denote
\begin{eqnarray}\label{eq.16}
\Gamma_i &\equiv&  \dot{r}_{i} ~\cos \theta ~+~ 4.74~r_i ~ ( \mu_{\alpha,i}\cos \delta_{i} ~\cos\phi
~+~ \mu_{\delta,i} ~\sin\phi ) ~sin \theta
\nonumber \\
& & -~ X_S ~( \sin \alpha_{i} ~\cos\phi  ~\sin \theta
~+~ \cos \alpha_{i} ~ \sin \delta_{i} ~\sin \phi ~\sin \theta ~-~ \cos \alpha_{i} ~\cos \delta_{i} ~ \cos \theta)
\nonumber\\
& & +~Y_S~ ( \cos \alpha_{i} ~\cos\phi ~\sin\theta ~-~\sin \alpha_{i} ~\sin \delta_{i}  ~\sin \phi ~\sin \theta
~+~ \sin \alpha_{i} ~ \cos \delta_{i} ~\cos\theta )
\nonumber \\
& & +~ Z_S ~ ( \cos \delta_{i} ~\sin \phi ~\sin \theta ~+~\sin \delta_{i} ~\cos \theta )~.
\end{eqnarray}
Then rewriting Eq.~(\ref{eq.15}) by using Eqs. (12), (13), ~(\ref{eq.16})-~(\ref{eq.17}) we get:
\begin{eqnarray}\label{eq.17}
\frac{\partial S}{\partial X_S}&=&2~\sum_{i=1}^N ~ \Gamma_i ~( -~ \sin \alpha_{i} ~ \cos \phi ~ \sin \theta
~-~ \cos \alpha_{i} ~ \sin \delta_{i} ~\sin \phi ~\sin \theta
\nonumber\\
& & +~ \cos \alpha_{i} ~ \cos \delta_{i} ~\cos\theta ) = 0~,
\nonumber\\
\frac{\partial S}{\partial Y_S}&=& 2~ \sum_{i=1}^N ~\Gamma_i ~( \cos \alpha_{i} ~\cos \phi ~\sin \theta
~-~ \sin \alpha_{i} ~ \sin \delta_{i} ~\ sin \phi ~\sin \theta
\nonumber\\
& & +~\sin \alpha_{i} ~ \cos \delta_{i} ~\cos\theta ) = 0 ~
,\nonumber\\
\frac{\partial S}{\partial Z_S} &=& 2~ \sum_{i=1}^N~  \Gamma_i ~ ( \cos\delta_{i} ~\sin \phi ~\sin \theta
~+~\sin \delta_{i} ~\cos\theta )  = 0 ~.
\end{eqnarray}
$\phi, \theta$ are arbitrary angles, $\phi$ $\in$ $\langle 0, 2\pi )$, $\theta$ $\in$ $\langle 0,\pi \rangle$. 
Therefore, coefficients of the same combination of ($\sin \phi$, $\cos \phi$, $\sin \theta$, $\cos \theta$) are equal 
to zero for each equation in ~(\ref{eq.18}). From this we get six independent linear systems of three equations 
in unknowns ($X_{S}$,  $Y_{S}$, $Z_{S}$). These systems are used to
determine the solar motion using the Least square method and they are presented in Appendix A (Eqs. A1-A6).

Eqs. (A1) represent the way how to calculate the solar motion by using only the proper motions in right ascension. 
However, they allow us to determine the solar motion only in the $X_S$, $Y_S$ directions. 
The system of equations Eqs. (A1) is similar to the type of
Eqs. (\ref{eq.5}-\ref{eq.13})-(\ref{eq.5}-\ref{eq.14}) by \cite{mihalas}.
The system of equations Eqs. (A5) is similar to the type of
Eqs. (\ref{eq.5}-\ref{eq.10})-(\ref{eq.5}-\ref{eq.12}) in \cite{mihalas}.
Suprising is the following formulation in \cite{mihalas}:
``Finally, if we add equations (\ref{eq.5}-\ref{eq.10})
to (\ref{eq.5}-\ref{eq.13}) and (\ref{eq.5}-\ref{eq.11}) to (\ref{eq.5}-\ref{eq.14}) to utilize the proper motion information
as fully as possible, we obtain three equations of their form ...''
(equations for $X_{S}$, $Y_{S}$, $Z_{S}$). However, the access of the
authors yield result not consistent with our system represented by Eqs. (\ref{A2}).

Eqs. (\ref{A2})-(\ref{A6}) provide a complete information about the solar motion ($X_{S}$, $Y_{S}$, $Z_{S}$).
Thus, we determine the solar motion for each system.
In this Chapter, up to now we have used equatorial coordinate system with the right ascension
$\alpha$ and the declination $\delta$.
Dealing with motions in Galaxy, we will use galactic coordinate system with the galactic longitude $l$ and
the galactic latitude $b$. Those transformations are done as is described in Sec.~\ref{sec.gal}.

\section{Results}
We selected stars from the database SIMBAD \cite{simbad}.
At first, we chose stars with the heliocentric distance
less than 100 $pc$, but only stars with complete observational data
(radial velocities, proper motion, parallax). The number of stars to that
distance was 24167. Then we selected stars with the best quality index (A)
in parallax, proper motions and radial velocities. This selection
reduced the number of stars to 769. These stars are used in our calculations
of the solar motion. We remind that Eqs. (\ref{eq.3}), (\ref{eq.7}) and (\ref{eq.18}) are given in the
equatorial coordinates, but the results are shown in the galactic coordinates.

\subsection{For $r_{i}$ $<$ 100 $pc$}\label{s.1}
For the set of 769 stars with $r_{i}$ $<$ 100 $pc$
we compute the solar motion presented in Table ~\ref{t.1}.

\begin{table}[h]
\begin{center}
\begin{tabular}{|c|c|c|c|c|} \hline
 Method & $X_S~[km/s]$ & $Y_S~[km/s]$ & $Z_S~[km/s]$ & $|{\bf V_S}|~[km/s]$\\ \hline
\lower.3ex\hbox{Eqs.~(\ref{eq.7})}&\lower.3ex\hbox{7.63}&\lower.3ex\hbox{16.65}&\lower.3ex\hbox{7.43}&\lower.3ex\hbox{19.76}\\ \hline
\lower.3ex\hbox{Eqs.~(\ref{A2})}&\lower.3ex\hbox{74.7}&\lower.3ex\hbox{-34.0}&\lower.3ex\hbox{159.5}&\lower.3ex\hbox{179.4}\\ \hline
\lower.3ex\hbox{Eqs.~(\ref{A3})}&\lower.3ex\hbox{76.2}&\lower.3ex\hbox{106.6}&\lower.3ex\hbox{-0.7}&\lower.3ex\hbox{131.1}\\ \hline
\lower.3ex\hbox{Eqs.~(\ref{A4})}&\lower.3ex\hbox{59.3}&\lower.3ex\hbox{-53.4}&\lower.3ex\hbox{38.5}&\lower.3ex\hbox{88.6}\\ \hline
\lower.3ex\hbox{Eqs.~(\ref{A5})}&\lower.3ex\hbox{8.82}&\lower.3ex\hbox{17.64}&\lower.3ex\hbox{8.07}&\lower.3ex\hbox{21.31}\\ \hline
\lower.3ex\hbox{Eqs.~(\ref{A6})}&\lower.3ex\hbox{10.59}&\lower.3ex\hbox{17.86}&\lower.3ex\hbox{6.90}&\lower.3ex\hbox{21.88}\\ \hline
\end{tabular}
\end{center}
\caption{{\small The solar motion in the galactic coordinates for our sample of stars with $r_i$ $<$ 100 $pc$ calculated using 
the direct method (Eqs. (\ref{eq.7})) and the Least square method (Eqs. (\ref{A2})-(\ref{A6})).}}
\label{t.1}
\end{table}

Table ~\ref{t.1} shows that the results $\{$ $Z_S$, $V_{S}$ $\equiv$ $|{\bf V}_S|$ $\}$ 
from the direct method (without approximations) for this case are in accord with
the standard solar motion presented by \cite{simbad}
(see Table ~\ref{t.4}). The solar motions calculated from Eqs. (\ref{A1}), (\ref{A5})-(\ref{A6}) are similar to results
of the direct method. The rest of equations deduced from the Least square method (Eqs. \ref{A2}-\ref{A4})
give quite different solutions (see Table ~\ref{t.1}) from the direct calculation. Thus, these
equations are useless in determining the solar motion.

\subsection{For $r_i$ $<$40 $pc$}\label{s.2}
Here we consider stars with $r_{i}$ $<$ 40 $pc$.
This set contains 360 stars. Calculated values of the solar motion in the
galactic coordinates are shown in Table ~\ref{t.2}.

\begin{table}[h]
\begin{center}
\begin{tabular}{|c|c|c|c|c|} \hline
 Method & $X_S~[km/s]$ & $Y_S~[km/s]$ & $Z_S~[km/s]$ & $|{\bf V_S}|~[km/s]$\\ \hline
\lower.3ex\hbox{Eqs.~(\ref{eq.7})}&\lower.3ex\hbox{7.09}&\lower.3ex\hbox{18.53}&\lower.3ex\hbox{7.70}&\lower.3ex\hbox{21.28}\\ \hline
\lower.3ex\hbox{Eqs.~(\ref{A2})}&\lower.3ex\hbox{11.5}&\lower.3ex\hbox{10.0}&\lower.3ex\hbox{50.5}&\lower.3ex\hbox{52.8}\\ \hline
\lower.3ex\hbox{Eqs.~(\ref{A3})}&\lower.3ex\hbox{150.9}&\lower.3ex\hbox{-10.7}&\lower.3ex\hbox{-49.1}&\lower.3ex\hbox{159.1}\\ \hline
\lower.3ex\hbox{Eqs.~(\ref{A4})}&\lower.3ex\hbox{1678.8}&\lower.3ex\hbox{-1314.3}&\lower.3ex\hbox{-142.4}&\lower.3ex\hbox{2136.9}\\ \hline
\lower.3ex\hbox{Eqs.~(\ref{A5})}&\lower.3ex\hbox{5.92}&\lower.3ex\hbox{18.28}&\lower.3ex\hbox{10.53}&\lower.3ex\hbox{21.91}\\ \hline
\lower.3ex\hbox{Eqs.~(\ref{A6})}&\lower.3ex\hbox{13.18}&\lower.3ex\hbox{20.95}&\lower.3ex\hbox{4.37}&\lower.3ex\hbox{25.13}\\ \hline
\end{tabular}
\end{center}
\caption{{\small The solar motion in the galactic coordinates for our set of stars with $r_i$ $<$ 40 $pc$ calculated 
using the direct method (Eqs. (\ref{eq.7})) and the Least square method (Eqs. (\ref{A2})-(\ref{A6})).}}
\label{t.2}
\end{table}

The solutions of Eqs. (\ref{A2})-(\ref{A4}) are quite different from the solar motion calculated by the direct method again.
Eqs. (\ref{A5}) give correct absolute value of the velocity, but the components of the solar motion are slightly
different from our direct calculations. The solution of Eqs. (\ref{A6}) is more different than the solution of Eqs. (\ref{A5}), 
but both of them are still good estimations of the solar motion.

\subsection{For $r_{i}$ $<$ 15 $pc$}\label{s.3}
The nearest hundred stars in the solar neighborhood have the heliocentric
distance less than 15 $pc$. Our results are presented in Table ~\ref{t.3}.

\begin{table}[h]
\begin{center}
\begin{tabular}{|c|c|c|c|c|} \hline
 Method & $X_S~[km/s]$ & $Y_S~[km/s]$ & $Z_S~[km/s]$ & $|{\bf V_S}|~[km/s]$\\ \hline
\lower.3ex\hbox{Eqs.~(\ref{eq.7})}&\lower.3ex\hbox{14.16}&\lower.3ex\hbox{17.59}&\lower.3ex\hbox{7.09}&\lower.3ex\hbox{23.68}\\ \hline
\lower.3ex\hbox{Eqs.~(\ref{A2})}&\lower.3ex\hbox{-8.3}&\lower.3ex\hbox{15.7}&\lower.3ex\hbox{30.4}&\lower.3ex\hbox{35.2}\\ \hline
\lower.3ex\hbox{Eqs.~(\ref{A3})}&\lower.3ex\hbox{120.0}&\lower.3ex\hbox{-43.7}&\lower.3ex\hbox{25.3}&\lower.3ex\hbox{130.2}\\ \hline
\lower.3ex\hbox{Eqs.~(\ref{A4})}&\lower.3ex\hbox{41.77}&\lower.3ex\hbox{-56.5}&\lower.3ex\hbox{20.75}&\lower.3ex\hbox{73.3}\\ \hline
\lower.3ex\hbox{Eqs.~(\ref{A5})}&\lower.3ex\hbox{7.02}&\lower.3ex\hbox{18.42}&\lower.3ex\hbox{2.37}&\lower.3ex\hbox{19.85}\\ \hline
\lower.3ex\hbox{Eqs.~(\ref{A6})}&\lower.3ex\hbox{24.08}&\lower.3ex\hbox{20.58}&\lower.3ex\hbox{3.77}&\lower.3ex\hbox{31.9}\\ \hline
\end{tabular}
\end{center}
\caption{{\small The solar motion in the galactic coordinates for our set of stars with $r_i$ $<$ 15 $pc$ 
calculated using the direct method (Eqs. (\ref{eq.7})) and the Least square method (Eqs. (\ref{A2})-(\ref{A6})).}}
\label{t.3}
\end{table}

The solutions of Eqs. (\ref{A2})-(\ref{A4}) are useless in
comparison with the direct method (Eqs. (\ref{eq.7})). 
Eqs. (\ref{A5}) give relatively good approximation of the
solar motion in respect of this LSR. But the solution of
Eqs. (\ref{A6}) differ from the direct method
(Eqs. (\ref{eq.7})) especially in the absolute value of the velocity and
$Z_S$ component. However, these results may be influenced by using
small solar neighborhood.

\subsection{The solution of Eqs.~(\ref{eq.7}) for different distances}
Now we use the direct method (Eqs.~(\ref{eq.7})) to calculate the
solar motion for the Solar neighborhoods with various radii. The
results are presented in Table ~\ref{t.4}.

\begin{figure}[h!]
\includegraphics[width=16cm,height=12cm,bb=0 0 842 595]{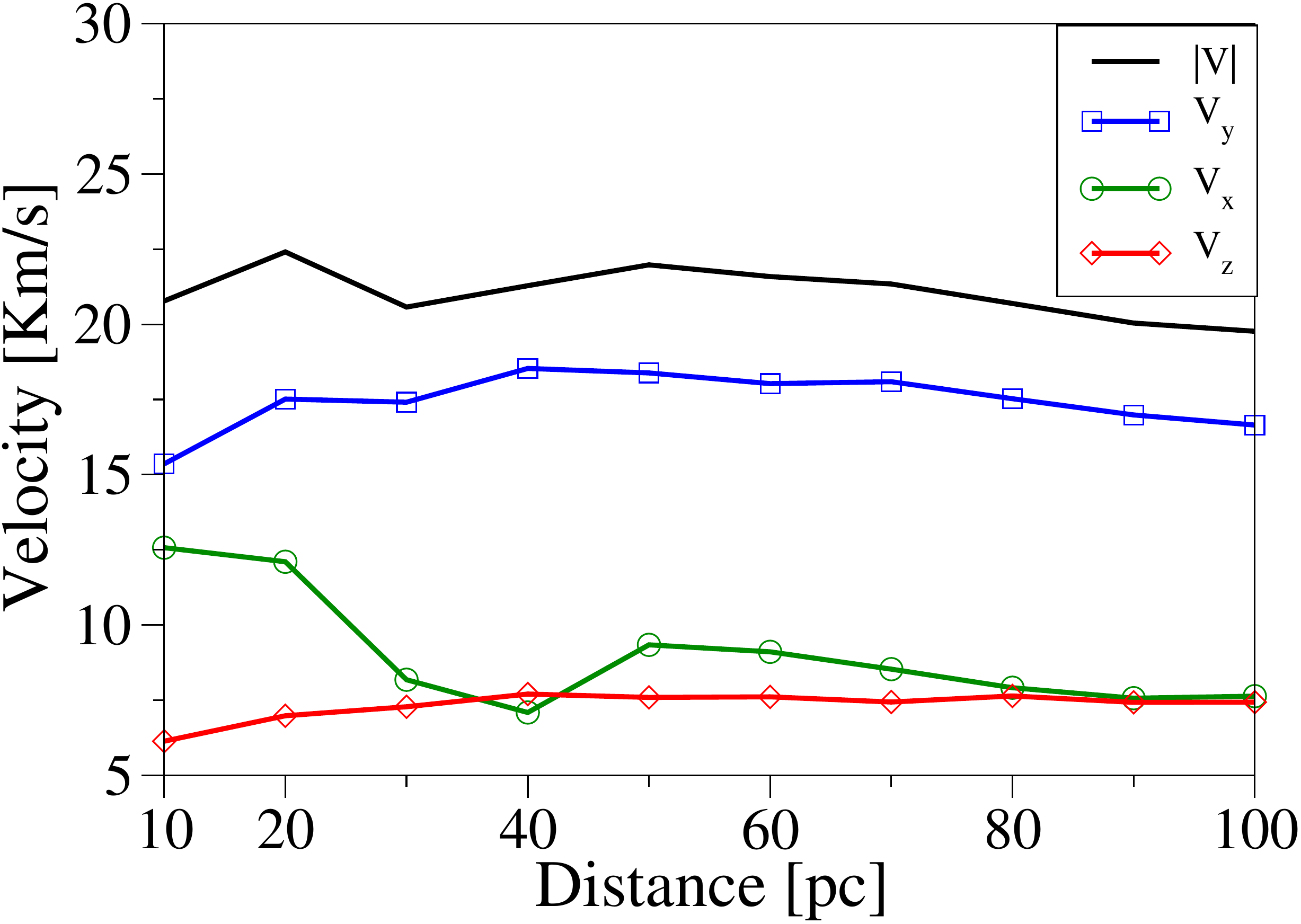}
\caption{{\small The relation between the velocity components of the solar
motion and radius of the solar neighborhood.} }
\label{fig.1}
\end{figure}

\begin{table}[h]
\begin{center} 
\begin{tabular}{|c|c|c|c|c|c|} \hline
Distance  & $X_S$ & $Y_S$ & $Z_S$ & $|{\bf V_S}|$ &Number\\
$[pc]$&$[km/s]$&$[km/s]$&$[km/s]$&$[km/s]$&of stars\\ \hline
\lower.3ex\hbox{10}&\lower.3ex\hbox{12.57}&\lower.3ex\hbox{15.36}&\lower.3ex\hbox{6.14}&\lower.3ex\hbox{20.77}&\lower.3ex\hbox{44}\\ \hline
\lower.3ex\hbox{20}&\lower.3ex\hbox{12.10}&\lower.3ex\hbox{17.51}&\lower.3ex\hbox{6.98}&\lower.3ex\hbox{22.41}&\lower.3ex\hbox{165}\\ \hline
\lower.3ex\hbox{30}&\lower.3ex\hbox{8.18}&\lower.3ex\hbox{17.41}&\lower.3ex\hbox{7.28}&\lower.3ex\hbox{20.57}&\lower.3ex\hbox{273}\\ \hline
\lower.3ex\hbox{40}&\lower.3ex\hbox{7.08}&\lower.3ex\hbox{18.53}&\lower.3ex\hbox{7.70}&\lower.3ex\hbox{21.28}&\lower.3ex\hbox{360}\\ \hline
\lower.3ex\hbox{50}&\lower.3ex\hbox{9.34}&\lower.3ex\hbox{18.38}&\lower.3ex\hbox{7.59}&\lower.3ex\hbox{21.97}&\lower.3ex\hbox{445}\\ \hline
\lower.3ex\hbox{60}&\lower.3ex\hbox{9.11}&\lower.3ex\hbox{18.03}&\lower.3ex\hbox{7.61}&\lower.3ex\hbox{21.59}&\lower.3ex\hbox{532}\\ \hline
\lower.3ex\hbox{70}&\lower.3ex\hbox{8.53}&\lower.3ex\hbox{18.09}&\lower.3ex\hbox{7.44}&\lower.3ex\hbox{21.34}&\lower.3ex\hbox{603}\\ \hline
\lower.3ex\hbox{80}&\lower.3ex\hbox{7.91}&\lower.3ex\hbox{17.52}&\lower.3ex\hbox{7.64}&\lower.3ex\hbox{20.69}&\lower.3ex\hbox{669}\\ \hline
\lower.3ex\hbox{90}&\lower.3ex\hbox{7.57}&\lower.3ex\hbox{16.99}&\lower.3ex\hbox{7.43}&\lower.3ex\hbox{20.04}&\lower.3ex\hbox{725}\\ \hline
\lower.3ex\hbox{100}&\lower.3ex\hbox{7.63}&\lower.3ex\hbox{16.65}&\lower.3ex\hbox{7.43}&\lower.3ex\hbox{19.77}&\lower.3ex\hbox{769}\\ \hline
\end{tabular}
\end{center}
\caption{{\small The solar motion calculated by direct method
(Eqs. (\ref{eq.7})) for different distances. Velocity components are given in galactic coordinates.}}
\label{t.4}
\end{table}
The relation between the velocity components of the solar motion and
radius of the solar neighborhood is illustrated in Fig.~\ref{fig.1}.
\newpage
\section{Application}
The knowledge of the solar motion, especially $Z_S$ component, allows us to calculate 
the solar oscillations in direction perpendicular to the galactic equator. 
In this case, we can use solar equation of motion presented in Kla\v{c}ka (2009) \cite{galaxy}:
\begin{eqnarray}\label{eq.19}
\ddot{z} = - ~ [ 4\pi G \rho (z)  ~+~ 2 ( A^2 ~-~B^2 ) ] ~z ~
\end{eqnarray}
where $A$, $B$ are the Oort constants and $\rho$ is the mass density
($z$ $=$ 0 in the galactic equatorial plane) in a galactocentric distance equal
to the galactocentric distance of the Sun. The relevant values are:
\begin{eqnarray}
A &=& 14.2~ km s^{-1} ~kpc^{-1} ~, 
\nonumber \\
B&=& -~12.4~  km s^{-1} ~kpc^{-1} ~, 
\nonumber \\
\rho &\equiv& \rho ( z = 0 ) = 0.13~ M_{\odot} ~pc^{-3} ~.
\end{eqnarray}
\subsection{The solar oscillation -- simple access}\label{5.1}
If the density $\rho (z)$ in Eq. (\ref{eq.19}) is a constant $\rho=\rho_{disk}+\rho_{halo}$, then  Eq. (\ref{eq.19}) 
is the equation of motion of a 
linear harmonic oscillator with angular frequency $\omega$. In our case 
$\omega ^2 = 4 \pi G \rho  + 2 (A^2 - B^2)$. The solution of Eq. (\ref{eq.19}) is:
\begin{eqnarray}\label{eq.20}
z &=& C_1 ~\cos ( \omega ~t ) ~+~ C_2 ~\sin ( \omega ~t ) ~,
\nonumber \\
C_{1} &=& z ( t = 0 ) ~, ~~~ C_{2} = \dot{z} (t=0) / \omega ~,  
\nonumber \\
z_{max} &=&  \sqrt{\left [ z(t=0) \right ]^2+\left [ \dot{z}(t=0) / \omega \right ] ^2} ~, 
\end{eqnarray}
since the constants $C_1,~ C_2$ are given by initial conditions.
Our results of the solar motion in $z-$direction are used for various initial values of $\dot{z}(t=0)$ and for $z(t=0) =$ 30 $pc$.
\begin{table}[h!]
\begin{center}
\begin{tabular}{|c|c|c|c|} \hline
&$Z_S [km/s]$&$z_{max} [pc]$&P [Myrs]\\ \hline
min. value of $Z_S$ &6.14& $78.71$ & $73.89$  \\\hline
max. value of $Z_S$ &7.70& $96.06$ & $73.89$ \\\hline
average $Z_S$ value&$7.32\pm 0.15$ & $91.79\pm1.66$ &$73.89$\\\hline
\end{tabular}
\end{center}
\caption{{\small Amplitude $z_{max}$ and periods of the solar oscillations  with the
initial position $z(t=0) =$ 30 $pc$. Various initial velocities $Z_S$ and constant mass density are used.}}
\label{t.5}
\end{table}

As we can see in Table ~\ref{t.5}, maximal distance of the Sun from
the galactic equator relevantly depends on the $z-$compontent of the
solar motion.

\subsection{The solar oscillation -- improved access}\label{5.2}
Using $Z_S$ component and a better approximation of a mass density
as a function of the coordinate $z$, $\rho (z)$ $=$ $\rho_{disk}$ ( 1
$-$ $u$ $| z |$ ) $+$ $\rho_{halo}$, $u$ $=$ 3.3 $kpc^{-1}$ (Kla\v{c}ka 2009), 
numerical calculation of the solar oscillations in
the direction perpendicular to the galactic equator (Eq.~\ref{eq.19}) yields the results
summarized in Table ~\ref{t.6}.

\begin{table}[h!]
\begin{center}
\begin{tabular}{|c|c|c|c|} \hline
&$Z_S [km/s]$&$z_{max} [pc]$&P [Myrs]\\ \hline
minimum value of $Z_S$ &$6.14$& $81.8$ & $77.0$  \\\hline
maximum value of $Z_S$ &$7.70$& $101.1$ & $77.9$ \\\hline
average $Z_S$ value&$7.32$ & $96.2$ &$77.8$\\ \hline
\end{tabular}
\caption{{\small Amplitude $z_{max}$ and periods of the solar oscillations with the
initial position $z(t=0)$ $=$ 30 $pc$. Various initial velocities
$Z_S$ are used. Mass density as a function of $z$ is considered.}}
\label{t.6}
\end{center}
\end{table}

\subsection{The Oort constants}
There are a couple of numbers $A$ and $B$ that describe the relative
orbital motions of the Sun and stars in our neighborhood of the
Galaxy. These are called Oort constants. The values of $A$ and $B$
can be calculated from the observational data based on radial
velocities and proper motions.

The original results come back to Bottlinger, who derived the
relevant equations in 1924-1925 (see \cite{kuli}).
Simplification of the Bottlinger's equations are known
as the Oort equations containing the Oort's constants $A$ and $B$.
Since the Oort's equations hold only for galactic latitude $b$ $=$ 0,, we will
consider more general equations holding for arbitrary $b$.

The following subsections present the relevant equations for finding
the values of $A$ and $B$ using various observational data. The
subsections are based on the values of the observed radial
velocities, proper motions in right ascension, and, proper motions
in declination, respectively.

\subsubsection{Radial velocities}
The radial velocity for the $i$-th star is given by
\begin{equation}
 v_{r, i} = A~r ~\cos^{2}b ~\sin ( 2~l ) ~,
\end{equation}
where $A$ is the first Oort constant. By using Eq.~(\ref{eq.1}) and
the unit vector
${\bf e}_{r, i}$ $=$ ($\cos l_{i} ~\cos b_{i}$, $\sin l_{i} ~\cos b_{i}$, $\sin b_{i}$)
we  project the vector	velocity $V_{i}$ to the radial direction as
\begin{eqnarray}
( {\bf v}_{i} + {\bf V}_S ) \cdot \frac{{\bf r}_{i} + {\bf R}_{S}}{
|{\bf r}_{i} + {\bf R}_{S}| } = A ~|{\bf r}_{i} + {\bf R}_{S} |
~\cos^{2} B_{i} ~\sin  ( 2 ~L_{i} ) ~,
\end{eqnarray}
then
\begin{eqnarray}
 ({\bf v}_{i} + {\bf V}_{S}) \cdot ( {\bf r}_{i} + {\bf R}_{S} )
 = 2 A ( {\bf r}_{i} + {\bf R}_{S} ) _{x} ( {\bf r}_{i} + {\bf R}_{S} )_{y} ~,
\end{eqnarray}
where ${\bf r}_{i}$, ${\bf v}_{i}$, and  ${\bf V}_{S}$ are known
values. But ${\bf R}_{S}$ $\neq$ 0, in general. We will assume, as it
is usual, that the LSR is at the Sun (the LSR has its origin at the
Sun's location, \cite{ryden}). Therefore
${\bf R}_{S}$ $=$ 0 and the unique unknown parameter is $A$.
 This constant is found by using the Least Square method.

\subsubsection{Proper motions $\mu_{l}$}
Here we do the same steps as for the previous case.
The velocity $v_{l, i}$ and the proper motion $\mu_{l, i}$ for the $i$-th star
is given by
\begin{eqnarray}\label{eq.bb}
v_{l,i}&=& 4.74~ r_{i}~\mu_{l,i} ~\cos b_{i} ~, 
\nonumber \\
\mu_{l,i}&=&  4.74^{-1} ~\left [ A~\cos ( 2~ l_{i} )  ~+~ B  \right ] ~,
\nonumber \\
v_{l,i}&=& [ A~ \cos ( 2~l_{i} )  ~+~ B ] ~ r_{i} ~\cos b_{i} ~,
\end{eqnarray}
where $A$ and $B$ are the Oort constants which we want to find.
Also, the velocity $v_{l,i}$ can be found by using the projection of
${\bf v}_{i}$ $+$ ${\bf V}_{S}$ onto $l$ as
\begin{eqnarray}
\label{eq.aa}
 v_{l,i} = ( {\bf v}_{i} + {\bf V}_{S} ) \cdot {\bf e}_{l, i} =
 ( {\bf v}_{i}  ~+~ {\bf V}_{S} )_{x} ~ (-~\sin l_{i}) ~+~ ( {\bf v_{i} ~+~ {\bf V}_{S}})_{y} ~( \cos l_{i} ) ~,
\end{eqnarray}
where we used ${\bf e}_{l,i}$ $=$ ( $-$ $\sin l_{i}$, $\cos l_{i}$, 0). From
equations ~(\ref{eq.bb}) and ~(\ref{eq.aa}) we can calculate the
Oort constants by using the Least Square method. Note that by using
proper motions $\mu_{l, i}$ we can find $A$ and $B$.

\subsubsection{Proper motions $\mu_{b}$}
 The velocity $v_{b, i}$ and the proper motion $\mu_{b, i}$ for the $i$-th star
is given by
\begin{eqnarray}\label{eq.cc}
v_{b,i} &=&  4.74 ~r_{i} ~\mu_{b,i} ~,
\nonumber \\
\mu_{b,i}&=& -~ 4.74^{-1} ~A ~\sin ( 2~ l_{i} ) ~\sin b_{i} ~\cos b_{i} ~,
\nonumber \\
v_{b,i}&=& -~\frac{1}{2} ~A ~r_{i} ~\sin ( 2~l_{i} )  ~\sin ( 2 ~b_{i} ) ~.
\end{eqnarray}
The velocity $v_{b,i}$ is replaced by using the projection of
${\bf v}_{i}$ $+$ ${\bf V}_{S}$ onto $b$ as
\begin{eqnarray}\label{eq.dd} 
v_{b,i} = ( {\bf v}_{i} + {\bf V}_{S}) \cdot {\bf e}_{b,i}~,
\end{eqnarray}
where ${\bf e}_{b,i}$ $= $ ( $-$ $\cos l_{i}~\sin b_{i}$, $-$ $\sin l_{i} ~\sin b_{i}$, $\cos b_{i}$) 
is the unit vector on the $b$ direction.
From equations ~(\ref{eq.cc}) and ~(\ref{eq.dd}) we  can find $A$ by using the Least Square method.

\subsection{Oort cloud of comets}
As we have already presented in sections ~\ref{5.1} and \ref{5.2}, the
found values of $Z_{S}$ lead to various values of maximal distance
between the Sun and the galactic equatorial plane. The result
suggests that $Z_{S}$ may play an important role in orbital evolution
of comets of the Oort cloud. As a consequence, the uncertainty in $Z_{S}$ generates 
an uncertainty in the mass of the Oort cloud.

\begin{figure}
\includegraphics[scale=1.5]{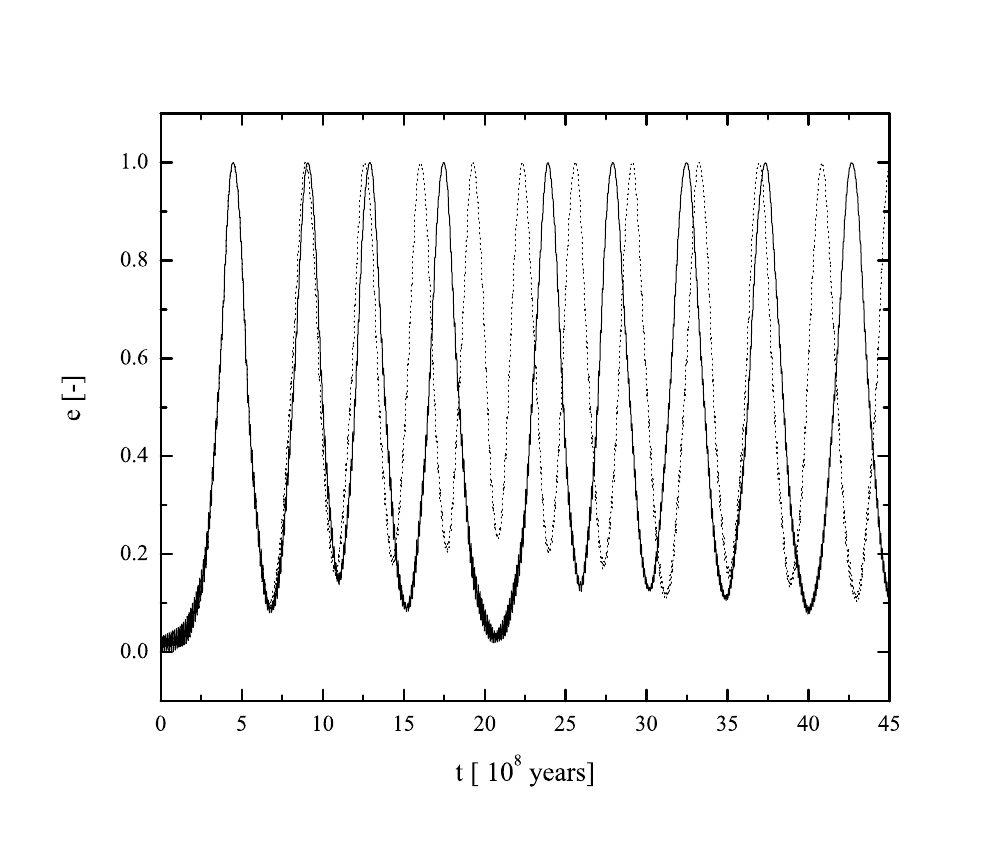}
\caption{{\small Evolution of eccentricity under the action of gravity of the Sun and Galaxy.
The model by \cite{galaxy} is used. Two values of $Z_{S}$ are used: the dotted line 
holds for $Z_{S}$ $=$ 6.14 $km/s$, the solid line holds for $Z_{S}$ $=$ 7.70 $km/s$.}}
\label{fig.2}
\end{figure}

The standard model of the Oort cloud considers that the Sun is situated
in the galactic equatorial plane and $Z_{S}$ $\equiv$ 0. Let us
consider, as an example, that a comet is characterized by the following
initial orbital elements: semi-major axis $a_{in}$ $=$ 5 $\times$ 10$^{4}$ $AU$,
eccentricity $e_{in}$ $\approx$ 0, inclination to galactic equatorial plane
$i_{in}$ $=$ 90 degrees.
The conventional models yield 6 returns of the comet into the inner part
of the Solar System during the existence of the Solar System
(see Fig. 2 in \cite{laco}- conventional models $=$ standard or simple models).
Using the physical model by \cite{galaxy}, we obtain the following number
of returns (see also Fig.~\ref{fig.2}): 
\begin{itemize}
\item 9 for $Z_{S}$ $=$ 7.70 $km/s$ ~,
\item  12 for $Z_{S}$ $=$ 6.14 $km/s$ ~.
\end{itemize}
Thus, the frequency of cometary returns into the inner part of the Solar System is 2-times
greater than the frequency of the conventional models (6 returns for the conventional models,
see \cite{laco}, if the effect of the Sun and Galaxy are considered.

\section{Discussion}
The presented method of determining ${\bf V}_{S}$ by the least
square method is based on the definition of a new general unit
vector defined in Eq. (\ref{eq.9}). Various values of
$\phi$ and $\theta$ produce six independent systems of
linear equation determining components of the vector ${\bf V}_{S}$.
None of the systems, represented by Eqs. (\ref{A1})-(\ref{A6}), contains the set
of quantities $\left \{ \dot{r}_{i}, \mu_{\alpha~i},
\mu_{\delta~i}\right \}$ simultaneously. One, or maximally two of the components 
of the set $\left \{ \dot{r}_{i}, \mu_{\alpha~i}, \mu_{\delta~i}\right \}$ 
are present in each of the
final systems. This can be explained by the usage of the least square
method. The exponent in Eq. (\ref{eq.14}) equals to 2. If the exponent
would be equal to 3, 4, ..., then also a set of equations with all three
measured quantities $\dot{r}_{i}$, $\mu_{\alpha~i}$, $\mu_{\delta~i}$
would exist.

Mihalas and McRae Routly \cite[p.~97]{mihalas} state that ``Finally, if we
add equations (\ref{eq.5}-\ref{eq.10}) to (\ref{eq.5}-\ref{eq.13}) and (\ref{eq.5}-\ref{eq.11}) to (\ref{eq.5}-\ref{eq.14}) to utilize the
proper motion information as fully as possible, we obtain three
equations of their form ...'' (equations for $X_{S}$, $Y_{S}$,
$Z_{S}$). This access of the authors yield result not consistent
with our Eqs. (\ref{A2}). It is not possible to
combine various equations in an arbitrary manner unless the equations are exact.
Our access represented by Eqs. (\ref{eq.8})-(\ref{eq.16})
yields combination of the measured
quantities $\dot{r}_{i}$, $\mu_{\alpha~i}$, $\mu_{\delta~i}$ in a
unique way.

As Tables ~\ref{t.1}-~\ref{t.3}  
show, the systems of equations
Eqs. (\ref{A2}) - (\ref{A4}) do not yield results consistent with the direct method represented by
Eqs. (\ref{eq.6})-(\ref{eq.7}). 

On the other side, the systems of Eqs. (\ref{A1}), (\ref{A5}) and (\ref{A6}) produce results
which are much better consistent with the direct method.
As Tables ~\ref{t.1}-~\ref{t.3} show, the error is less than 90\%. If we compare
Eqs. (\ref{A1})-(\ref{A6}), then we can find one important property.
Eqs. (\ref{A1}), (\ref{A5})-(\ref{A6}) contain parts where all the terms of the sums
are of the same sign for all stars. Thus, these terms dominate 
in Eqs. (\ref{A1}), (\ref{A5})-(\ref{A6}). Moreover, these terms ensure that the systems
(\ref{A1}), (\ref{A5})-(\ref{A6}) can produce Eqs. (\ref{eq.7})  [a simple addition of the $j-$th equations
of the systems produces the $j-$th equation of Eqs. (\ref{eq.7}), $j$ $=$ 1, 2, 3].
Nothing like this exists for Eqs. (\ref{A2})-(\ref{A4}). This could explain the poor consistency   
of the results of Eqs. (\ref{A2})-(\ref{A4}) with the direct method represented by Eqs. (\ref{eq.7}), 
and, a much better consistency of the results of  Eqs. (\ref{A1}), (\ref{A5})-(\ref{A6}) with the 
direct method. In any case, our results show that the Least square method is not
a good method for finding the solar motion.

The results for the direct method are summarized in Table ~\ref{t.4}. On
the basis of Table ~\ref{t.4} we obtain the following results:
\begin{eqnarray}\label{eq.new}
Z_{S} &=& (7.32 \pm 0.15) ~km s^{-1} ~,
\nonumber \\
V_{S} &=& (21.04 \pm 0.26) ~km s^{-1} ~.
\end{eqnarray}
Our values correspond to the standard solar motion. The value of
$Z_{S}$ is consistent with the results published by other authors
(see values in Table ~\ref{t.7}). However, our value of $V_{S}$ is greater than
the values of other authors and presented in Table ~\ref{t.7}. Especially the
value presented by Binney and Merrifield \cite[p.~624--628]{binney} and
Dehnen and Binney \cite{other} is very small.

\begin{table}[!h]
\begin{center}\footnotesize
\begin{tabular}{|c|c|c|c|c|c|c|} \hline
source: &SM1&SM2&SM3&SM4 \\ \hline
$Z_S [km/s]$ & $7.3$& $6.0$  & $7.17\pm0,09$ & $7.2\pm0.4$  \\ \hline
$|{\bf V_S}|~[km/s]$ & $19.5$& $15.4$ & $13.4$ & $13.4$ \\ \hline
\end{tabular}
\caption{{\small The solar motion.
SM1 corresponds to the standard motion presented by
\cite{mihalas,ryden}.
SM2 corresponds to the basic motion presented by
\cite{mihalas}.
SM3 is presented by \cite{other}.
SM4 is presented by \cite{binney} and \cite{choudhuri}.}}
\label{t.7}
\end{center}
\end{table}

\subsection{More on the Least square method}
We have already discussed the approach of the Least square method. 
We have pointed out errors of the results for the systems represented by Eqs. (\ref{A1}), (\ref{A5})-(\ref{A6}). 
Moreover, Eqs. (\ref{A2})-(\ref{A4}) do not yield satisfactory results. In general, we can conclude
that one should not use the Least square method. However, let us look in a more detail
into the method. The fundamental approach is given by Eqs. (\ref{eq.9})-(\ref{eq.15})
and the result is represented by Eqs. (\ref{A1})-(\ref{A6}). While Eqs. (\ref{A2})-(\ref{A4}) yield completely 
incorrect results, Eqs. (\ref{A1}), (\ref{A5})-(\ref{A6}) yield results with errors less than $\approx$ 100\%.
So, the idea is that we do not need to solve the complete systems (A1), (\ref{A5})-(\ref{A6}), but it
is sufficient to take into account the most important parts on the right-hand sides of 
Eqs. (\ref{A1}), (\ref{A5})-(\ref{A6}). In doing this, we can substitute the terms at $X_{S}$, $Y_{S}$ and $Z_{S}$
of the right-hand sides of Eqs. (\ref{A1}), (\ref{A5})-(\ref{A6}) by the the average values of the type
\begin{eqnarray}
\frac{1}{N} ~\sum_{i=1}^{N} ~ 
\sin^{k} \alpha_{i} ~\cos^{l} \alpha_{i} ~\cos^{m} \delta_{i} ~ \sin^{n} \delta_{i} &\rightarrow& 
\frac{1}{4~\pi} ~\int_{4 \pi} ~ \sin^{k} \alpha ~\cos^{l} \alpha ~\cos^{m} \delta ~ 
\sin^{n} \delta ~d \Omega ~, 
\nonumber \\
d \Omega &=& \cos \delta ~d \delta ~d \alpha ~, 
\nonumber \\
& & \alpha \in \langle 0, 2 ~\pi )~, ~~~~\delta \in \langle -~\frac{\pi}{2}, +~\frac{\pi}{2} \rangle ~.
\end{eqnarray}    
Eqs. (\ref{A1}) reduce to
\begin{equation}\label{eq:lsm1}
X_S = 2~\frac{1}{N} ~4.74 ~\sum_{i=1}^{N} ~ r_{i} ~ ( \mu_{\alpha, i} ~ \cos \delta_{i} ) ~ \sin \alpha_{i}~,
\end{equation}
\begin{equation}\label{eq:lsm2}
Y_S = -~ 2 ~\frac{1}{N} ~4.74 ~\sum_{i=1}^{N} ~ r_{i} ~ ( \mu_{\alpha, i} ~ \cos \delta_{i} ) ~ \cos \alpha_{i}~.
\end{equation}
Eqs. (\ref{A5}) reduce to
\begin{equation}\label{eq:lsm3}
X_S = 6 ~\frac{1}{N} ~4.74 ~\sum_{i=1}^{N} ~ r_{i} ~ \mu_{\delta, i} ~ \sin \delta_{i}  ~ \cos \alpha_{i} ~,
\end{equation}
\begin{equation}\label{eq:lsm4}
Y_S = 6 ~\frac{1}{N} ~4.74 ~\sum_{i=1}^{N} ~ r_{i} ~ \mu_{\delta, i} ~ \sin \delta_{i}  ~ \sin \alpha_{i} ~,
\end{equation}
\begin{equation}\label{eq:lsm5}
Z_S = -~ \frac{3}{2} ~\frac{1}{N} ~4.74 ~\sum_{i=1}^{N} ~ r_{i} ~ \mu_{\delta, i} ~ \cos \delta_{i} ~,
\end{equation}
and, Eqs. (\ref{A6}) reduce to
\begin{equation}\label{eq:lsm6}
X_S = -~3 ~\frac{1}{N} ~\sum_{i=1}^{N} ~ \dot{r}_{i} ~ \cos \alpha_{i} ~ \cos \delta_{i} ~,
\end{equation}
\begin{equation}\label{eq:lsm7}
Y_S = -~3 ~\frac{1}{N} ~\sum_{i=1}^{N} ~ \dot{r}_{i} ~ \sin \alpha_{i} ~ \cos \delta_{i} ~,
\end{equation}
\begin{equation}\label{eq:lsm8}
Z_S = -~3 ~\frac{1}{N} ~\sum_{i=1}^{N} ~ \dot{r}_{i} ~ \sin \delta_{i} ~.
\end{equation}
Eqs. (\ref{eq:lsm1})-(\ref{eq:lsm8}) can be considered as a more straightforward approximation
to the direct calculation represented by Eqs.~(\ref{eq.7}).
Numerical calculations for the set of our stars with $r_{i}$ $<$ 100 $pc$, $i$ $=$ 1 to $N$,
yield the following results in the galactic coordinates: \\
$X_{S}$ (Eq. \ref{eq:lsm3}) $=$ 8.48 $km/s$,  
$Y_{S}$ (Eq. \ref{eq:lsm4}) $=$ 17.42 $km/s$,
$Z_{S}$ (Eq. \ref{eq:lsm5}) $=$ 6.89 $km/s$, with the value $V_{S}$ $=$ 20.56 $km/s$, and, \\ 
$X_{S}$ (Eq. \ref{eq:lsm6}) $=$ 10.08 $km/s$,  
$Y_{S}$ (Eq. \ref{eq:lsm7}) $=$ 17.87 $km/s$,  
$Z_{S}$ (Eq. \ref{eq:lsm8}) $=$ 8.54 $km/s$ with the value $V_{S}$ $=$ 22.22 $km/s$. \\
Of course, one may take various combinations of $X_{S}$, $Y_{S}$ and $Z_{S}$
in calculating $V_{S}$ $=$ $\sqrt{X_{S}^{2} + Y_{S}^{2} + Z_{S}^{2}}$. 
The values of the $X_{S}$, $Y_{S}$ and $Z_{S}$ components can be considered to be consistent with the correct 
values given in Table ~\ref{t.1} (Eq.~(\ref{eq.7} in Table ~\ref{t.1}). The found values of $V_{S}$ are greater than 
the values of Dehnen and Binney \cite{other}, Binney and Merrifield \cite[p.~628]{binney} and
Mihalas and McRae Routly \cite[p.~101]{mihalas}. However, the values of $V_{S}$ are less than the
speed of the neutral interstellar gas \cite{springerlink:10.1007/BF00170822, springerlink:10.1007/BF00170815}.
In any case, the systems of
Eqs. (\ref{A1}), (\ref{A5})-(\ref{A6}) yields that the most probable values should be averaged in the following form: \\      
$X_{S}$ $=$ (1/2) ~$X_{S}$ (Eq. \ref{eq:lsm1}) $+$ (1/6) $X_{S}$ (Eq. \ref{eq:lsm3}) $+$ (1/3) $X_{S}$ (Eq. \ref{eq:lsm6}), \\
$Y_{S}$ $=$ (1/2) ~$Y_{S}$ (Eq. \ref{eq:lsm2}) $+$ (1/6) $Y_{S}$ (Eq. \ref{eq:lsm4}) $+$ (1/3) $Y_{S}$ (Eq. \ref{eq:lsm7}), \\
$Z_{S}$ $=$ (2/3) ~$Z_{S}$ (Eq. \ref{eq:lsm5}) $+$ (1/3) $Z_{S}$ (Eq. \ref{eq:lsm8}). 

\subsection{More on the real motion of the Sun}
We have dealt with the solar motion. We have considered it in the conventional way, i.e., the solar motion 
has meant the motion of the Sun with respect to the surrounding stars, with respect to the LSR. 
Now, we have in disposal another approach. Interplanetary probes found flux of interstellar dust 
and gas streaming into the Solar System. 

The direct method for finding solar motion yields that the direction
of the solar motion is characterized by the ecliptic longitude 
$\lambda_{solar~ motion}$ $=$ 277.5$^{\circ}$ and latitude $\beta_{solar~ motion}$ $=$ 60.3$^{\circ}$.
Eqs. (\ref{eq:lsm3})- (\ref{eq:lsm5}) yield 
$\lambda_{solar~ motion}$ $=$ 265.2$^{\circ}$ and $\beta_{solar~ motion}$ $=$ 58.4$^{\circ}$.
Eqs. (\ref{eq:lsm6})-(\ref{eq:lsm8}) yield 
$\lambda_{solar~ motion}$ $=$ 258.2$^{\circ}$ and $\beta_{solar~ motion}$ $=$ 56.1$^{\circ}$.
The solar motion was determined by the motion of the Sun with respect 
to the surrounding stars. The obtained direction may be compared
with the direction of the interstellar gas streaming into the Solar System:
$\lambda_{0}$ $=$ 259$^{\circ}$ and $\beta_{0}$ $=$ 8$^{\circ}$ \cite{landgraf}.
Similarly, the flow of interstellar grains arrives from the direction with heliocentric
ecliptic longitude 259$^{\circ}$ and heliocentric ecliptic latitude
$+$ 8$^{\circ}$ \cite{frisch}. It is wise to stress that the motion
of the surrounding stars with respect to the Sun is given by the direction 
$\lambda$ $=$ $\lambda_{solar~ motion}$ $-$ 180$^{\circ}$ and
$\beta$ $=$ $-$ $\beta_{solar~ motion}$. 
The result [ $\lambda_{solar~ motion}$ $=$ 277.5$^{\circ}$, $\beta_{solar~ motion}$ $=$ 60.3$^{\circ}$] 
significantly differs from the direction of motion of the interstellar gas and dust
[$\lambda_{0}$ $=$ 259$^{\circ}$, $\beta_{0}$ $=$ 8$^{\circ}$].

The interplanetary dust and gas are coming from the direction 
$\lambda_{0}$ $=$ 259$^{\circ}$ and $\beta_{0}$ $=$ 8$^{\circ}$.
This direction corresponds to the values of the galactic latitude
$l_{0}$ $=$ 8.9$^{\circ}$ and the galactic latitude $b_{0}$ $=$ 13.5$^{\circ}$.
Thus, the velocity components of the Sun with respect to the dust and gas
are given by the equations 
$V_{x}$ $=$ $V_{S}$ $\cos l_{0}$ $\cos b_{0}$, 
$V_{y}$ $=$ $V_{S}$ $\sin l_{0}$ $\cos b_{0}$,
$V_{z}$ $=$ $V_{S}$ $\sin b_{0}$ and the values are
presented in Table ~\ref{t.8}. The values significantly differ from the values presented in Table ~\ref{t.8} 
and from the conventional values presented in Table ~\ref{t.7}. Only $z-$component
of the solar motion is characterized by a consistency among the results obtained from
various approaches, $V_{z}$ $\equiv$ $Z_{S}$ $\in$ $\langle$ 6.1 $km/s$, 7.7 $km/s$
$\rangle$ (direct method; another presented result is $Z_{S}$ (Eq. \ref{eq:lsm8}) $=$ 8.54 $km/s$) . 

\begin{table}[!h]
\begin{center}\footnotesize
\begin{tabular}{|c|c|c|c|} \hline
$V_{S} [km/s]$ & $V_{x} [km/s]$ &  $V_{y} [km/s]$ & $V_{z} [km/s]$   \\ 
\hline \hline
26.0 & 25.0 & 3.9 & 6.1 \\ 
\hline
28.0 & 26.9 & 4.2 & 6.5 \\ 
\hline
\end{tabular}
\caption{{\small Velocity components of the Sun
with respect to the surrounding interstellar gas and dust.
Two values of the speed $V_{S}$ are considered.
The components are measured in galactic coordinates.}}
\label{t.8}
\end{center}
\end{table}

Let us consider that the dust and gas components of matter are situated
in the galactic disk. We can determine the motion of the Sun with respect to
these two components.  This is the relevant solar motion in the direction normal to the
galactic equatorial plane. So, also on the basis of the values presented in Table ~\ref{t.8},
we can obtain an orbital evolution of a comet from the Oort cloud as it is presented
in Fig.~\ref{fig.1}. However, the lower value of $V_{z}$ has to be preferred, i.e.
$Z_{S}$ $\approx$ 6.1 $km/s$ and not $Z_{S}$ $\approx$ 7.7 $km/s$.

\begin{figure}
\includegraphics[scale=1.5]{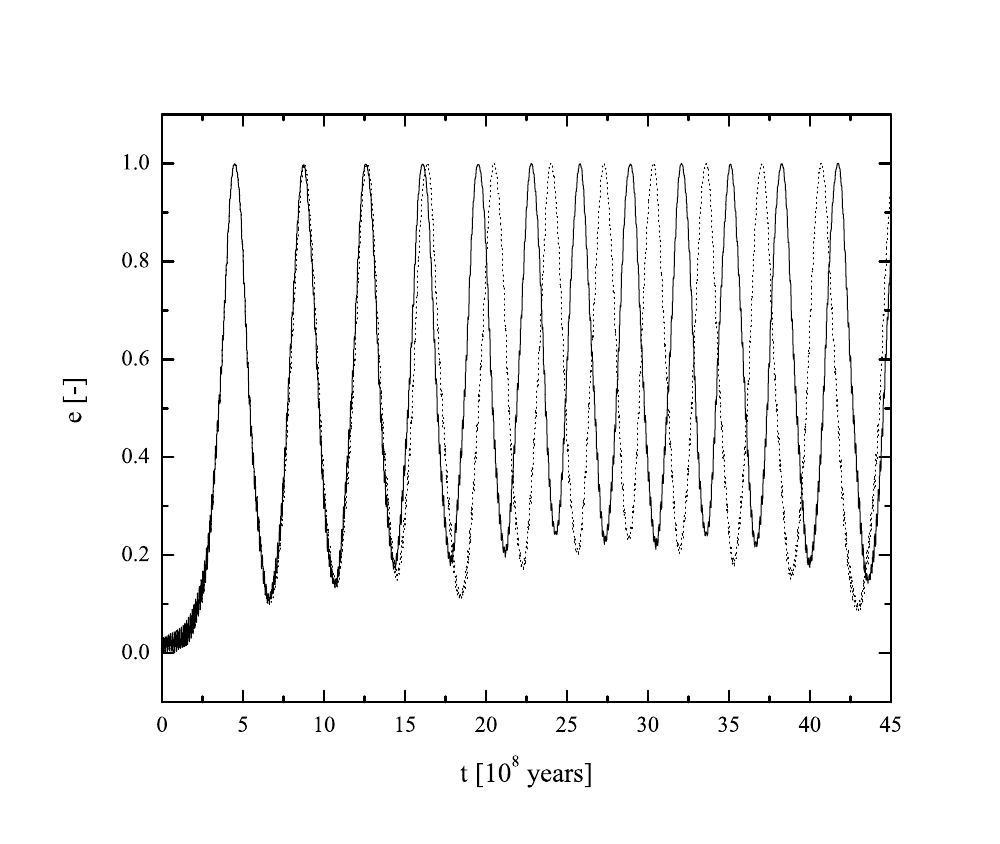}
\caption{{\small Evolution of eccentricity under the action of gravity of the Sun and Galaxy.
The model by Kla\v{c}ka (2009) is used. Two values of $Z_{S}$ are used: the dotted line 
holds for $Z_{S}$ $=$ 6.14 $km/s$, the solid line holds for $Z_{S}$ $=$ 7.70 $km/s$.}}
\end{figure}

%% file: chapter7.tex
\chapter{Conclusions}
\label{chapter.6}
In this thesis we presented three ways to look at perihelion distances and we also presented a detailed analysis of 
the solar motion. In principle, these results can be useful for estimate the enhance of the number of comets entering 
the inner part of the Solar System. 

First, in Chapter \ref{chap.nonint} we showed that this approach gives results which are in agreement
with the literature (see \cite{comet}) for Barnard's (within 7.7 \% ), Gl 217.1 (within 0.7\%), Gl 729 
(within  14.0\%) and GJ 2046 (within 12.7\%) stars, only. Second, in Chapter \ref{chap.twobody} we consider 
the motion of the object in the gravitational field of the Sun as the two-body problem, and as we expected, in this 
case we obtain less values of perihelion distances than for the non-interacting approach. Third, in 
Chapter \ref{chapter.2} we proposed a simple model to describe the relative motion Sun-object, where in addition 
to the gravitational effects of the Galaxy, we consider oscillations (including anharmonic) 
of the Sun and the object with respect to the galactic equatorial plane. 

We solved analytically equation of motion for this anharmonic oscillations. 
Since anharmonic oscillations are not considered in the literature, 
it is reasonable to expect greater perihelion distances for selected stars \cite{comet, garcia}. 
However, for stars Gl 729 (within 37.0\%), Gl 54.1 (within 4.0\%) and LP 816-60 (within 9.5 \%) 
this simple model gives reasonable values in comparison with the literature.
This simple model may be improved by considering both the galactic effects and the two-body problem Sun-object.

In Chapter \ref{chapter.5} we presented a detailed study of the solar motion. 
We have calculated the solar motion with respect to three different
sets of stars using direct calculation and the approximative Least
square method. We can see in subsections \ref{s.1}, \ref{s.2} and
\ref{s.3} that, the solar motion depends on the number of stars in
our sample and their heliocentric distances. Our results correspond
to the standard solar motion.

Our Least square method offered six independent systems of linear
equations for the components $X_{S}$, $Y_{S}$ and $Z_{S}$ of the
solar motion. The systems containing two of the three velocity
components of the stars yield results for $X_{S}$, $Y_{S}$ and
$Z_{S}$ not consistent with the values found by the direct method.
However, also the systems of linear equations containing only one of the
observed velocity components (radial velocity of the $i-$th star,
its proper motions $\mu_{\alpha, i}$ and $\mu_{\delta, i}$) yield result
which may differ in more than 80\% from the direct method.
This is probably caused by the approximation given by Eq.~(\ref{eq.16}). 
However, the same approximation yields acceptable results for the cases when
only one of the velocity components of the stars (either radial
velocity, or proper motion in right ascension, or proper motion in
declination) is used. In any case, we can conclude that the Least square method 
is not a physical method for finding the solar motion. 

The value of $Z_{S}$ which we have found from the direct calculation
fulfills the condition $Z_{S}$ $\in$ $\langle 6.14, 7.70 \rangle$
$km~s^{-1}$. The values are consistent with the values presented by
other authors. However, our value of $V_{S}$ is larger than the
presented by other authors.

The nonzero values of $Z_{S}$ are important in determining the
largest distances of the Sun from the galactic equatorial plane. The
results are summarized in Tables ~\ref{t.5} and ~\ref{t.1}. Moreover, the values of
$Z_{S}$ are relevant in long-term evolution of comets of the Oort
cloud. Using the newest form of the galactic tides, the orbital
evolution of cometary orbital elements strongly depends on the value
of $Z_{S}$. The cometary orbital evolution significantly differs
from the results obtained by standard approach assuming a fixed
Sun's position in the galactic equatorial plane.


We have chosen only stars with the best quality index (A) in
parallax, proper motions and radial velocities \cite{simbad}, that
minimized the influences of measurement errors. As we can see in
Table ~\ref{t.1}, our results from direct calculation are in
agreement with the standard solar motion presented in Table ~\ref{t.7}. Next, 
results in Table ~\ref{t.7} obviously ensure that the solar motion depends
on the stellar type. However, we want to know the solar
motion regardless of the stellar type. Hence we did not separate
stellar types in our calculations.

The direct method for finding solar motion yields that the direction
of the solar motion is characterized by the ecliptic longitude 
$\lambda$ $=$ 277.5$^{\circ}$ and latitude $\beta$ $=$ 60.3$^{\circ}$
(based on the values of Table ~\ref{t.4} for stars with distances less than 100 $pc$).
The solar motion was determined by the motion of the Sun with respect 
to the surrounding stars. The obtained direction may be compared
with the direction of the interstellar gas and dust streaming into the Solar System:
$\lambda$ $=$ 259$^{\circ}$ and $\beta$ $=$ 8$^{\circ}$.
The solar motion based on the surrounding stars yields $Z_{S}$
$\in$ $\langle$ 6.14 $km/s$, 7.70 $km/s$ $\rangle$. The solar motion
based on the surrounding dust and gas yields, approximately, 
$Z_{S}$ $\in$ $\langle$ 6.1 $km/s$, 6.5 $km/s$ $\rangle$.
In application to orbital evolution of a comet from the Oort cloud
we come to the conclusion that the number of cometary returns
into the inner part of the Solar System is twice the conventional value.

The knowledge of the solar motion allows to simulate the solar
motion with respect to galactic center. Especially $Z_S$, which
represents the solar motion in direction perpendicular to the
galactic equator, is used as a starting condition in simulation of
solar oscillations.

The uncertainty in the value of $Z_{S}$, $Z_{S}$ $\in$ $\langle$ 6.14 $km/s$, 7.70 $km/s$ $\rangle$,
would lead to the uncertainty in the mass of the Oort cloud of comets.
The uncertainty of the latter is greater than the uncertainty in $Z_{S}$.
A decrease in $Z_{S}$ in 20\% leads to the decrease of mass of the Oort cloud
in more than 30\%.
However, if also the results for interstellar gas and dust are taken into account, then
the lower value of $Z_{S}$ should be preferred.
This case yields that the frequency of cometary returns into the inner part 
of the Solar System is 1.33-times greater than the number for $Z_{S} \geq$
7.3 $km.s^{-1}$ (approximately), and 2-times greater than the frequency of the 
conventional models of the galactic tides. 

%% file: resumeSK.tex
\chapter{Zhrnutie diplomovej pr\'{a}ce}
\label{chapter.7}
V tejto diplomovej pr\'{a}ci sme sk\'{u}mali met\'{o}dy na po\v{c}\'{\i}tanie perih\'{e}liov\'{y}ch vzdialenost\'{\i}  
a z\'{a}mern\'{y}ch vzdialenost\'{\i} pre vybran\'{e}
bl\'{\i}zke hviezdy. Tie\v{z} sme sa venovali v\'{y}skumu pohybu Slnka. 
Na\v{s}e teoretick\'{e} v\'{y}sledky boli aplikovan\'{e} na bl\'{\i}zke hviezdy
\v{s}tudovan\'{e} v \cite{comet, garcia}. Observa\v{c}n\'{e} d\'{a}ta sme z\'{\i}skali 
z r\^{o}znych zdrojov ako \emph{ARICNS} \cite{ari}, \emph{NStED Data Base} \cite{Stauffer}, 
\emph{SIMBAD Data Base} \cite{simbad}, a z \cite{wilson}. V nasleduj\'{u}cich sekci\'{a}ch vysvet\v{l}ujeme, 
\v{c}o sme robili v jednotliv\'{y}ch kapitol\'{a}ch tejto diplomovej pr\'{a}ci.

\section{Syst\'{e}m bez interakcie}
V tomto pr\'{\i}pade sme po\v{c}\'{\i}tali perih\'{e}liov\'{u} vzdialenos\v{t} pre hviezdu, 
ktor\'{a} sa pohybuje po priamke vzh\v{l}adom na inerci\'{a}lnu s\'{u}stavu a neinteraguje so Slnkom. 
V kapitole \ref{chap.nonint} sme op\'{\i}sali n\'{a}\v{s} postup, kde sme minimalizovali rovnicu ~(\ref{eq.3.1}). 
Po malej \'{u}prave sme dostali vz\v{t}ah ~(\ref{eq.sininteracion}), ktor\'{y} reprezentuje 
perih\'{e}liov\'{u} vzdialenos\v{t} a \v{c}as, za ktor\'{y} sa hviezda do takej polohy dostane.
Pre neinteraguj\'{u}ci syst\'{e}m plat\'{\i}, \v{z}e perih\'{e}liov\'{a} vzdialenos\v{t} a 
z\'{a}mern\'{a} vzdialenos\v{t} nadob\'{u}daj\'{u} tie ist\'{e} hodnoty.

\section{Probl\'{e}m dvoch telies}
Na v\'{y}po\v{c}et perih\'{e}liovej vzdialenosti tu uva\v{z}ujeme hviezdu pohybuj\'{u}cu sa v 
gravita\v{c}nom potenci\'{a}li Slnka. \v{C}i\v{z}e sk\'{u}mame probl\'{e}m dvoch telies. 
Pre presn\'{y} postup odpor\'{u}\v{c}ame pozrie\v{t} sa do kapitoly \ref{chap.twobody}. 
V\'{y}sledn\'{e} vz\v{t}ahy s\'{u} dan\'{e} rovnicami (\ref{eq.impactone}) 
(na po\v{c}\'{\i}tanie z\'{a}mernej vzdialenosti) a (\ref{eq.3.20.1}) 
(na po\v{c}\'{\i}tanie perih\'{e}liovej vzdialenosti). 
Tu, tak ako sme o\v{c}ak\'{a}vali, sme dostali men\v{s}ie perih\'{e}liov\'{e} vzdialenosti
ako v syst\'{e}me bez interakcie. Nasved\v{c}uje to tomu, \v{z}e po\v{c}\'{\i}tame fyzik\'{a}lne korektne.

\section{Galaktick\'{e} slapy}
Cie\v{l}om kapitoly \ref{chapter.2} je tie\v{z} po\v{c}\'{\i}ta\v{t} perih\'{e}liov\'{u} vzdialenos\v{t} 
pre vybran\'{u} hviezdu bl\'{i}zko Slnka. Na relat\'{\i}vny pohyb Slnko-hviezda v tomto pr\'{i}pade 
p\^{o}osobia gravita\v{c}n\'{e} efekty Galaxie. Uva\v{z}ujeme oscila\v{c}n\'{e} pohyby Slnka a hviezdy 
vzh\v{l}adom na galaktick\'{y} rovn\'{\i}k, vr\'{a}tane anharmonick\'{y}ch oscil\'{a}ci\'{\i}.  
\v{Z}iadne z tak\'{y}chto oscil\'{a}ci\'{\i} sa doteraz neuvažovali.
\subsection{Oscil\'{a}cie Slnka a hviezdy}
Rie\v{s}ime analytick\'{u} rovnicu pohybu Slnka a hviezdy v $z-$ smere, kde sa predpoklad\'{a}, 
\v{z}e Slnko i hviezda osciluj\'{u} okolo galaktick\'{e}ho rovn\'{i}ka. 
Teraj\v{s}ia poloha (30 pc) a r\'{y}chlost (7.3 km/s) s\'{u} \'{u}daje z literat\'{u}ry. 
Rie\v{s}enie je dan\'{e} rovnicou \ref{eq.3.3.21}.
\subsection{Jednoduch\'{y} model}
Pre u\v{l}ah\v{c}enie na\v{s}ich v\'{y}po\v{c}tov sme navrhli jednoduch\'{y} model, ktor\'{y} je
zalo\v{z}en\'{y} na tom, \v{z}e relat\'{i}vny pohyb v smeroch $x$ a $y$ 
(rovina galaktick\'{e}ho rovn\'{\i}ka) rovnomerne z\'{a}vis\'{\i} od \v{c}asu 
a v $z$ konaj\'{u} vo v\v{s}eobecnosti anharmonick\'{e} oscil\'{a}cie. Tieto predpoklady s\'{u}
op\'{\i}san\'{e} v rovniciach (\ref{eq.simple.1}). 
V po\v{c}iato\v{c}n\'{y}ch podmienkach pohybu uva\v{z}ujeme aj kone\v{c}n\'{u} r\'{y}chlos\v{t} svetla.
\section{Pohyb Slnka}
V tejto \v{c}asti sme analyzovali pohyb Slnka. 
Porovn\'{a}vali sme rozli\v{c}n\'{e} met\'{o}dy na v\'{y}po\v{c}et slne\v{c}n\'{e}ho pohybu vzh\v{l}adom na
bl\'{i}zke hviezdy (vzh\v{l}adom na LSR -Local Standard of Rest: Miestny \v{s}tandard pokoja-) 
priamou met\'{o}dou (Subsec.~\ref{sec.priama})
a met\'{o}dou najmen\v{s}\'{\i}ch \v{s}tvorcov (\ref{sec.najmestv}).

Vedomosti o pohybe Slnka n\'{a}m umo\v{z}\v{n}uj\'{u} sk\'{u}ma\v{t} dynamiku Galaxie. 
Slne\v{c}n\'{e} oscil\'{a}cie m\^{o}\v{z}u ma\v{t} svoje d\^{o}sledky
p\^{o}sobenia na Oortov oblak, \v{c}i\v{z}e každá inform\'{a}cia o tak\'{y}ch oscil\'{a}ci\'{a}ch 
je d\^{o}le\v{z}it\'{a}.  Pri or\v{c}ovan\'{\i} pohybu Slnka sme vybrali hviezdy do vzdialenosti 100 pc.
Pou\v{z}ili sme datab\'{a}zu SIMBAD \cite{simbad}. 
Hviezdy boli vybran\'{e} len tak, aby mali v\v{s}etky zn\'{a}me parametre s najvy\v{s}\v{s}ou kvalitou. 
Vyu\v{z}it\'{y}ch bolo 769 hviezd. Na v\'{y}po\v{c}et met\'{o}dou najmen\v{s}\'{\i}ch \v{s}tvorcov 
sme pou\v{z}ili zjednodu\v{s}enie dan\'{e} v \cite{mihalas}, ale n\'{a}\v{s} pr\'{\i}stup je v\v{s}eobecnej\v{s}\'{\i}
ako v \cite{mihalas}. Z\'{i}skali sme 6 nez\'{a}visl\'{y}ch line\'{a}rnych s\'{u}stav s troma rovnicami.

Met\'{o}dy boli aplikovan\'{e} na hviezdy v slne\v{c}nom okol\'{i} do 100, 40 a 15 pc. 
Konzistentn\'{e} v\'{y}sledky s priamou met\'{o}dou s\'{u} tie, kde
vystupuje $v_{r}$ a $\mu_{\delta}$. Rovnice s kombinovan\'{y}m $v_{r}$, $\mu_{\delta}$, a $\mu_{\alpha}$ 
ned\'{a}vaj\'{u} v\'{y}sledky konzistentn\'{e} s priamou met\'{o}dou. 
Tie\v{z} sme uk\'{a}zali existuj\'{u}cu z\'{a}vislos\v{t} r\'{y}chlosti slne\v{c}n\'{e}ho pohybu 
vzh\v{l}adom na LSR od ve\v{l}kosti slne\v{c}n\'{e}ho okolia (z priamej met\'{o}dy). 
V\'{y}sledn\'{e} r\'{y}chlosti s\'{u}  v intervale od 6,1 po 7,7 km/s. 
Porovnan\'{\i}m s literat\'{u}rou vid\'{\i}me, \v{z}e to, \v{c}o sa norm\'{a}lne pou\v{z}\'{\i}va ako 
r\'{y}chlos\'{t} v $z-$ smere sa nach\'{a}dza v nami n\'{a}jdenom intervale.

Presnej\v{s}ie modely galaktickej dynamiky by mali obsahova\v{t} aj oscil\'{a}cie Slnka v rovine kolmej na rovinu 
galaktick\'{e}ho rovn\'{\i}ka, ak chceme uva\v{z}ova\v{t}  aj vplyv galaktick\'{y}ch slapov na Oortov oblak. 
V z\'{a}vislosti na tom, \v{c}i uva\v{z}ujeme harmonick\'{e} alebo anharmonick\'{e} oscil\'{a}cie,
a r\^{o}zne po\v{c}iato\v{c}n\'{e} r\'{y}chlosti, dostaneme r\^{o}zne  amplit\'{u}dy oscil\'{a}ci\'{\i}. 
A ke\v{d} to aplikujeme na kom\'{e}ty, tak dostaneme rozdielny v\'{y}voj
orbit\'{a}lnych elementov kom\'{e}t. Po\v{c}et n\'{a}vratov kom\'{e}ty do slne\v{c}nej s\'{u}stavy 
za dobu jej existencie: pre r\'{y}chlos\v{t} 6,1 km/s m\'{a}me 12 n\'{a}vratov,
pre 7,7 km/s m\'{a}me 9 n\'{a}vratov, a pre nulov\'{u} r\'{y}chlos\v{t} len 6 n\'{a}vratov. 
Z toho vych\'{a}dza, \v{z}e hmotnos\v{t} Oortovho oblaku je (15-60)-kr\'{a}t men\v{s}ia ako sa predpoklad\'{a}.
V dodatku \ref{appendix.A} s\'{u} rovnice z\'{\i}skan\'{e} met\'{o}dou najmen\v{s}\'{\i}ch \v{s}tvorcov.

%% file: appa.tex
\chapter{ Equations deduced from the Least square method}
\label{appendix.A}
\setcounter{equation}{0}

For arbitrary choice of $\phi$, and $\theta$, we get

\subsection*{Coefficients at $\sin^{2} \theta ~\cos^{2} \phi$}
\begin{eqnarray}\label{A1}
4.74~\sum_{i=1}^{N} ~r_{i} ~( \mu_{\alpha, i}~\cos \delta_{i}) ~\sin \alpha_{i} &=& X_{S}~\sum_{i=1}^{N}~ \sin^{2} \alpha_{i}
\nonumber\\ 
& &~-~ Y_{S}~\sum_{i=1}^{N} ~\cos \alpha_{i} ~\sin \alpha_{i} ~,
\nonumber \\
-~4.74 ~\sum_{i=1}^{N} ~r_{i} ~( \mu_{\alpha, i} ~\cos \delta_{i}) ~\cos \alpha_{i} &=& -~X_{S} ~\sum_{i=1}^{N} ~\sin \alpha_{i} ~ 
\cos \alpha_{i}
\nonumber\\ 
& &~+~Y_{S}~\sum_{i=1}^{N}~\cos^{2} \alpha_{i} ~.
\end{eqnarray}

\subsection*{Coefficients at $\sin^{2} \theta ~\sin \phi ~\cos \phi$}
\begin{eqnarray}\label{A2}
LHS(2,1) &=& RHS(2,1) ~,
\nonumber \\
LHS(2,1) &=& 4.74 ~\sum_{i=1}^{N} ~r_{i} ~( \mu_{\alpha, i} \cos \delta_{i} )  ~\cos \alpha_{i} ~\sin \delta_{i} 
\nonumber \\
& & ~+~ 4.74~\sum_{i=1}^{N} ~r_{i} ~\mu_{\delta, i} ~\sin \alpha_{i} ~,
\nonumber \\
RHS(2,1) &=& 2~X_{S} ~\sum_{i=1}^{N} ~\sin \alpha_{i} ~\cos \alpha_{i} ~\sin \delta_{i}
\nonumber \\
& & +~ Y_{S} ~\sum_{i=1}^{N} ~\left ( \sin^{2} \alpha_{i} ~-~ \cos^{2} \alpha_{i} \right ) ~\sin \delta_{i}
\nonumber \\
& & -~Z_{S} ~\sum_{i=1}^{N} ~\sin \alpha_{i} ~\cos \delta_{i}~,
\nonumber \\
LHS(2,2) &=& RHS(2,2) ~,
\nonumber \\
LHS(2,2) &=& -~4.74 ~\sum_{i=1}^{N} ~r_{i} ( \mu_{\alpha, i} ~\cos \delta_{i} ) ~\sin \alpha_{i} ~\sin \delta_{i} 
\nonumber \\
& & +~ 4.74~\sum_{i=1}^{N}~r_{i}~ \mu_{\delta, i} ~\cos \alpha_{i} ~,
\nonumber \\
RHS(2,2) &=& -~ X_{S} ~\sum_{i=1}^{N} ~\left ( \sin^{2} \alpha_{i} ~-~ \cos^{2} \alpha_{i} \right) ~\sin \delta_{i}
\nonumber \\
& & +~2~Y_{S}~\sum_{i=1}^{N} ~\sin \alpha_{i} ~\cos \alpha_{i} ~\sin \delta_{i} 
\nonumber \\
& & -~Z_{S}~\sum_{i=1}^{N} ~\cos \alpha_{i} ~\cos \delta_{i} ~,
\nonumber \\
4.74~\sum_{i=1}^{N} ~r_{i} ~( \mu_{\alpha, i} ~\cos \delta_{i} ) ~\cos \delta_{i} &=& X_{S} ~\sum_{i=1}^{N}~\sin \alpha_{i} ~\cos \delta_{i}
\nonumber \\
& & -~Y_{S} ~\sum_{i=1}^{N}~\cos \alpha_{i} ~\cos \delta_{i}~.
\end{eqnarray}


\subsection*{Coefficients at  $\sin \theta ~\cos \theta~\sin \phi$}
\begin{eqnarray}\label{A3}
LHS(3,1) &=& RHS(3,1) 
\nonumber \\
LHS (3,1) &=& - ~\sum_{i=1}^{N} ~\dot{r}_{i} ~\cos \alpha_{i} ~\sin \delta_{i} 
\nonumber \\
& & +~  4.74 ~\sum_{i=1}^{N} ~r_{i} ~\mu_{\delta, i} ~\cos \alpha_{i}  ~\cos \delta_{i} 
\nonumber \\
RHS(3,1) &=& 2~X_{S}~\sum_{i=1}^{N}~\cos^{2} \alpha_{i} ~\sin \delta_{i} ~\cos \delta_{i}
\nonumber \\
& & +~2~Y_{S}~\sum_{i=1}^{N} ~\sin \alpha_{i} ~\cos \alpha_{i} ~\sin \delta_{i} ~\cos \delta_{i}
\nonumber \\
& & +~Z_{S}~\sum_{i=1}^{N} ~\left ( \sin^{2} \delta_{i} ~-~ \cos^{2} \delta_{i} \right ) ~\cos \alpha_{i}~,
\nonumber \\
LHS(3,2) &=& RHS(3,2) 
\nonumber \\
LHS(3,2) &=& -~ \sum_{i=1}^{N} ~\dot{r}_{i} ~\sin \alpha_{i} ~\sin \delta_{i} 
\nonumber \\
& & +~ 4.74~ \sum_{i=1}^{N} ~r_{i} ~\mu_{\delta, i} ~\sin \alpha_{i} ~\cos \delta_{i} 
\nonumber \\
RHS(3,2) &=& 2 ~X_{S} ~\sum_{i=1}^{N} ~\sin \alpha_{i} ~\cos \alpha_{i} ~\sin \delta_{i} ~\cos \delta_{i}
\nonumber \\
& & +~2 ~Y_{S} ~\sum_{i=1}^{N} ~\sin^{2} \alpha_{i} ~\sin \delta_{i} ~\cos \delta_{i} 
\nonumber \\
& & +~Z_{S} ~\sum_{i=1}^{N} ~( \sin^2 \delta_{i}  ~-~ \cos^2 \delta_{i} ) ~\sin \alpha_{i} ~,
\nonumber \\
\sum_{i=1}^{N} ~\dot{r}_{i} ~\cos \delta_{i} ~+~ 4.74 ~\sum_{i=1}^{N} ~r_{i} ~\mu_{\delta, i} \sin \delta_{i} &=& -~ X_{S} ~\sum_{i=1}^{N} 
~\left ( \cos^{2} \delta_{i} ~-~ \sin^{2}  \delta_{i} \right ) ~\cos \alpha_{i}
\nonumber \\
& & -~Y_{S}~\sum_{i=1}^{N} ~\left ( \cos^{2} \delta_{i} ~-~ \sin^{2} \delta_{i} \right ) ~\sin \alpha_{i}
\nonumber \\
& & -~2~Z_{S}~\sum_{i=1}^{N} ~\sin \delta_{i} ~\cos \delta_{i} ~.
\end{eqnarray}

\subsection*{Coefficients at $\sin \theta ~\cos \theta ~\cos\phi$}
\begin{eqnarray}\label{A4}
LHS(4,1) &=& RHS(4,1) 
\nonumber \\
LHS(4,1) &=& -~\sum_{i=1}^N ~\dot{r}_i ~\sin \alpha_i 
\nonumber \\
& & +~ 4.74~\sum_{i=1}^N ~r_{i} ~ ( \mu_{\alpha, i} ~\cos \delta_{i} ) ~\cos \alpha_{i} ~\cos \delta_{i} 
\nonumber \\
RHS(4,1) &=& 2~ X_S ~\sum_{i=1}^N ~\sin \alpha_{i} ~\cos \alpha_{i} ~\cos \delta_{i}
\nonumber \\
& & +~Y_S ~\sum_{i=1}^N ~ ( \sin^2 \alpha_{i} ~-~ \cos^2 \alpha_{i} ) ~\cos \delta_{i}
\nonumber \\
& & +~ Z_S ~\sum_{i=1}^N ~\sin \alpha_{i} ~\sin \delta_{i}~,
\nonumber \\
LHS(4,2) &=& RHS(4,2) 
\nonumber \\
LHS(4,2) &=& \sum_{i=1}^N ~\dot{r}_i ~\cos \alpha_{i} 
\nonumber \\
& & +~ 4.74 ~\sum_{i=1}^N ~r_{i} ~ ( \mu_{\alpha,i} ~\cos \delta_{i} ) ~\sin \alpha_{i} ~\cos \delta_{i}
\nonumber \\
RHS(4,2) &=&  X_S ~\sum_{i=1}^N \left ( \sin^2 \alpha_{i}  ~-~ \cos^2 \alpha_{i} \right ) ~\cos \delta_{i}
\nonumber \\
& & -~2~ Y_S ~\sum_{i=1}^N ~\sin \alpha_{i} ~\cos \alpha_{i} ~\cos \delta_{i}
\nonumber \\
& & -~ Z_S ~\sum_{i=1}^N ~\cos \alpha_{i} ~\sin \delta_{i} ~,
\nonumber \\
4.74~\sum_{i=1}^N ~r_i~ ( \mu_{\alpha,i} ~\cos \delta_{i} ) ~\sin \delta_{i} &=&
X_S ~\sum_{i=1}^N ~\sin \alpha_{i} ~\sin \delta_{i}
\nonumber \\
& &-~ Y_S ~\sum_{i=1}^N ~\cos \alpha_{i} ~\sin \delta_{i}~.
\end{eqnarray}


\subsection*{Coefficients at  $\sin^{2} \theta ~\sin^{2} \phi$}
\begin{eqnarray}\label{A5}
4.74 ~\sum_{i=1}^N ~r_i ~\mu_{\delta,i} ~\cos \alpha_{i} ~\sin \delta_{i} &=& X_S ~\sum_{i=1}^N ~\cos^2 \alpha_{i} ~\sin^2 \delta_{i}
\nonumber \\
& &+~Y_S ~\sum_{i=1}^N ~\sin \alpha_{i} ~\cos \alpha_{i} ~\sin^2 \delta_{i}
\nonumber \\
& & -~Z_S ~\sum_{i=1}^N ~\cos \alpha_{i} ~\sin \delta_{i} ~\cos \delta_{i}~,
\nonumber \\
4.74~\sum_{i=1}^N ~r_i ~\mu_{\delta,i} ~\sin \alpha_{i} ~\sin \delta_{i} &=& X_S ~\sum_{i=1}^N ~\sin \alpha_{i} ~\cos \alpha_{i} 
~\sin^2 \delta_{i}
\nonumber \\
& & +~Y_S~\sum_{i=1}^N ~\sin^2 \alpha_{i} ~\sin^2 \delta_{i}
\nonumber \\
& & -~Z_S ~\sum_{i=1}^N ~\sin \alpha_{i} ~\sin \delta_{i} ~\cos \delta_{i}~,
\nonumber \\
-~4.74~\sum_{i=1}^N ~r_{i} ~ \mu_{\delta,i} ~\cos \delta_{i} &=& -~X_S ~\sum_{i=1}^N ~\cos \alpha_{i} ~\sin \delta_{i} ~\cos \delta_{i}
\nonumber \\
& & -~Y_S ~\sum_{i=1}^N ~\sin \alpha_{i} ~\sin \delta_{i} ~\cos\delta_{i}
\nonumber \\
& & +~Z_S ~\sum_{i=1}^N ~\cos^2 \delta_i~.
\end{eqnarray}

\subsection*{Coefficients at $\cos^{2} \theta$}
\begin{eqnarray}\label{A6}
-~\sum_{i=1}^N ~\dot{r}_i ~\cos \alpha_{i} ~\cos \delta_{i} &=& X_S ~\sum_{i=1}^N ~\cos^2 \alpha_{i} ~\cos^2 \delta_{i}
~+~ Y_S ~\sum_{i=1}^N ~\sin \alpha_{i} ~\cos \alpha_{i} ~\cos^2 \delta_i
\nonumber \\
& & +~Z_S ~\sum_{i=1}^N ~\cos \alpha_{i} ~\sin \delta_{i} ~\cos \delta_{i}~,
\nonumber \\
-~\sum_{i=1}^N \dot{r}_i ~\sin \alpha_{i} ~\cos \delta_{i} & = & X_{S}~ \sum_{i=1}^N ~\sin \alpha_{i} ~\cos \alpha_{i} ~
\cos^{2} \delta_{i}
\nonumber \\
& & +~Y_S ~\sum_{i=1}^N ~\sin^{2} \alpha_{i} ~\cos^{2} \delta_{i}
~+~ Z_S~\sum_{i=1}^N ~\sin \alpha_{i} ~\sin \delta_{i} ~\cos \delta_{i}~,
\nonumber \\
-~\sum_{i=1}^N ~\dot{r}_{i} ~\sin \delta_{i} &=& X_S~\sum_{i=1}^N ~\cos \alpha_{i} ~\sin \delta_{i} ~\cos \delta_{i}
~+~ Y_S~\sum_{i=1}^N ~\sin \alpha_{i} ~\sin \delta_{i} ~\cos \delta_{i}
\nonumber \\
& & +~Z_S ~\sum_{i=1}^N ~\sin^2 \delta_{i}~.
\end{eqnarray}

\clearpage
\newpage